%% file: main.tex
\RequirePackage{fix-cm}

\documentclass[twocolumn,11pt]{svjour3}
\smartqed % flush right qed marks, e.g. at end of proof

\usepackage[switch]{lineno} 

\usepackage{setspace}
\usepackage{paralist}%
\usepackage{appendix}
\usepackage{csvsimple}
\usepackage{subcaption}
\usepackage[normalem]{ulem}
\usepackage{tikz}
\usetikzlibrary{shapes.geometric, arrows.meta, positioning, calc}
\usepackage{siunitx}
\usepackage{mathtools}
\smartqed  % flush right qed marks, e.g. at end of proof

\usepackage{graphicx}
\usepackage{framed}
\usepackage{adjustbox}
\usepackage{fancyvrb} % for Verbatim
\usepackage[dvipsnames]{xcolor}

\usepackage[T1]{fontenc}

\usepackage{url}
\usepackage{geometry}
\usepackage{amsmath,amssymb,amsfonts}

\usepackage{algpseudocode}
\usepackage{algcompatible}
\usepackage{makecell} 
\usepackage{array}
\usepackage[linesnumbered,ruled,vlined ]{algorithm2e}
\usepackage{etoolbox}
\algnewcommand{\algorithmicand}{\textbf{ and }}
\algnewcommand{\algorithmicor}{\textbf{ or }}
\algnewcommand{\algorithmicnot}{\textbf{ not }}
\algnewcommand{\NOT}{\algorithmicnot}
\algnewcommand{\OR}{\algorithmicor}
\algnewcommand{\AND}{\algorithmicand}
\algnewcommand{\var}{\texttt}

\usepackage{tikz}

\makeatletter
\newcommand*{\algrule}[1][\algorithmicindent]{%
  \makebox[#1][l]{%
    \hspace*{.1em}% <------------- This is where the rule starts from
    \vrule height .75\baselineskip depth .25\baselineskip
  }
}
\newcount\ALG@printindent@tempcnta
\def\ALG@printindent{%
    \ifnum \theALG@nested>0% is there anything to print
    \ifx\ALG@text\ALG@x@notext% is this an end group without any text?
    \else
    \unskip
    \ALG@printindent@tempcnta=1
    \loop
    \algrule[\csname ALG@ind@\the\ALG@printindent@tempcnta\endcsname]%
    \advance \ALG@printindent@tempcnta 1
    \ifnum \ALG@printindent@tempcnta<\numexpr\theALG@nested+1\relax
    \repeat
    \fi
    \fi
}

\patchcmd{\ALG@doentity}{\item[]\nointerlineskip}{}{}{} % no spurious vertical space
\makeatother
\algrenewcommand\algorithmicindent{1.1em}

\def\BibTeX{{\rm B\kern-.05em{\sc i\kern-.025em b}\kern-.08em
    T\kern-.1667em\lower.7ex\hbox{E}\kern-.125emX}}

\usepackage{orcidlink}
\usepackage{listings}
\usepackage{textcomp} % upquote
\usepackage{lstautogobble}
\usepackage[inline]{enumitem}
\usepackage[acronym]{glossaries}
\usepackage{stmaryrd}
\usepackage{isabelle-listings}
\usepackage{microtype}

\newcommand{\RC}[1]{{\sffamily #1}}

\definecolor{eclipseBlue}{RGB}{42,0.0,255}
\definecolor{eclipseGreen}{RGB}{63,127,95}
\definecolor{eclipsePurple}{RGB}{127,0,85}

\definecolor{IsabelleKeyWordBlue}{RGB}{0, 115, 173}
\definecolor{IsabelleVarBlue}{RGB}{16,12,243}
\definecolor{IsabelleGreen}{RGB}{27,148,92}

\lstdefinestyle{IsabelleStyle}{
    aboveskip=2pt,
    belowskip=2pt,
    language={Isabelle},
    breaklines=true, 
    breakatwhitespace=true, 
    mathescape=true,
    escapeinside=~~,
    showstringspaces=false,
    numbers=none, % Remove line numbers
    frame=none, 
xleftmargin=0pt, 
xrightmargin=0pt, 
     linewidth=\columnwidth
}

\newcommand{\isalogo}{\includegraphics[width=9pt]{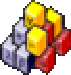}}
\newcommand{\isalink}[1]{\hfill \href{#1}{\isalogo}}
\usepackage{hyperref}
\hypersetup{
    colorlinks,
    linkcolor={red!50!black},
    citecolor={blue!50!black},
    urlcolor={blue!80!black},
    pdfborder={0 0 0}
}

\let\fun\rightarrow
\let\inj\rightarrowtail
\def\surj{\mathrel{\ooalign{$\fun$\hfil\cr$\mkern4mu\fun$}}}
\def\bij{\mathrel{\ooalign{$\inj$\hfil\cr$\mkern5mu\fun$}}}
\def\pfun{\@p\fun}
\def\pinj{\@p\inj}
\def\psurj{\@p\surj}
\def\pbij{\@p\bij}
\def\ffun{\@f\fun}
\def\finj{\@f\inj}

\usepackage{graphbox} % provide option align=c to \includegraphics
\usepackage{float} % for the H option to place the figures HERE
\usepackage{multirow} % 
\usepackage{booktabs}
\usepackage{cite}
\usepackage{tikz}
\usetikzlibrary{arrows,positioning,shapes.geometric,shapes.misc,fit}

\usepackage[none]{hyphenat} 

\usepackage{pifont}
\newcommand{\cmark}{\ding{51}} % ✓
\newcommand{\xmark}{\ding{55}} % ✗
\newcommand{\cA}{\ding{51}\textsuperscript{A}}
\newcommand{\cM}{\ding{51}\textsuperscript{M}}
\newcommand{\pA}{\textit{Partial}\textsuperscript{A}}
\newcommand{\pM}{\textit{Partial}\textsuperscript{M}}

\usepackage{tabularx}

\DeclareMathSymbol{\zseqcat}{\mathbin}{AMSa}{"61}  % \frown
\def \cat {\mathbin{\raise 0.8ex\hbox{$\mathchar\zseqcat$}}}

\usepackage[skins, breakable]{tcolorbox}
\usepackage[subfigure]{tocloft}
\usepackage{fmtcount}
\usepackage{imakeidx}
\makeindex[title=Index of Changes, name=changes]

\ifdefined\CHANGES



\newlistof{changes}{s}{\listchangename}	
\newtcolorbox{echanged}[1]{enhanced,title=#1, sharp corners,colback=blue!20,
  attach boxed title to top right=
  {xshift=0mm,yshift=0mm},
  boxed title style={size=small,colback=blue},
  size=minimal,
  finish={
    \refstepcounter{changes}
    \addcontentsline{s}{changes}
    {\protect\numberline{\thechanges}#1}
  }
}
\fi
\newcommand{\notinpaper}[1]{%
  \index[changes]{Comment \textbf{C\expandafter\sortgref#1\sortgref} not reflected in the text.\removecomma|HIDE}%
}

\newrobustcmd{\removecomma}[1]{}
\newcommand{\HIDE}[1]{}

\newcommand{\C}[1]{%
  \index[changes]{Comment \textbf{C\expandafter\sortgref#1\sortgref}|BH{\arabic{changes}}}{C#1}%
}%

\def\sortgref#1\sortgref{%
  \ignoresort{\ifnum#1<10 00\else\ifnum #1<100 0\fi\fi#1}#1%
}
\protected\def\ignoresort#1{}
\ifdefined\CHANGES	

\else
\newcommand{\listchangename}{List of Changes}
\newlistof{changes}{s}{\listchangename}

\newtcolorbox{echanged}[1]{size=minimal}	
\fi
\lstdefinelanguage{Epsilon}{
  morekeywords={
      if, then, else, operation, var, and, or, not, new,
      @lazy, rule, transform, to, self, in, for, while, void,
      int, double, removeAt, remove, add, instanceof,
  },
  sensitive=true, % keywords are not case-sensitive
  morecomment=[l]{//}, % l is for line comment
  morecomment=[is]{/*}{*/}, % s is for start and end delimiter
  morestring=[b]" % defines that strings are enclosed in double quotes
} %

\definecolor{eclipseBlue}{RGB}{42,0.0,255}
\definecolor{eclipseGreen}{RGB}{63,127,95}
\definecolor{eclipsePurple}{RGB}{127,0,85}

\definecolor{epsilonred}{rgb}{0.6,0,0} % for strings
\definecolor{epsilongreen}{rgb}{0.25,0.5,0.35} % comments
\definecolor{epsilonpurple}{rgb}{0.5,0,0.35} % keywords
\definecolor{epsilondocblue}{rgb}{0.25,0.35,0.75}

\lstdefinestyle{epsilonStyle}{
  language={Epsilon},
  basicstyle=\small\ttfamily,
  captionpos=b, % Position of the Caption (t for top, b for bottom)
  extendedchars=true, % Allows 256 instead of 128 ASCII characters
  tabsize=2, % number of spaces indented when discovering a tab
  columns=fullflexible, % make all characters equal width
  keepspaces=false, % does not ignore spaces to fit width, convert tabs to spaces
  showspaces=false,
  breaklines=true, % wrap lines if they don't fit
  frame=trbl, % draw a frame at the top, right, left and bottom of the listing
  frameround=tttt, % make the frame round at all four corners
  framesep=4pt, % quarter circle size of the round corners
  numbers=left, % show line numbers at the left
  numberstyle=\tiny, % style of the line numbers
  linewidth=1.0\linewidth,
  xleftmargin=0.5cm,
  commentstyle=\color{eclipseGreen}, % style of comments
  keywordstyle=\color{epsilonpurple}\bfseries, % style of keywords
  stringstyle=\color{eclipseBlue}, % style of strings
  upquote=true, % straight single quotes in lstlistings and need \usepackage{textcomp}
}

\lstdefinelanguage{PRISM}{
  morekeywords={
      dtmc, mdp, ctmc, const, int, float, bool, global, module, init, endmodule, true, false, double,
      rewards, endrewards, formula, label
  },
  sensitive=true, % keywords are not case-sensitive
  morecomment=[l]{//}, % l is for line comment
  morecomment=[is]{/*}{*/}, % s is for start and end delimiter
  morestring=[b]" % defines that strings are enclosed in double quotes
} %

\definecolor{eclipseBlue}{RGB}{42,0.0,255}
\definecolor{eclipseGreen}{RGB}{63,127,95}
\definecolor{eclipsePurple}{RGB}{127,0,85}
 
\lstdefinestyle{prismStyle}{
  language={PRISM},
  basicstyle=\footnotesize\ttfamily, % Global Code Style
  captionpos=b, % Position of the Caption (t for top, b for bottom)
  extendedchars=true, % Allows 256 instead of 128 ASCII characters
  tabsize=2, % number of spaces indented when discovering a tab 
  columns=fullflexible,
  keepspaces=false, % does not ignore spaces to fit width, convert tabs to spaces
  showstringspaces=false, % lets spaces in strings appear as real spaces
  breaklines=true, % wrap lines if they don't fit
  breakatwhitespace=true,
  postbreak=\mbox{\hspace*{-2em}\textcolor{red}{$\hookrightarrow$}\space},
  frame=trbl, % draw a frame at the top, right, left and bottom of the listing
  frameround=tttt, % make the frame round at all four corners
  framesep=4pt, % quarter circle size of the round corners
  numbers=left, % show line numbers at the left
  numberstyle=\tiny\ttfamily, % style of the line numbers
  numbersep=5pt,
  linewidth=1.0\linewidth,
  xleftmargin=0.2cm,
  commentstyle=\color{eclipseGreen}, % style of comments
  keywordstyle=\color{eclipsePurple}, % style of keywords
  stringstyle=\color{eclipseBlue}, % style of strings
  upquote=true, % straight single quotes in lstlistings and need \usepackage{textcomp}
}

\lstdefinelanguage{ADProperty}{
  morekeywords={
      reaches, at, terminated, successfully, failed, terminated, on, fail
  },
  sensitive=true, % keywords are not case-sensitive
  morecomment=[l]{//}, % l is for line comment
  morecomment=[is]{/*}{*/}, % s is for start and end delimiter
  morestring=[b]" % defines that strings are enclosed in double quotes
} %

\lstdefinestyle{ADPropertyStyle}{
  language={ADProperty},
  basicstyle=\footnotesize\ttfamily, % Global Code Style
  captionpos=b, % Position of the Caption (t for top, b for bottom)
  extendedchars=true, % Allows 256 instead of 128 ASCII characters
  tabsize=2, % number of spaces indented when discovering a tab 
  columns=fullflexible,
  keepspaces=false, % does not ignore spaces to fit width, convert tabs to spaces
  showstringspaces=false, % lets spaces in strings appear as real spaces
  breaklines=true, % wrap lines if they don't fit
  breakatwhitespace=true,
  postbreak=\mbox{\hspace*{-2em}\textcolor{red}{$\hookrightarrow$}\space},
  frame=trbl, % draw a frame at the top, right, left and bottom of the listing
  frameround=tttt, % make the frame round at all four corners
  framesep=4pt, % quarter circle size of the round corners
  numbers=left, % show line numbers at the left
  numberstyle=\tiny\ttfamily, % style of the line numbers
  numbersep=5pt,
  linewidth=1.0\linewidth,
  xleftmargin=0.2cm,
  commentstyle=\color{eclipseGreen}, % style of comments
  keywordstyle=\color{eclipsePurple}, % style of keywords
  stringstyle=\color{eclipseBlue}, % style of strings
  upquote=true, % straight single quotes in lstlistings and need   \usepackage{textcomp},
    mathescape=true,
     frame=single, % Set a single-line frame with straight corners
     frameround=ffff, % Ensure straight corners (f = flat, no rounding)
     numbers=none, % Remove line numbers
}
\definecolor{darkblue}{RGB}{0,0,139}
\definecolor{darkpurple}{RGB}{100,0,150}
\newif\ifreplyblue
\replybluefalse   % Turn off revision marking
\newcommand{\rc}[2]{%
  \ifreplyblue
     {\color{darkpurple}#2\textsuperscript{\textbf{C#1}}}%
  \else
     \renewcommand{\sout}[1]{}% \sout{...} becomes empty
     #2%
  \fi
}

\newif\ifreplyblueUpd
\replyblueUpdfalse       % Toggle for General Update: true = blue, false = black

\newcommand{\upd}[1]{%
  \ifreplyblueUpd
     {\color{darkpurple}#1}%
  \else
     #1%
  \fi
}
\begin{document}

\flushbottom
% Relax float placement restrictions to prevent figures/code blocks from being easily deferred to subsequent pages
\renewcommand{\topfraction}{0.85}
\renewcommand{\bottomfraction}{0.8}
\renewcommand{\textfraction}{0.15}
\renewcommand{\floatpagefraction}{0.7}
 
\title{Formal Evidence Generation for Assurance Cases \\ \upd{of Control Software for Robots}}

\author{Fang Yan \and Simon Foster  \and Ana Cavalcanti \and Ibrahim Habli \and\\ James Baxter 
}
\institute{University of York, York, UK\\
%\and  Lancaster University, Lancaster, UK \and D-RisQ Software Systems, Malvern, UK\\ 
\email{\{fang.yan, simon.foster,  ana.cavalcanti, Ibrahim.habli, james.baxter\}@york.ac.uk} 
}
\authorrunning{Fang Yan \and Simon Foster  \and Ana Cavalcanti \and Ibrahim Habli \and James Baxter}

\maketitle
% \linenumbers

\begin{abstract}

Robotics and Autonomous Systems are increasingly deployed in safety-critical domains, so that demonstrating their safety is essential. Assurance Cases\ (ACs) provide structured arguments supported by evidence, but generating and maintaining this evidence is labour-intensive, error-prone, and difficult to keep consistent as systems evolve.
We present a model-based approach to systematically generating AC evidence by embedding formal verification into the assurance workflow. The approach addresses three challenges:~systematically deriving formal assertions from natural language requirements using templates, orchestrating multiple formal verification tools to handle diverse property types, and integrating formal evidence production into the workflow. Leveraging Robo\-Chart, a domain-specific modelling language with formal semantics, we combine model checking and theorem proving in our approach. 
Structured requirements are automatically transformed into formal assertions using predefined templates, and verification results are automatically integrated as evidence.
Case studies demonstrate the effectiveness of our approach.
    
\end{abstract}
\input{introduction}
\input{background}
\input{AC_approach}
\input{Requirement_structuring}
\input{Assertion_generation}

\input{Evidence_model_generation_and_integration}

\input{case_study}

\input{relwork}

\input{concl}

%{\mycomment Acknowledge}
\section*{Acknowledgements}
This work was partially funded by the European Union’s Horizon 2020 research and innovation programme under the Marie Skłodowska-Curie grant agreement No.~812788 (MSCA-ETN SAS).
It was also supported in part by the UK EPSRC under grants EP/R025479/1 and EP/V02 6801/2, and by the Royal Academy of Engineering under grants CiET1719/45 and IF2122$\backslash$183.
Additional support was provided by the RoboSAPIENS project, funded by the European Commission’s Horizon programme under grant agreement No.~101133807.
We would like to thank Nick Tudor and Colin O'Halloran from D-RisQ for kindly providing free access to the Kapture tool used in this work.
\begin{appendices}
\input{Appendix}

\end{appendices}
%%%%%%%%%%%%%%%%%%%%%%%%%%%%%%%%%%%%%%%%%%%%%%%%%%%%%%%%%%%%%%%%%%%
%%%%%%%%%%%%%%%%%%%%%%%%%%%%%%%%%%%%%%%%%%%%%%%%%%%%%%%%%%%%%%%%%%%

% this is to avoid a url in the bibliography to break in page break.
% otherwise, we I get an error when compiling.
\interlinepenalty=10000

\bibliographystyle{spmpsci}%

\bibliography{reference,publications}%

% name your BibTeX data base

%%%%%%%%%%%%%%%%%%%%%%%%%%%%%%%%%%%%%%%%%%%%%%%%%%%%%%%%%%%%%%%%%%%
%%%%%%%%%%%%%%%%%%%%%%%%%%%%%%%%%%%%%%%%%%%%%%%%%%%%%%%%%%%%%%%%%%%

\ifdefined \CHANGES \indexprologue{%
  This index lists for each comment the pages where the text has been modified to address the comment. Since the same page may contain multiple changes, the page number contains the index of the change in superscript to identify different changes. Finally, the page number contains a hyperlink that takes the reader to the corresponding change.%
}%
\printindex[changes] \fi

\end{document}

%% file: introduction.tex
\section{Introduction}
\label{sec:intro}

As Robotics and Autonomous Systems~(RAS) increasingly permeate safety-critical sectors like healthcare, automotive, and aerospace, ensuring the safety of these systems is paramount. Given their complexity and autonomy, RAS must undergo rigorous verification throughout their development and operational lifecycle.

In safety-critical industries, the development of assurance Cases~(ACs), commonly required by safety standards such as MOD Def Stan 00-55~\cite{standard1standard} and ISO26262\ \cite{standard2011iso}, has become common practice. ACs are living documents that evolve alongside the systems they support. A key component of ACs is evidence, which substantiates claims about the system's compliance with key properties, such as the safety requirements. However, generating and updating this evidence is often labour-intensive, prone to errors, and difficult to trace, especially as systems undergo changes or adaptations in uncertain environments. These challenges highlight the need for systematically generating and integrating evidence through automation to ensure ACs remain consistent and aligned with system updates~\cite{sorokin2024towards, meng2021automating, odu2025automatic}. 

Formal verification techniques, such as model checking and theorem proving, can provide evidence that system models satisfy properties by verifying behaviours against formal specifications. Formal verification can generate rigorous evidence to support AC claims. 
While various studies have explored different ways of linking verification results to assurance arguments, full automation, providing traceability, maintainability, and adaptability, ensuring consistency of ACs throughout the lifetime of the system, remains elusive and poses significant difficulties.

% A range of studies~\cite{Denney2018, Hawkins2015a, Gacek2014a, sljivo2023guided, foster2020formal} have explored the use of formal verification in ACs, from generating evidence fragments to integrating verification results through mechanised assurance cases and structured argument patterns with tool support.
% Despite these research efforts, an automated approach for generating AC evidence using formal verification remains unavailable due to the following challenges.

The first challenge is integrating the production of verification evidence into the workflow for the development of ACs.
Traditionally, formal verification is taken into account in ACs through verification results that are referenced at the evidence nodes~\cite{Calinescu2018b,jee2010assurance,Prokhorova2015}. So, the process of obtaining these results is usually left outside the workflow for the development of ACs.
% Similarly, Sorokin et al.~\cite{sorokin2024towards} integrated testing and simulation tools into the assurance case workflow %via the Evidential Tool Bus
% , so that evidence artefacts are generated automatically during workflow execution.
Existing approaches~\cite{Denney2018,sorokin2024towards} have sought to address this issue, with some integrating code-level formal verification or simulation results directly into the AC process. 
Further work~\cite{Bourbouh2021a} has demonstrated the integration of model-level formal verification into ACs. However, automation is only partially achieved.
%As a result, model-level formal verification has not been integrated into an automated AC process, leaving the automation gap unaddressed.
Consequently, when the system evolves, formal verification is not automatically triggered, and evidence may therefore fail to be updated, leading to inconsistencies between claims and their supporting evidence.
% Despite these advances, a fully automated approach for generating AC evidence using formal verification, which requires the automated formalisation of both system models and specifications, remains unavailable. 
% One major reason for this is that prior work primarily focuses on using verification results to instantiate evidence models, rather than embedding the formal verification process itself as an integral part of the AC workflow. 
% As a result, most AC automation efforts overlook the formalisation and verification steps, leaving the creation of claims and formalised system models as largely manual tasks. Moreover, the properties to be substantiated in ACs are often of different types, requiring diverse formal verification tools. To facilitate evidence gathering from multiple verification tools, a toolchain is needed to orchestrate and manage the delegation of evidence collection tasks.

The second challenge is the description of property specifications as formal tool-ready assertions.
In current practice, requirements in safety-critical domains are usually expressed in natural language or semi-structured formats. So, the use of formal verification tools requires the involvement of formal methods experts to formalise assertions.  This is a major obstacle to automation.

A substantial body of work has targeted automation of the generation of temporal logic assertions from property specifications written in natural language. Early approaches~\cite{bouskela2022formal,bolton2014automatically, Bourbouh2021a, Lindoso2021a, ghosh2016arsenal} formalise requirements into LTL using templates, controlled natural languages, or ontology-based parsing.  They rely on requirements being rewritten in specialised notations or supported by domain-specific ontologies, and in some cases, can only generate relatively simple assertions. More recent work has used large language models~(LLMs) to translate natural language into LTL specifications~\cite{zhao2024nl2ctl,li2025automatic,cosler2023nl2spec,xu2024learning}, but with limited accuracy.

Very few approaches have addressed probabilistic temporal logics, such as PCTL, which is used in probabilistic model checking via tools like PRISM~\cite{kwiatkowska2002prism}. For CSP~\cite{roscoe1998theory}, the complexity of specifications, covering concurrency, communication, and refinement, makes the automation of formalisation particularly challenging. Some attempts are based on diagrammatic notations for properties of components of RoboChart~\cite{Miyazawa2019a} models~\cite{Lindoso2021a,WC22}, where requirements are manually specified as diagrams and then translated into CSP assertions for FDR~\cite{gibson2014fdr3}.

The third challenge is orchestrating multiple formal verification tools.
The need for combining different formal methods has long been recognised in the literature on integrated formal methods~\cite{gleirscher2019new}. %Different types of properties naturally call for different verification techniques: some require refinement checking, others probabilistic analysis, and unbounded behaviours may call for theorem proving. 
Prokhorova et~al.~\cite{Prokhorova2015}, for example, demonstrate how different requirement types can be covered in an AC argument by combining different formal verification tools. However, that work does not automate the coordination of these tools. Some approaches embed verification into the AC development process, but usually integrate only one type of formal-verification engine
%, such as Denney and Pai’s AdvoCATE framework with AutoCert
~\cite{Denney2018}.
Consequently, providing a unified framework that can automatically delegate requirements to suitable verification backends, collect their results, and feed them back as assurance evidence remains an open challenge.

In this work, we leverage RoboChart~\cite{Miyazawa2019a}, a domain-specific modelling language for robotic systems, to support evidence generation through formal verification. RoboChart’s formal semantics, combined with verification tools such as FDR and PRISM, facilitate the automatic production of verification evidence, which can then be systematically integrated into ACs. To address the above challenges, we employ both model checking and theorem proving to generate verification results, and adopt a template-based approach to derive formal assertions from natural language requirements.

Our main contribution in this paper is an approach for creating and evolving AC evidence by embedding formal verification into a model-based workflow. 
With the level of automation that we achieve, not only we can support the initial development and offline maintenance of ACs. but also their evolution in the context of adaptive systems.  For instance, the work in~\cite{LABCGLMOPPTTZ24} presents an architecture for adaptive safety-critical systems that includes a legitimisation component.  Using our work, it is possible to include the update of an AC as part of the functionality of that component to both preclude software adaptations that compromise safety and provide evidence of safety if the adaptation can go ahead. If requirements change, describing those requirements using our controlled language requires manual intervention, but otherwise the update of the AC can be done fully automatically. 

In detail, we present:
\begin{enumerate*}[label=(\arabic*)]

\item a traceable assurance case evidence generation approach, which embeds formal verification of software models within the assurance case development process and integrates multiple verification techniques to support diverse property types;

\item a template-based technique for systematically structuring 
requirements and automatically generating formal assertions, 
bridging informal requirements and tool-ready verification 
properties while reducing the need for formal methods expertise;

\item an evaluation through multiple robotic case studies that demonstrate the applicability and effectiveness of our approach. Our case studies include an unscrewing robot whose behaviour adapts during operation when new types of screw are found, a safety monitor for an autonomous underwater vehicle that adapts its operation depending on safety-related conditions, a mail delivery robot that runs on a rechargeable battery, and a high-voltage controller with a risk of causing a fire.
\end{enumerate*}

\begin{figure}[t]
    \centering
    \includegraphics[scale=0.7]{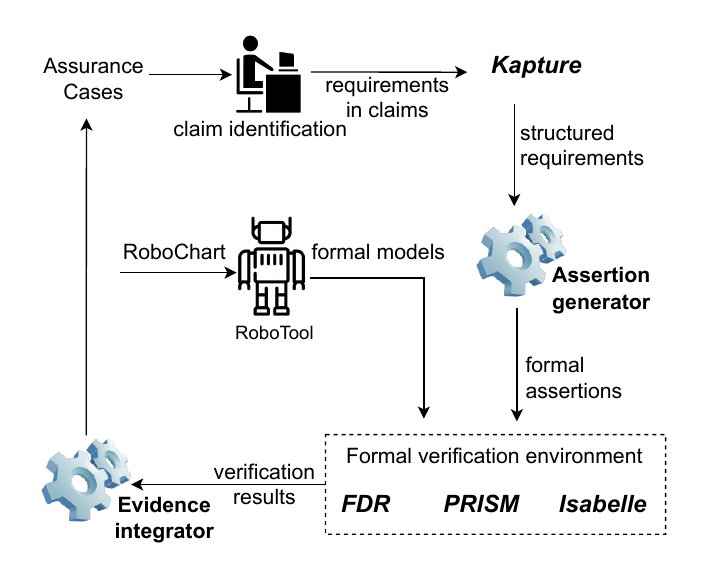}
    \caption{Approach for evidence generation.}
    \label{fig:toolchain}
\end{figure}

Our approach is described in Fig.~\ref{fig:toolchain}. It is supported by five existing tools:~(1)~Kapture\footnote{\url{https://www.drisq.com/product-kapture}}, a software requirement management tool designed to help engineers create precise and unambiguous requirements; (2)~RoboTool\footnote{\url{https://robostar.cs.york.ac.uk/robotool/}}, which supports modelling, validation, and automatic generation of CSP and PRISM semantics for RoboChart models; (3)~FDR; (4)~PRISM; and (5)~Isabelle~\cite{nipkow2002isabelle}.  We also include a novel assertion generator and an evidence integrator.  Development of an AC using our approach starts by identifying the software requirements referenced in the AC claims that require formal verification~(model checking or theorem proving). Then, the evidence is generated from the verification results of the requirements to support the claims. 

We have established a classification for requirements to be verified using formal verification, and customised and refined requirement templates used by Kapture. Using our Kapture templates, we obtain a structured requirement, written in a controlled natural language. From such requirements, an assertion generator produces formal assertions based on assertion templates we have designed for each class of requirement. \rc{1.2a}{These assertion templates equip the structured  requirements with a formal semantics, and so provide the basis for verification}. They are verified against RoboChart models (or more precisely, their semantics) using the appropriate formal verification tools, and their verification results are transformed into evidence models. These are subsequently integrated into the model-based AC using the evidence integrator. 

The remainder of this paper is organised as follows. Sect.~\ref{sec:background} briefly introduces ACs, RoboChart, and our formal notations. %CSP, the PRISM language, and the Z-Machine notation. 
In Sect.~\ref{sec:approach}, we give an overview of our approach for model-based AC evidence generation. A running example is introduced in Sect.~\ref{section:example}.
Sect.~\ref{sec: claim in CNL} presents the controlled natural language used for structuring software requirements, namely the requirement templates for RoboChart that we have developed based on the Kapture templates.
Sect.~\ref{model checking assertion generation} discusses the generation of formalised assertions from the structured requirements.
Sect.~\ref{evidence model gen for model checking} introduces the method for generating and integrating the evidence from verification results.
%The detailed method is presented in Sect.~\ref{sec:model_checking}.
%The tool implementation is discussed in Sect.~\ref{impl}.
The case studies are presented in Sect.~\ref{sec:cases}. We review related work in Sect.~\ref{sec:relwork}. 
Finally, we conclude and discuss future work in Sect.~\ref{sec:concl}.

%% file: background.tex
\section{Background} 
\label{sec:background}
This section briefly introduces ACs (Sect.~\ref{sec:background:AC}), the Robo\-Chart language (Sect.~\ref{sec:background:RoboChart}), describes the FDR (CSP) and PRISM assertion languages~(Sects.~\ref{sec:background:CSP},~\ref{sec:background:PRISM}), and provides an overview of the Z-Machine notation used in the Isabelle/HOL theorem prover (Sect.~\ref{sec:background:z-machine}).

\subsection{Assurance Cases}
\label{sec:background:AC}
An AC is a structured argument that decomposes top-level claims, that is, properties, into supporting sub-claims.
 A safety property is often addressed through a safety case - a specialised form of assurance case~\cite{kelly1999arguing}.
ACs are crucial for systems operating in high-risk industries like automotive and healthcare. While not mandatory in all safety-related fields, they are commonly used for stakeholder communication and are often required by safety standards, such as ISO26262~\cite{standard2011iso}.

ACs can be represented in different forms, including textual, graphical, and machine-readable models.
The most widely used are the structured graphical notations, e.g., the Goal Structuring Notation (GSN) \cite{AssuranceCaseWorkingGroup} and Claims-Arguments-Evidence~(CAE) \cite{Bloomfield1998}.
The two notations use various shapes to graphically represent AC elements and their relationships in a structured manner. This helps stakeholders comprehend the range of information necessary to support arguments about system properties. 

\begin{comment} 
\begin{figure}[h]
	\centering
	\includegraphics[width=0.4\linewidth]{images/CAE.png}
	\caption{An abstract example of AC in CAE.\label{fig:CAE elements}}
\end{figure}
\end{comment}

%Although CAE and GSN share similar notation methods~\cite{Tokoro2015}, 
% GSN, developed by Kelly~\cite{kelly1999arguing} in 1998, is the dominant notation used in engineering and safety-critical industries. In GSN, claims are documented as goals, and items of evidence are cited in solutions. In addition to goals and solutions, the other main elements of GSN are strategy (that is, a description of the approach used to support a goal), context, justification, and assumptions. Two types of linkages between elements are also used:~SupportedBy and InContextOf, as shown in Fig.~\ref{fig:GSN elements}. 

\begin{figure}
	\centering
	\includegraphics[width=0.9\linewidth]{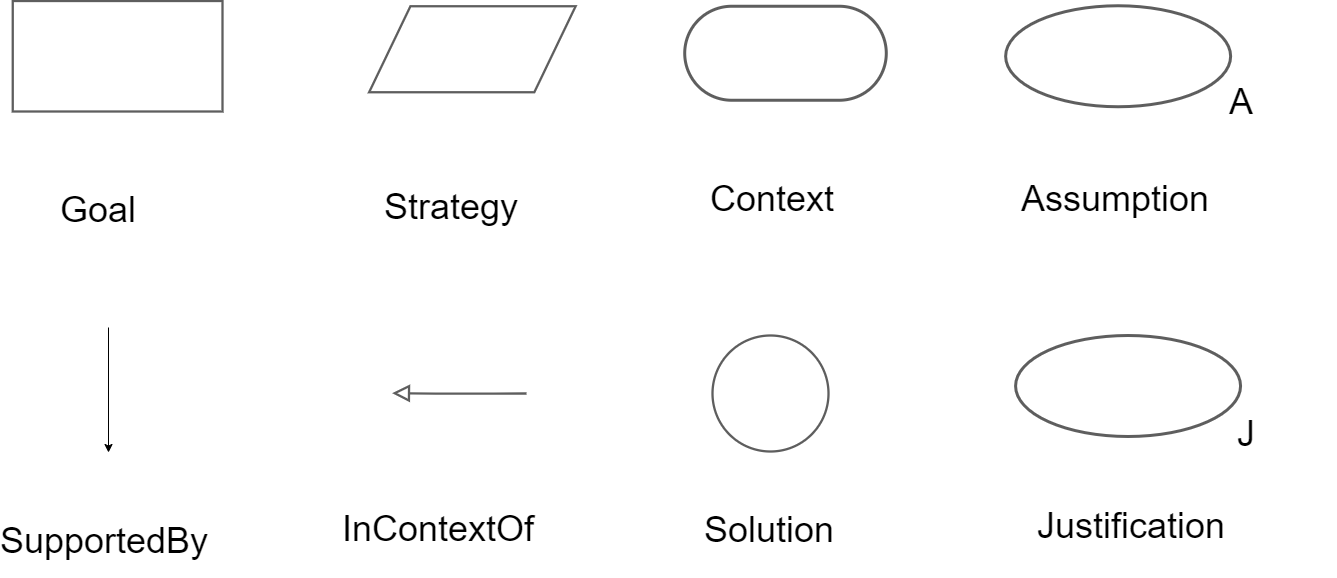}
	\caption{Main elements of GSN.\label{fig:GSN elements}}
\end{figure}

GSN was developed by Kelly~\cite{kelly1999arguing} in 1998. A GSN model is structured as a hierarchical argument, typically starting with a top-level goal (a claim about the system) that is progressively decomposed into sub-goals until a point is reached where claims can be supported by direct evidence.
GSN defines six main elements: (i) \emph{Goals} represent claims or properties to be established; (ii) \emph{Strategies} describe the reasoning approach used to decompose a goal into sub-goals; (iii) \emph{Solutions} provide the evidence that directly supports a goal; (iv) \emph{Context} frames how a goal or strategy should be interpreted; (v) \emph{Assumptions} document what is assumed for a strategy, goal, or solution to hold; and (vi) \emph{Justifications} explain why a particular claim or argument is acceptable. These elements are connected by relationships: goals are decomposed via strategies into sub-goals, and ultimately linked to solutions that provide the supporting evidence. The main GSN elements and their relationships are illustrated in Fig.~\ref{fig:GSN elements}.

% \begin{figure}
% 	\centering
% 	\includegraphics[width=1.0\linewidth]{images/GSN_example.png}
% 	\caption{GSN example~(adapted from ~\cite{yan2022model}).\label{fig:GSN elements}}
% \end{figure}

% \rc{2.5}{Fig.~\ref{fig:gsn_pattern} shows a concrete example of a GSN argument pattern for hazard mitigation. The top-level goal---``Hazard \{Hazard\_ID\} is mitigated''---is decomposed via a strategy into sub-goals addressing the causes and effects of the hazard. Each cause is then further decomposed via strategies into safety measure goals, which are eventually supported by evidence (solutions) after verification and validation activities. This pattern---from top-level goal through strategies and sub-goals to evidence---illustrates how GSN structures a safety argument.}

% \rc{2.5}{The figure also introduces notation conventions commonly used in GSN patterns. It distinguishes between elements that are instantiated, uninstantiated, and undeveloped/uninstantiated. Uninstantiated elements represent placeholders whose specific content is not yet defined, while undeveloped and uninstantiated elements indicate parts of the argument that have not yet been elaborated. The legend summarises the notation for these various element types. This pattern represents a reusable argument skeleton; a concrete instantiation of such a structure is shown later in the mail-delivery robot case study (Fig.~\ref{fig:ac_mail_robot}).}
\begin{figure}
	\centering
	\includegraphics[width=1.0\linewidth]{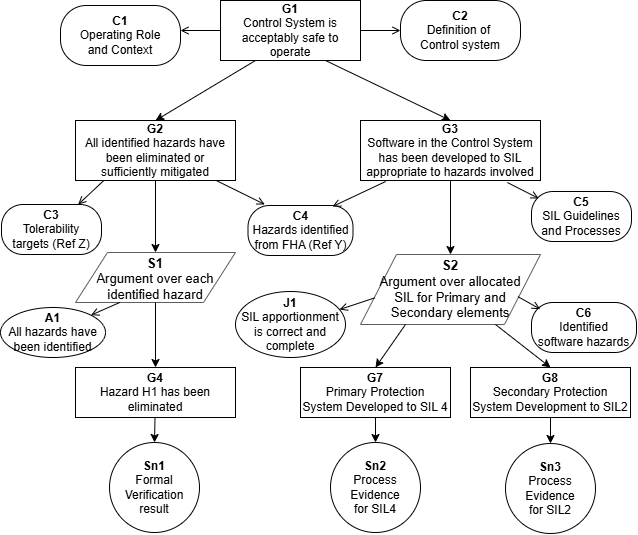}
	\caption{GSN example (adapted from~\cite{AssuranceCaseWorkingGroup}).\label{fig:gsn example}}
\end{figure}

\rc{2.5}{Fig.~\ref{fig:gsn example} shows an example of a GSN structure. In this example, the top-level goal is ``Control System is acceptably safe to operate'' (G1). This goal is qualified by the contexts C1 (Operating Role and Context) and C2 (Definition of Control System). The top-level goal is then decomposed into two sub-goals: G2 (``All identified hazards have been eliminated or sufficiently mitigated'') and G3 (``Software in the Control System has been developed to SIL appropriate to hazards involved''). G2 is further supported by context C3 (Tolerability targets) and C4 (Hazards identified from FHA), and is decomposed via strategy S1 (``Argument over each identified hazard'') into sub-goal G4 (``Hazard H1 has been eliminated''). G4 is then decomposed via strategy S2 (``Argument over allocated SIL for Primary and Secondary elements'') into goals G7 and G8, which are supported by solutions Sn1, Sn2, and Sn3. 
The argument also relies on assumption A1 (``All hazards have been identified'') and justification J1 (``SIL apportionment is correct and complete''). This GSN example shows how a top-level claim is systematically decomposed through strategies into lower-level claims, ultimately supported by evidence (solutions).}

GSN did not originally have a formal metamodel. Therefore, a number of meta-models have been proposed~\cite{Denney2018,Hawkins2015a}.
More recently, the Object Management Group released the Structured Assurance
Case Meta-Model~(SACM)~\cite{ObjectManagementGroupOMG}, which aims to improve standardisation and interoperability. Compared to other proposed AC metamodels, SACM provides unique features, such as fine-grained modularity, controlled terminology, and traceability from argument to evidence artifacts~\cite{Wei2019}. SACM can support a variety of notations, including GSN and CAE.

In this work, we use the SACM standard, but adopt the GSN notation to present results graphically, as GSN is generally more familiar to practitioners.

\subsection{RoboChart}
\label{sec:background:RoboChart}

The core of RoboChart~\cite{Miyazawa2019a} is a subset of the UML state machine notation with a formal semantics. %that excludes non-essential features.% that lead to an increase in the complexity of the underlying mathematical models.
Robo\-Chart offers several features that align with the objectives of this research:~(i)~it is supported by an EMF metamodel, enabling easy manipulation when implementing model-based techniques, and (ii)~it possesses formal semantics and model-checking support, which are advantageous for automating the generation of formal evidence.

A RoboChart model captures the software architecture in a hierarchical format, where a RAS software design is modelled as a RoboChart module. The components of a module are a robotic platform and multiple controllers. %It deﬁnes the boundaries of the robotic application modelled and speciﬁes the interface of such an application through a robotic platform.

A robotic platform abstracts a physical robot in terms of the functionalities it provides to controllers. These may be of three types:~variables, operations, or events. These elements can be either declared directly in the robotic platform, or grouped in interfaces. An interface groups variable lists, operations, events, and clocks.  An event may or not have a type~\cite{Miyazawa2016}, depending on whether they communicate data or are just signals. %, which is an expression that defines the values that can be communicated via the connection, if any~\cite{Miyazawa2016}.
 
A controller represents software components that interact with the robotic platform. The behaviour of a controller is defined using state machines.
RoboChart's %includes a profile of UML 
state machines are similar to those of UML, but are enriched with facilities to define time properties, and with probabilistic junctions~\cite{ye2022probabilistic}. 
Detailed information on the syntax and semantics of RoboChart can be found in~\cite{Miyazawa2016}.

RoboChart's development environment is RoboTool\footnote{\url{www.cs.york.ac.uk/circus/RoboCalc/robotool/.}}.
%RoboTool is a set of Eclipse\footnote{www.eclipse.org.} plug-ins implemented using the Xtext\footnote{www.eclipse.org/xtext.} and Sirius\footnote{www.eclipse.org/sirius} frameworks.
%It enables modelling, performs type checking and analysis of well-formedness, automatically calculates CSP models and PRISM models, parses the properties written in the form of controlled English, calls the FDR and PRISM model checkers, and returns the verification reports.
%The RoboTool uses Eclipse Modeling Framework (EMF) to implement RoboChart metamodel as a basis to generate a textual editor using Xtext and a graphical editor using Sirius.
%RoboChart metamodel is accompanied by a set of well-formedness conditions.
%These conditions are implemented through the validation mechanism provided by Xtext. Each condition is associated with one or more validation rules.
%The graphical editor constructs a RoboChart model and automatically veriﬁes all the well-formedness conditions, and type compatibility in expressions and statements.
It calculates %both the untimed and timed 
a formal semantics for RoboChart models described using the ASCII version of CSP %Communicating Sequential Processes~(CSP) 
called CSP{$_\mathrm{M}$}.
This semantics supports  automatic verification of
properties using FDR.
RoboTool also converts the RoboChart models into PRISM models for verification using PRISM.

RoboChart is equipped with assertion languages: controlled English to define properties to be verified in FDR and PRISM.
%The properties in these assertion languages are written in assertion files (with extension .assertions).
RoboTool can transform these assertions into CSP assertions for FDR, or PCTL %Probabilistic Computation Tree Logic~(PCTL) 
specifications for PRISM. 
RoboTool calls the model checkers directly and produces a verification report with results.
%The advantage of the assertions ﬁles to generate the verification report is to avoid the need for users to interact directly with the model checkers~\cite{Miyazawa2016}.

%RoboTool has other features that support RoboChart modelling and verification which are not used in this work. This information can be found in the RoboTool Manual~\cite{robotool2023}.

\subsection{CSP and the assertion language for CSP}
\label{sec:background:CSP}

RoboChart has an untimed semantics defined using CSP, and a timed semantics defined using its timed dialect, tock-CSP~\cite{BRC22}.
CSP, a specification language for concurrent systems, has processes and channels as its main constructs. Processes describe interaction patterns, such as choice, deadlock, and termination, and can be composed in parallel, interacting through channels. CSP has three core denotational models:~traces, stable failures, and failures/divergences. The traces model represents sequences of observable events, the stable-failures model adds refusal sets, and the failures/divergences model addresses livelock. More details on CSP semantics and operators can be found in~\cite{Roscoe2010}.

%Expertise is required to write the assertions in CSP{$_\mathrm{M}$} for RoboChart models.
The RoboChart assertion DSL for CSP is based on CSP{$_\mathrm{M}$}. %, a machine readable format of CSP, 
%to simplify the assertion presentation.
The rest of this section provides a brief overview of this assertion language.
For more comprehensive information, we refer to Sect.~5.1 of the RoboChart manual~\cite{Miyazawa2016}.

The syntax of an assertion is described below, using an EBNF notation. When defining assertions, the semantics to be checked can be selected by labelling the assertion with \lstinline[style=ADPropertyStyle]|untimed| or \lstinline[style=ADPropertyStyle]|timed|.
%For example, the label \lstinline[style=ADPropertyStyle]|timed| ensures that the timed semantics of the model are checked.
If no label is given, both semantics are checked by default, leaving it to the user’s discretion to decide which result is most relevant.
% when we check an assertion for RoboChart model in FDR, the verification is conducted on both the timed and untimed CSP{$_\mathrm{M}$} semantics.
% However, we could specify CSP{$_\mathrm{M}$} semantics to be used in the verification as timed or untimed through the label \lstinline[style=
% ADPropertyStyle]|timed|  and \lstinline[style=
% ADPropertyStyle]|untimed| when needed.
% \textcolor{orange}{the rationale for check timed, untimed?}

\lstset{style=ADPropertyStyle}{\footnotesize
\begin{lstlisting}[caption={}, label={lst:syntax_FDR_assert}]
Assertion ::= 
    ('timed' $\vert$ 'untimed')? 'assertion' N ':' SPEC
    ('in' 'the' MODEL)?
    ('with' ('constant' $\vert$ 'constants') CONSTANTS)?
\end{lstlisting}}
%In RoboTool, compiling a RoboChart model into CSP{$_\mathrm{M}$} produces two variants of the semantics:~untimed and timed.

\noindent%
An assertion has a name \lstinline[style=
ADPropertyStyle]|N| and a property specification \lstinline[style=
ADPropertyStyle]|SPEC|, and optionally the specification of a model \lstinline[style=
ADPropertyStyle]|MODEL| (for instance, traces model), and  of values for constants used in the RoboChart model.
%\lstinline[style=ADPropertyStyle]|N|  denotes the name category, here it represents the name of the assertion.
The syntax for \lstinline[style=
ADPropertyStyle]|SPEC| is  sketched below; we describe just the elements used in this paper.

\lstset{style=ADPropertyStyle}{\small
\begin{lstlisting}[caption={}, label={lst:syntax_FDR_assert_spec}]
SPEC ::= 
    N 'is' ('not')? ('deadlock-free' 
                        $\vert$ 'divergence-free')
    $\vert$ N ('does' 'not' 'terminate' $\vert$ 'terminates')
    $\vert$ N 'is' ('not')? 'reachable' 'in' N
    $\vert$ N (('does' 'not' 'refine') $\vert$ ('refines') ) N
    $\vert$ ...
\end{lstlisting}
}

\noindent%
\lstinline[style=
ADPropertyStyle]|SPEC| can be unary or binary. Unary assertions describe general properties of RoboChart elements, such as, deadlock freedom, divergence freedom, termination, and state reachability. %determinism, clock initialisation, and timelock freedom (where no events occur and time cannot progress). 
Binary assertions compare two RoboChart elements. In the definition of \lstinline[style=
ADPropertyStyle]|SPEC| above, we show the binary assertion for comparison based on refinement.  %using refinement and equality relations. Our method addresses several types of assertions, including termination, state reachability, deadlock freedom, divergence freedom, and refinement.

%For the refinement assertions, the specification and implementation of RoboChart models are represented by two CSP{$_\mathrm{M}$} modules. These modules are defined using the following CSP{$_\mathrm{M}$} block:

% \noindent
% \begin{minipage}{\columnwidth}
% \footnotesize
% \begin{lstlisting}[caption={}, label={lst:syntax_FDR_assert_csp}]
% CSP module ::= 
%     ('timed' | 'untimed')? 'csp' N
%      'csp-begin' CSPM 'csp-end'
% \end{lstlisting}
% \end{minipage}

% In the list above, \lstinline[style=
% ADPropertyStyle]|CSPM|  is the formalisation of the specifications or the implementations of the system using CSP processes in CSP{$_\mathrm{M}$}.

%The RoboChart assertion language simplifies the process of writing simple assertions, such as deadlock freedom, without requiring users' in-depth knowledge of the naming conventions of CSP semantics of RoboChart. 
%However, for more complex refinement properties, it is necessary to utilize the CSP blocks and have a deeper understanding of the structure and naming conventions of the CSP semantics, in addition to the CSP{$_\mathrm{M}$} syntax. 

The RoboChart assertion language facilitates writing simple assertions, such as deadlock freedom, without requiring detailed knowledge of RoboChart's CSP semantics naming conventions. However, more complex refinement properties necessitate the use of CSP blocks, along with a deeper understanding of the CSP semantics structure, naming conventions, and CSP{$_\mathrm{M}$} syntax.

\subsection{PRISM and the assertion language for PRISM}
\label{sec:background:PRISM}

RoboChart also has a formal semantics defined using the reactive modules notation (of PRISM and many other probabilistic model checkers).
RoboTool can automatically calculate this semantics as well.

The properties to be checked in the PRISM model checker are specified as formal assertions in PCTL. The RoboChart assertion language for PRISM provides a more readable syntax,
% In the rest of this section, a brief introduction to the syntax of the RoboChart assertion language for PRISM is provided.
sketched below.

\noindent
\begin{minipage}{\columnwidth}
% \footnotesize
\begin{lstlisting}[caption={}, label={lst:syntax_assert}，basicstyle=\footnotesize]
ProbProperty ::= 
    'prob' 'property' N ':' pExpr
    ('with' 'constants' (ConstConfig+ $\vert$ N))?
    ...
\end{lstlisting}
\end{minipage}

\noindent%
A probabilistic property starts with the keyword \lstinline[style=
ADPropertyStyle]|prob|. It has a name 
\lstinline[style=
ADPropertyStyle]|N| with a
body given by an expression that denotes the property to be verified.
In our work, this expression must have a Boolean type. Optionally, 
an assertion has constant configurations~(given by \lstinline[style=
ADPropertyStyle]|ConstConfigs|). These configurations can be used to define values of constants declared in the RoboChart model. %, function and operation definitions (given by \lstinline[style=ADPropertyStyle]|pFunction| or \lstinline[style=ADPropertyStyle]|pOperation|), PRISM module (given by \lstinline[style=ADPropertyStyle]|pModule|), and customised PRISM command line options.
For details of the full syntax of probabilistic assertions, we refer to Sect.~5.2 of the RoboChart manual~\cite{Miyazawa2016}.

We also consider reward assertions as follows. 

\noindent
\begin{minipage}{\columnwidth}
\begin{small}
\begin{lstlisting}[caption={}, label={lst:syntax_reward}]
Reward ::= ('[' pEvent ']')? pExpr ':' pExpr ';'
\end{lstlisting}
\end{small}
\end{minipage}
In a reward-based assumption, a reward value is defined using a guard condition \lstinline[style=
ADPropertyStyle]|pExpr| of boolean type, an expression \lstinline[style=
ADPropertyStyle]|pExpr| defining the value, and an optional event \lstinline[style=ADPropertyStyle]|pEvent|, which is an action label. In PRISM, action labels are names attached to transitions, allowing rewards to be associated with specific types of transitions rather than all transitions satisfying the guard.

Our work automates the generation of assertions from requirements by providing a set of templates for CSP-based and PCTL-based RoboChart assertions.

\subsection{Isabelle and Z-Machines}
\label{sec:background:z-machine}
Z-Machine formalism is used to give a simplified semantics to RoboChart state machines~\cite{yan2023automated} in the Isabelle/HOL theorem prover.
Z-Machines is a formal modelling language and tool in the style of the Z specification language~\cite{Spivey89} and B method~\cite{Abrial96BBook}. 
It is a form of abstract machine and consists of (1)~a state space including invariants; (2)~a set of operations acting over that state space; and (3)~the machine itself, which initialises the state and groups together the operations. Z-Machines can make use of any data structures available in HOL, such as sets, functions, records, algebraic data types, and real numbers.

Z-Machines is supplied with a proof method called \lstinline{deadlock_free}, which automates deadlock checking in Isabelle~\cite{yan2023automated}. The method calculates a deadlock-freedom condition  for the whole Z-Machine that can typically be discharged using Isabelle's powerful tools that automate the process of finding proofs, such as \textit{auto} or \textit{sledgehammer}.

%% file: AC_approach.tex
\section{Overview of our approach}
\label{sec:approach}

\begin{figure*}[t]
	\centering
	\includegraphics[width=1.0\linewidth]{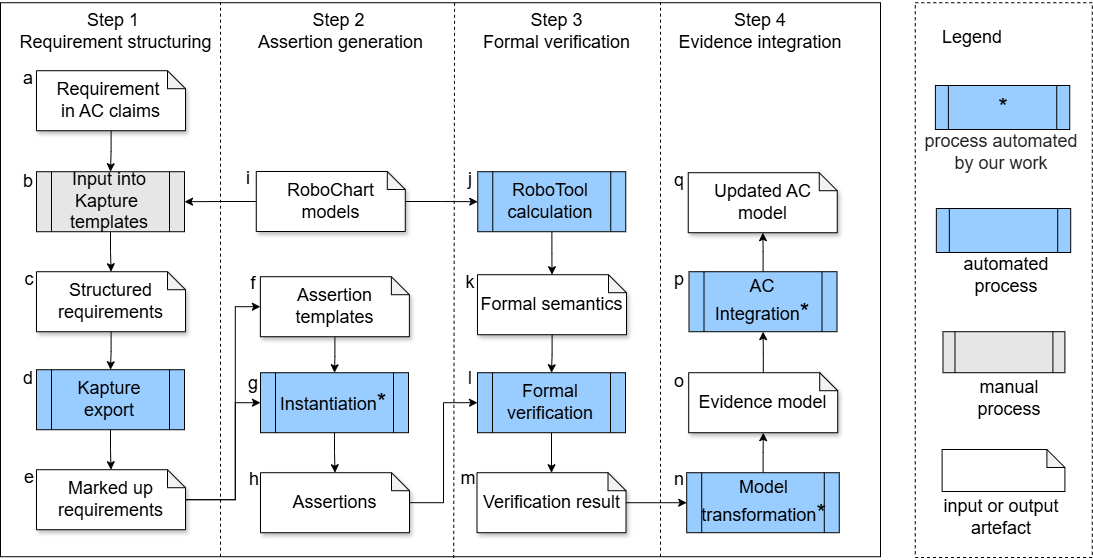}
	\caption{Overview of our four-step approach to evidence generation by formal verification. Activities and artefacts within these steps are labelled \emph{a} - \emph{q} to allow cross-referencing in the text.\label{fig:evidence_generation_method}}
\end{figure*}

This section describes our approach to automating the generation of evidence through formal verification.  
Fig.~\ref{fig:evidence_generation_method} provides an overview of the four-step process, from requirement structuring~(Step~1 Fig.~\ref{fig:evidence_generation_method}) to the integration of verification results into an AC~(Step 4).  
%The degree of automation and the tools supporting each step are summarised in Table~\ref{table_tool}.  Together, Fig.~\ref{fig:evidence_generation_method} and Table~\ref{table_tool} give a high-level overview of our proposed workflow and its tool support detailed later.  

We assume that a partially instantiated model-based AC is provided as the starting point, where some of the evidence has not yet been concretised with verification results.  
\rc{2.5}{In our workflow, we focus on the lower levels of the AC hierarchy, where a requirement-level claim (e.g., a software requirement has been implemented) is specified into a verification-method claim (e.g., the requirement is verified by model checking), which is then directly supported by the verification evidence. This pattern is central to our approach, as it connects software requirements to the formal verification method and the formal verification results.}

The AC may contain claims of various forms; we focus on claims such as ``the software requirement $R_{\mathit{ID}}$ has been implemented'', which refer to software-related concerns of a broader assurance argument, and for which $R_{ID}$ is a  property to be verified using formal methods.

The evidence supporting the claim typically takes the form ``the result is true,'' indicating that the software satisfies the requirement.  
Based on this structure, the automation of evidence generation requires the formalisation of $R_{ID}$.  

The four steps of our process~(see Fig.~\ref{fig:evidence_generation_method}) are as follows. 
\begin{enumerate}
  \item {\it Structuring the requirements $R_{ID}$ (labelled~(a) in Fig.~\ref{fig:evidence_generation_method}) referenced in the AC claim using predefined templates in the Kapture tool.} Each requirement $R_{ID}$ is described precisely as a property of a RoboChart model of the software~(\emph{i}) using Kapture templates~(\emph{b}).
  Kapture can export~(\emph{d}) structured requirements~(\emph{c}) into markup formats such as XML~(\emph{e}), thereby enabling their use in assertion generation.

 \item {\it Assertion generation by instantiating~(\emph{g}) novel predefined assertion templates~(f) with the markup requirements~(e)}. 
 This step leads to automatically generated formal assertions~(\emph{h}) traceable to requirements.

 \item {\it Using formal verification~(l) based on the RoboChart model~(i) to check the assertions~(h).} RoboTool is used to calculate~(\emph{j}) a formal semantics~(\emph{k}) for the Robo\-Chart model (\emph{i}), which is subsequently verified~(\emph{l}) against the assertions~(\emph{h}) from the previous step. This step yields the verification results~(\emph{m}).

  %These models have %formal semantics (e.g., CSP{$_\mathrm{M}$}, PRISM or Z-Machines), automatically generated from the RoboChart models.  

  \item {\it Using the verification results (m) to generate an SACM-based evidence model (o)}. This is achieved through model-transformation~(\emph{n}), after which the model is integrated~(\emph{p}) into the AC model~(\emph{q}).  
\end{enumerate}
The steps of our approach are automated, with the exception only of the input of software requirements into Kapture templates~(\emph{b}). 
%we authree main activities are carried out automatically by existing tools. These activities are:~(1)~requirement markup in Step~1 using Kapture, (2)~RoboChart model formalisation in Step~3 into CSP{$_\mathrm{M}$} and PRISM semantics using RoboTool, and into Z-Machines using our previous work~\cite{yan2023automated}, and (3)~formal verification in Step~3. 
Three key activities (namely, (\emph{g}), (\emph{n}), (\emph{p})) are automated by the work presented in this paper. They are highlighted in Fig.~\ref{fig:evidence_generation_method} with a blue background and a star, and described in later sections.  

%Table~\ref{table_tool} summarises the automation status and supporting tools for each step shown in Fig.~\ref{fig:evidence_generation_method}.  
The implementation of the approach is based on the Eclipse EMF framework and relies on three tools:~Robo\-Tool, Epsilon\footnote{\url{http://download.eclipse.org/epsilon/updates/2.2/}}, and Kapture. %\footnote{\url{https://www.drisq.com/product-kapture}}.  
%RoboTool, an Eclipse plugin, serves as the modelling environment for RoboChart and integrates the FDR and PRISM model checkers to enable formal verification.   It supports the generation of formal semantics from RoboChart models, the translation of assertions written in RoboChart assertion languages into formal assertions, and the retrieval of verification results by invoking the model checkers.  
Epsilon is employed to facilitate the transformation process, converting XML-encoded claims into RoboChart DSL assertions and translating model checking results into SACM evidence models.  
%Finally, Kapture is used to organise requirements and export them into XML files.  
%Kapture is a requirement management tool developed by D-RisQ\footnote{\url{http://www.drisq.com/}}, designed to help engineers create precise and unambiguous requirements. 

To illustrate the steps and the supporting tools of our approach, we introduce a running example next.

\section{Running example}
\label{section:example}

\begin{figure*}[h]
    \centering
    \includegraphics[width=\textwidth]{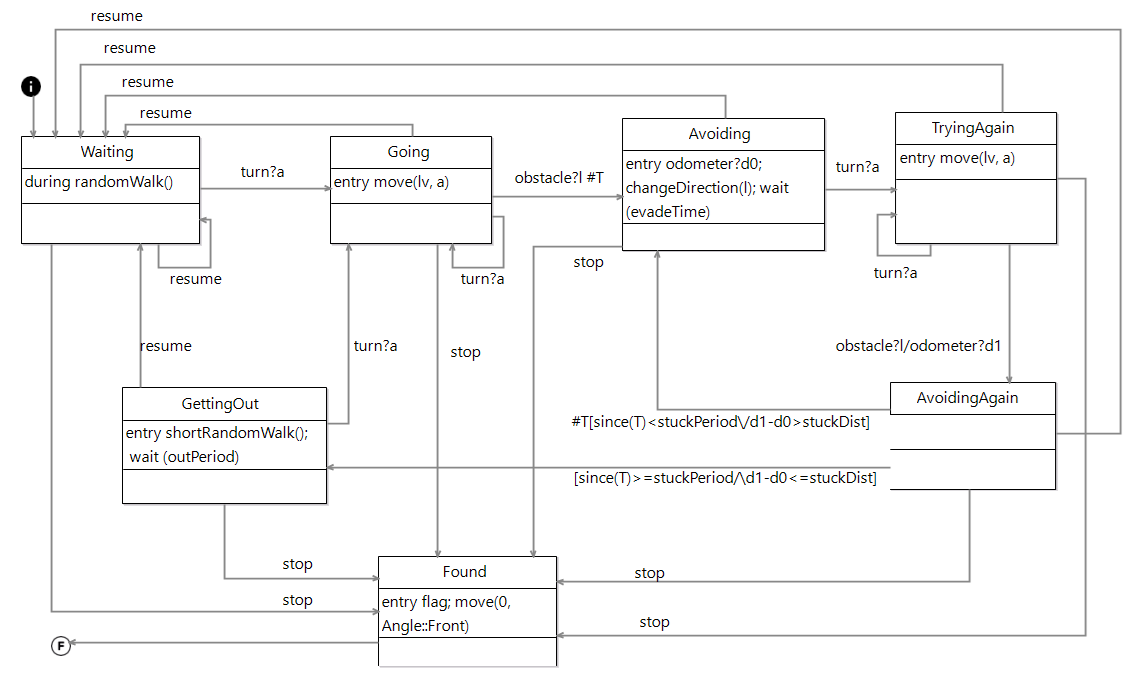}
    \caption{RoboChart state machine \RC{Movement}.}
    \label{fig:mvmt}
\end{figure*}

Our example is based on a mobile and autonomous chemical detector presented in~\cite{miyazawa2019robochart,hilder2012chemical}, which performs a random walk, while avoiding obstacles and analysing the air to detect dangerous gases.
Once a dangerous chemical is detected at a high intensity, the robot drops a flag to report the presence of the gas, and stops.

Gas detection is modelled by a RoboChart state machine called \RC{GasAnalysis}, and the random walk with obstacle avoidance is modelled in another state machine \RC{Movement}, shown in Fig.~\ref{fig:mvmt}. We use \RC{Movement} as the basis of our running example.  \RC{Movement} specifies obstacle avoidance using events \RC{obstacle} and \RC{odometer}, specifies the movement behaviours using events \RC{turn}, \RC{stop}, and \RC{resume}, sent from \RC{GasAnalysis}, and drops a flag using the event \RC{flag}.  All these events represent services, of the robotic platform~(a small trolley carrying an artificial nose and a flag) or of the parallel \RC{GasAnalysis} machine, used by the software implementing \RC{Movement}. 

A RoboChart block for a state machine defines a context of events, variables, and operations used in its actions. For brevity, we omit the definition of that context in Fig.~\ref{fig:mvmt}. 

\RC{Movement} starts in the state \RC{Waiting} (as indicated by the black circle with an \RC{i})  during which the robot performs a random walk. We omit here the definition of the operation \RC{randomWalk()} and of all other operations used by \RC{Movement}; they are all declared in the omitted context.
When a gas is identified by the machine \RC{GasAnalysis}, \RC{Movement} receives a \RC{turn} event and transitions to the state \RC{Going}.
\RC{Movement} may also receive a \RC{resume} event, indicating that no gas has been detected and that random walking should continue.
When a dangerous gas is found, \RC{Movement} receives a \RC{stop} event and transitions to \RC{Found}, where a \RC{flag} is dropped and the robot is stopped, using the \RC{move} operation, with argument 0 for the linear velocity.

In the state \RC{Going}, a \RC{turn} event does not lead to a change of state. The event \RC{resume} leads \RC{Movement} back to the state \RC{Waiting}. The event \RC{stop} leads to \RC{Found}. In addition, if obstacles appear along the path while \RC{Movement} is in the state \RC{Going}, then an
 avoidance mechanism is triggered. It uses the event \RC{odometer} and a clock \RC{T} to detect situations where the trolley becomes stuck. %In such cases, the robot takes special measures to leave the area before resuming its main behaviour.
When an obstacle is first detected, via the event \RC{obstacle}, guarding the transition from \RC{Going} to \RC{Avoiding}, the clock \RC{T} is reset~(\RC{$\#$T}). 

In the \RC{entry} action of \RC{Avoiding}, the distance travelled so far is obtained using the event \RC{odometer}, and recorded in a variable \RC{d0}. An operation \RC{changeDirection()} is then called so that the robot moves away from the obstacle. The \RC{wait(evadeTime)} action gives the robot an amount of time defined by the constant \RC{evadeTime} to move away. In \RC{Avoiding}, a \RC{resume} event leads back to \RC{Waiting}, a \RC{stop} event leads to \RC{Found}, and \RC{turn} to the state \RC{TryingAgain}. 

In \RC{TryingAgain}, the robot, having changed direction, will (attempt to) \RC{move} again (The \RC{entry} action calls the operation \RC{move} with arguments given by constants \RC{lv} and \RC{a} defining a linear velocity and an angle.).
If, in this state, another \RC{obstacle} is detected, \RC{Movement} enters the state \RC{AvoidingAgain}, getting a new \RC{odometer} measure \RC{d1}.

In \RC{AvoidingAgain}, occurrence of an event \RC{resume} leads back to \RC{Waiting}, and \RC{stop} leads to \RC{Found}. Another two transitions have guards that check whether enough time has elapsed since the first \RC{obstacle} was detected~(that is, the value \RC{since(T)} of the clock \RC{T} is greater than or equal to a value \RC{stuckPeriod}) but little progress has been made~(that is the difference between \RC{d1} and \RC{d0} is less than or equal to a value \RC{stuckDist}). If so, the robot is deemed to be stuck, and \RC{Movement} transitions to the state  \RC{GettingOut}. Otherwise, it returns to \RC{Avoiding} to continue the standard avoidance process.

%In most states, except when actively avoiding an obstacle (in the state \RC{AvoidingAgain}), \RC{Movement} can respond to \RC{turn} requests, receive a \RC{resume} command to initiate a random walk, or a \RC{stop} command indicating that a high-intensity gas has been found. In the latter case, the machine transits to \RC{Found}, drops a flag, and terminates movement.

%\RC{Movement} also makes explicit use of time primitives, such as clocks and wait actions, to capture the duration of operations.
In the \RC{entry} action of \RC{GettingOut}, the call to the operation \RC{shortRandomWalk()} is an attempt to move away from the stuck position, \RC{wait(outPeriod)} specifying a minimum time budget for this attempt.
%The transition from \RC{AvoidingAgain} to \RC{Avoiding} is guarded by timing constraints based on clock resets \RC{$\sharp$T} and the expression \RC{since(T)}, which yields the elapsed time since the most recent reset of clock \RC{T}.
Similar to other states, events \RC{resume}, \RC{turn}, and \RC{stop} are accepted. 

\begin{table}[t]
\caption{Requirements for the Chemical Detector}
\label{table:mvmt_requirements}\centering
\begin{small}
\begin{tabular}{c p{6.5cm}}
\toprule
No. & Requirement \\
\hline
R1 & If a dangerous gas is detected, the robot shall flag to report the detection. \\
R2 & Once an obstacle is detected, the location of the obstacle shall be recorded within one time unit. \\
R3 & The chemical detector shall be deadlock-free. \\

\bottomrule
\end{tabular}
\end{small}
\end{table}

\begin{table}[t]
\caption{Example claims for the chemical detector}
\label{table:mvmt_claims}\centering
\begin{small}
\begin{tabular}{c p{4.5cm} p{2.0cm}}
\toprule
No. & Claim & Verification Support \\
\hline
C1 & Requirement R1 is implemented. & FDR \\
C2 & Requirement R2 is implemented. & FDR \\
C5 & Requirement R3 is implemented. & FDR, Isabelle \\
\bottomrule
\end{tabular}
\end{small}
\end{table}

The set of requirements for the chemical detector shown in Table~\ref{table:mvmt_requirements} has been identified.
They occur in an AC as part of claims summarised in Table~\ref{table:mvmt_claims}.
Each claim asserts that the corresponding requirement has been implemented, and is linked to a specific verification method.
In our case, the FDR model checker and the Isabelle theorem prover provide the required verification support.

%Applying the approach outlined in Fig.~\ref{fig:evidence_generation_method}, the process begins with Kapture, which converts the requirements referenced in the claims into structured requirements.Formal assertions are then generated using our assertion generator.  The software is subsequently verified against these assertions, and the verification results are used to generate AC evidence, which is integrated into the AC structure.  

%In the remainder of this paper, Sect.~\ref{sec: claim in CNL} discusses the activities in Step~1,  Sect.~\ref{model checking assertion generation} addresses the activities in Step~2,  Step~3 is automated by the verification tools FDR, PRISM, and Isabelle and is therefore not discussed further,  and Step~4 is described in Sect.~\ref{evidence model gen for model checking}.  

In the next sections, we describe the steps of our process~(see Fig.~\ref{fig:evidence_generation_method}) in detail, and use this example to illustrate the application of the techniques. 

%% file: Requirement_structuring.tex
\section{Requirement structuring\label{sec: claim in CNL}}
This section describes Step~1 of our process, in which predefined templates are used to structure requirements written in natural language to enable their formalisation and verification. 
The remainder of this section is structured as follows. 
Sect.~\ref{claim Classification} presents a classification of software requirements covered by our framework. This provides a conceptual foundation for our new templates.  Sect.~\ref{Kapture Template Design} describes Kapture templates used in the design of our templates, which are then presented in 
Sect.~\ref{subsec:Claim Template Design}. % describes the software requirement templates, following the classification provided in Sect.~\ref{claim Classification}.
Finally, Sect.~\ref{Claim Structuring} describes how to use our templates.

\subsection{Requirement coverage and classification}\label{claim Classification}
The software requirements referenced in the claims, and subject to formal verification, can be classified according to their behavioural characteristics.  
This classification does not constitute a general taxonomy of robotic software requirements, but rather a coverage-oriented classification of the types of requirements that can be formally verified within our framework.  
Each category captures an aspect of a software behaviour, reflecting the diversity of properties that can be expressed and verified for RoboChart models.

Table~\ref{table:req_coverage} summarises the types of requirements and provides a concise overview of the kinds of properties that can be verified.  
The classification distinguishes between untimed and timed safety, structural correctness, quantitative probabilistic aspects, and qualitative temporal relationships, each representing a different dimension of behavioural assurance for a software.

\begin{table*}[t]
\caption{Requirement types supported by the proposed framework for RoboChart models.}
\label{table:req_coverage}
\centering
\begin{small}
\begin{tabular}{p{3.6cm} p{13cm}}
\toprule
\makecell[l]{\textbf{Requirement category}} &
\makecell[l]{\textbf{Description}} \\
\midrule
Untimed safety &
Event-based safety constraints defined over untimed behaviour. \\[3pt]

Timed safety &
Safety requirements involving explicit model-level time, such as bounded-response or deadline constraints. \\[3pt]

General correctness &
Structural and behavioural correctness properties, including state reachability, software component termination, divergence-freedom, and deadlock-freedom. \\[3pt]

Probabilistic properties &
Quantitative constraints on event likelihoods or performance measures, including probability thresholds and reward-based expectations. \\[3pt]

Temporal properties &
Qualitative properties expressed in temporal logics, capturing ordering or eventuality relations between events. \\
\bottomrule
\end{tabular}
\end{small}
\end{table*}

For illustration, we map the requirements of the chemical detector in Table~\ref{table:mvmt_requirements} to the categories in Table~\ref{table:req_coverage}.
Requirement~R1 is an untimed property. %, specifying that whenever dangerous gas is detected, the robot shall flag the event to report it.
Requirement~R2 is an example of a timed property, as it constrains the time within which the robot shall record the obstacle location. % to one time unit after detection.
Finally, Requirement~R3 is a general correctness property. %, ensuring that the chemical detector’s behaviour is deadlock-free during operation. 
Although the chemical detector model does not incorporate probabilistic features, if it were extended with stochastic behaviours, a probabilistic property could specify, for instance, that a dangerous gas is detected with probability at least~0.95, or that the robot successfully escapes from a stuck situation with a probability above a given threshold.
A reward-based property could also evaluate the expected number of steps or time units required to recover from being stuck.
A temporal property could express that whenever the robot becomes stuck, it eventually resumes normal movement.

In summary, these categories capture behavioural aspects of a software addressed by our framework and provide a structured basis for associating each type of requirement with suitable formal semantics and verification tools.  
%They serve as the conceptual foundation for the requirement templates presented in the next section.

\subsection{Base templates in Kapture
\label{Kapture Template Design}}

This section introduces the Kapture templates that form the basis for our requirement-template design. %, together with the adaptations needed for RoboChart requirements.  
 A requirement in Kapture is defined using three %hierarchical 
 components.   
%
%\begin{itemize}
  %\item 
  A \textit{data dictionary}, provides the basic elements of requirements, such as events, time limits, probability, and reward values.  
  For example, an %obstacle detection 
  event \emph{obstacle} can be defined as a signal in the data dictionary and later used in requirements.  

  %\item 
  A \textit{definitions} component specifies functions. % that serve as an intermediary between the data dictionary and requirements.  
  For instance, we define a definition of \texttt{reachable(scope, state)} to check whether a given \texttt{state} is reachable within a given \texttt{scope}, that is, a composite state, machine, controller, and so on.  
  %Such definitions modularise and simplify requirement specifications.  

  %\item 
  Finally, \textit{requirements} are constructed using predefined templates for different types of constraints.  
%\end{itemize}
Kapture provides seven requirement templates. We use 
 three of them:~% are selected and customised including 
 \texttt{when}, \texttt{trigger on event}, and \texttt{every} described below.  
%These are sufficient to capture all requirement categories defined in Sect.~\ref{claim Classification}, while the rest of the templates overlap in functionality and are not needed here.  
Each template defines a structured form with optional sections; we describe those we use.
%We adapt these forms by retaining only the relevant sections needed for RoboChart requirements.   The three customised base templates are described below.  

\paragraph{The \texttt{when} template.}  
This template captures causality between events:~when one condition holds, another must eventually be observed until a termination condition occurs.  
We use \texttt{when} templates of the following form.   

\begin{tcolorbox}[colback=white, boxrule=0.5pt, arc=0pt, halign=flush left, left=2pt, right=2pt, top=2pt, bottom=2pt, boxsep=2pt]
\begin{footnotesize}
When \textcolor{blue}{\textit{GuardCondition}} occurs/is true,\\\vspace{0.5ex}
until \textcolor{blue}{\textit{UntilCondition}} occurs,\\\vspace{0.5ex}
\textcolor{blue}{\textit{RequiredCondition}} shall be seen/true.\\
\end{footnotesize}
\end{tcolorbox}
\noindent
In the templates, we show placeholders, for instance,  \textit{GuardCondition}, in italics. These placeholders can be instantiated using the elements defined in the data dictionary. The template captures a causal relationship between events:~once the \emph{GuardCondition} becomes true, the software shall satisfy the \emph{RequiredCondition} until the specified \emph{UntilCondition} occurs. %(the termination or stopping condition).
The \emph{Until} section is optional.

\paragraph{The \texttt{trigger on event} template.}  
This template specifies time-bounded responses:~if a triggering event occurs, a required condition must hold within a given duration.  

\begin{tcolorbox}[colback=white, boxrule=0.5pt, arc=0pt, halign=flush left, left=2pt, right=2pt, top=2pt, bottom=2pt, boxsep=2pt]
\begin{footnotesize}
If ever \textcolor{blue}{\textit{TriggerCondition}} occurs,\\\vspace{0.5ex}
then at some point,
but within \textcolor{blue}{\textit{WithinDuration}},\\\vspace{0.5ex}
\textcolor{blue}{\textit{RequiredCondition}} shall be true.
\end{footnotesize}
\end{tcolorbox}

\paragraph{The \texttt{every} template.}  This template expresses invariants, stating that a condition must always or never hold.  

\begin{tcolorbox}[colback=white, boxrule=0.5pt, arc=0pt, halign=flush left, 
  left=2pt, right=2pt, top=2pt, bottom=2pt, boxsep=2pt]
\begin{footnotesize}
The \textcolor{blue}{\emph{Condition}} shall \textcolor{blue}{\emph{AlwaysOrNever}} be true. \\[4pt]
$\ast$ \emph{AlwaysOrNever} ::= \texttt{always} ~|~ \texttt{never}
\end{footnotesize}
\end{tcolorbox}
%
%In summary, the three customised base templates of \texttt{when}, \texttt{trigger on event}, and \texttt{every} provide sufficient expressive power to capture the full spectrum of requirement categories defined in Sect.~\ref{claim Classification}.  
\noindent%
Each individual requirement template we have designed is derived from one of these base templates through further specialisation.  
Our templates and their mapping to requirement categories are presented next.  

\subsection{Requirement-template design\label{subsec:Claim Template Design}}

%This section defines the requirement templates~(RTs) derived from the three Kapture bases introduced in Sect.~\ref{Kapture Template Design}.  
We have designed nine templates:~one for each requirement category in Table~\ref{table:req_coverage} and aligned with the corresponding verification tool~(FDR, PRISM, or Isabelle). We show here on four representative templates:~untimed safety, timed safety with deadlines, deadlock-freedom, and reward-based, which illustrate the main patterns and their instantiations.  
The remaining templates are in Appendix~\ref{sec:appendix:RTs}.
%Each template is accompanied by a minimal example to illustrate its instantiation.  

\paragraph{Placeholders and event naming.}  Templates use placeholders for RoboChart components.  
The most common placeholder is a RoboChart event, whose names are elements of the syntactic category \textit{fqEvent} defined as follows.

\begin{tcolorbox}[colback=white, boxrule=0.5pt, arc=0pt,
  left=2pt, right=2pt, top=2pt, bottom=2pt, boxsep=2pt]
\begin{footnotesize}
\begin{tabularx}{\linewidth}{ @{} l @{\hspace{0.1em}} l @{\hspace{0.35em}} X @{} }
\textit{fqEvent} & ::= &
  \textcolor{blue}{\textit{qualifiedScope}} \textcolor{blue}{\textit{eventName}}\\
 & & 
       [\,\texttt{.in}\, |\, \texttt{.out}\,]
       [\,\textcolor{blue}{\textit{valueOrVariable}}\,] \\[-1pt]\noalign{\vskip 8pt}

\textit{qualifiedScope} & ::= &
   [\,\textcolor{blue}{\textit{module}}\,\texttt{::}\,] 
   [\,\textcolor{blue}{\textit{controller}}\,\texttt{::}\,] \\ & &
   [\,\textcolor{blue}{\textit{stateMachine}}\,\texttt{::}\,] \\[-1pt]\noalign{\vskip 8pt}

\textit{eventName} & ::= & 
   \textcolor{blue}{\textit{RoboChartEventIdentifier}} \\[-1pt]\noalign{\vskip 8pt}

\textit{valueOrVariable} & ::= & 
   \texttt{.}\textcolor{blue}{\textit{value}} ~|~ \texttt{?}\textcolor{blue}{\textit{variable}}\\
\end{tabularx}
\end{footnotesize}
\end{tcolorbox}
\noindent
For readability, we follow a typographic convention in all grammar-style definitions.  
Non-terminals appear in \emph{italic black}, while references to non-terminals appear in \textcolor{blue}{\emph{blue italic}}.  
Terminals %, such as fixed keywords or separators (e.g.\ \texttt{::}, \texttt{.in}, \texttt{.out}, \texttt{buffered\_}), 
are shown in \texttt{monospace}.
In addition, we use standard EBNF symbols. %:~square brackets $[\,\cdot\,]$  for optional parts, curly braces $\{\,\cdot\,\}$ for zero or more  repetitions, and the vertical bar `\,$|$\,` for alternatives.

In RoboChart models, events are usually referred to by simple names (such as, \RC{flag}).
For CSP assertions, however, they are expanded into fully qualified events following the syntax above.
For instance, the event \RC{flag} of the state machine \RC{Movement} of the controller \RC{ctrl} in the module \RC{sys} is referred to as
\texttt{sys::ctrl::Movement} \texttt{::flag.out}.%, and the event \RC{flag} at the RoboChart module level is \texttt{sys::flag.out}.

Here, \texttt{::} separates modules, controllers, and state machines in the scope; events may be annotated with \texttt{.in} or \texttt{.out} to indicate input/output direction, and with \texttt{.}\textit{value} or \texttt{?}\textit{variable} if data is exchanged.
% The optional prefix \texttt{buffered\_} denotes asynchronous communication, which is captured in the CSP semantics by an additional buffer process, more details are provided in Appendix~\ref{sec:appendix:ATs}.

Instantiating placeholders with fully qualified event names is the first step in producing formal assertions. This requires only knowledge of the structure of the RoboChart model and the names of its components, not of CSP. %, thereby lowering the entry barrier for users. As defined above, the syntactic category \textit{fqEvent} is that of fully qualified event names.
%Based on this notation, we now present our first requirement template.

\paragraph{Untimed requirements.}  
These requirements typically express that the occurrence of a \emph{guard event} must be eventually followed by the occurrence of a \emph{required event}.  
We capture this pattern in the untimed template (RT-UNTIMED), which is a specialisation of the Kapture \texttt{when} template using \texttt{When} and \texttt{Required} sections.

\begin{tcolorbox}[colback=white, boxrule=0.5pt, arc=0pt, halign=flush left, 
  left=2pt, right=2pt, top=2pt, bottom=2pt, boxsep=1pt]
\begin{footnotesize}
\textbf{RT-UNTIMED} \label{RT-UNTIMED}  \\[4pt]
When \textcolor{blue}{\emph{guardEvent}} occurs,\\\vspace{0.5ex}
\textcolor{blue}{\emph{requiredEvent}} shall also be seen.\\
\vspace{1.5ex}

$\ast$ \emph{guardEvent} \;\;\;\,::= \textcolor{blue}{\emph{fqEvent}}\\\vspace{0.5ex}
$\ast$ \emph{requiredEvent} ::= \textcolor{blue}{\emph{fqEvent}}\\
\end{footnotesize}
\end{tcolorbox}
\noindent

%%%%%%%%%%%%%% explanation about eventually and immediately %%%%%%%%%%%
% \textit{Remark.} In this work, the phrase ``shall be seen'' in RT-UNTIMED 
% is interpreted as ``shall eventually occur'': after the guard event takes 
% place, the required event is guaranteed to occur at some later point, but 
% not necessarily as the next event. 
% Kapture also supports a stricter interpretation by adding the field 
% \texttt{and only when} after \texttt{when}, meaning that the required event 
% must follow immediately after the guard event with no intervening events. 
% Since such strict requirements are uncommon in our case studies, we focus 
% here on the eventual interpretation.

% \noindent Example (R-UNTIMED):  
% \emph{If dangerous gas is detected, the robot shall flag to report the detection.}

\noindent%
Here, there are two placeholders, \emph{guardEvent} and \emph{requiredEvent}, which both have the type \textit{fqEvent}. When instantiating such a template, every placeholder must be assigned a correctly typed expression.
The requirement R1 of our running example~(see Table~\ref{table:mvmt_requirements}) can be expressed using RT-UNTIMED as follows.
\begin{tcolorbox}[colback=white, boxrule=0.5pt, arc=0pt, halign=flush left, 
  left=2pt, right=2pt, top=2pt, bottom=2pt, boxsep=2pt]
\begin{footnotesize}
When \texttt{sys::stop.in} occurs,\\
\texttt{sys::flag.out} shall also be seen.
\end{footnotesize}
\end{tcolorbox}
\noindent %
In this example, “dangerous gas is detected” is represented by the input event \RC{stop} in module \RC{sys}, while “the robot shall flag to report the detection” is represented by the output event \RC{flag}.  
Here, the placeholders \emph{guardEvent} and \emph{requiredEvent} from the template are instantiated with the concrete RoboChart events \texttt{sys::stop.in} and \texttt{sys::flag.out}, identified by their fully qualified names. %ollowing the naming convention of the CSP semantics of RoboChart.  
%This instantiation establishes a direct link between the informal requirement and its CSP assertion, thereby enabling automated generation as detailed in Sect.~\ref{Assertion Template design}.

\paragraph{Timed safety requirements}  
%Timed safety requirements are essential in real-time softwares, where guaranteed response times are critical for correctness and safety.  
A particularly common pattern is the deadline requirement, in which the occurrence of a trigger condition initiates a bounded-time obligation for a required condition to be fulfilled.  
Specifically, for each instance of the trigger event, the required event must occur within a specified time bound: it may happen earlier, but not later.  
%While this covers a wide range of practical cases, more complex timing scenarios fall outside the scope of this section and will be addressed in future work.  
We capture the deadline pattern in the timed functional template (RT-TIMED), which is based on the \texttt{trigger on event} template. 

\begin{tcolorbox}[colback=white, boxrule=0.5pt, arc=0pt, halign=flush left, 
  left=2pt, right=2pt, top=2pt, bottom=2pt, boxsep=2pt]
\begin{footnotesize}
\textbf{RT-TIMED} \label{RT-TIMED}\\[4pt]
If ever \textcolor{blue}{\emph{triggerEvent}} occurs,\\\vspace{0.5ex}
then at some point, but within \textcolor{blue}{\emph{deadline}},\\\vspace{0.5ex}
\textcolor{blue}{\emph{targetEvent}} shall be true.\\
\vspace{1.5ex}

$\ast$ \emph{triggerEvent} ::= \textcolor{blue}{\emph{fqEvent}}\\\vspace{0.5ex}
$\ast$ \emph{targetEvent} \,~::= \textcolor{blue}{\emph{fqEvent}}\\\vspace{0.5ex}
$\ast$ \emph{deadline} ~~~~~~::= \texttt{(}\emph{nat}\texttt{) rounds}\\
\end{footnotesize}
\end{tcolorbox}
%
%\noindent Example (R-TIMED):   \emph{Once an obstacle is detected, the location of the obstacle shall be recorded within one time unit.}
\noindent%
The requirement R2~(in Table~\ref{table:mvmt_requirements}) can be expressed using the RT-TIMED template as follows.

\begin{tcolorbox}[colback=white, boxrule=0.5pt, arc=0pt, halign=flush left, 
  left=2pt, right=2pt, top=2pt, bottom=2pt, boxsep=2pt]
\begin{footnotesize}
If ever \texttt{sys::obstacle.in} occurs,\\\vspace{0.5ex}
then at some point, but within \texttt{1 rounds},\\\vspace{0.5ex}
\texttt{sys::odometer.in} shall be true.\\
\end{footnotesize}
\end{tcolorbox}

\noindent In this instantiation, the placeholders \emph{triggerEvent}, \emph{deadline}, and \emph{targetEvent} are replaced by the concrete RoboChart events \texttt{sys::obstacle.in}, \texttt{1 rounds}, and \texttt{sys::odometer.in}, respectively.  
Since Kapture provides \emph{rounds} as a time unit, we adopt \texttt{1 rounds} as a practical encoding of one time unit.  
%This instantiation establishes that the time spent between the identification of the obstacle and the recording of the length travelled should not exceed 1 round, thereby directly connecting the informal deadline requirement with its CSP counterpart.  

We recall that general requirements encompass properties such as deadlock freedom, divergence freedom, reachability, and termination, which are verified using FDR or Isabelle.  
To develop a structured template for these general properties, the base template \texttt{every} is employed.

\paragraph{Deadlock-freedom requirements}  
%Deadlock freedom requirements specify whether a RoboChart component shall or shall not be deadlock-free.  
We capture this pattern in the deadlock-freedom template (RT-DDLK).  

\begin{tcolorbox}[colback=white, boxrule=0.5pt, arc=0pt, halign=flush left, 
  left=2pt, right=2pt, top=2pt, bottom=2pt, boxsep=2pt]
\begin{footnotesize}
\textbf{RT-DDLK} \label{RT-DDLK}\\[4pt]
The property \textcolor{blue}{\emph{function}}(\textcolor{blue}{\emph{scope}}) 
shall \textcolor{blue}{\emph{boolStatement}} be true.\\
\vspace{1.5ex}

$\ast$ \emph{function}\, ~~~~~~~
::= \,\texttt{deadlock\_free\_untimed} ~|\\ 
~~~~~~~~~~~~~~~~~~~~~~~~~~~~~\texttt{deadlock\_free\_timed} ~|\\ ~~~~~~~~~~~~~~~~~~~~~~~~~~~~~\texttt{deadlock\_free} ~|\\~~~~~~~~~~~~~~~~~~~~~~~~~~~~ \texttt{deadlock\_free\_isa}\\\vspace{0.5ex}
$\ast$ \emph{scope}\, ~~~~~~~~~~~~~~::= \textcolor{blue}{\emph{RoboChartComponentIdentifier}}\\
$\ast$ \emph{boolStatement} ~~::= \texttt{always} ~|~ \texttt{never}\\
\end{footnotesize}
\end{tcolorbox}

\noindent%
As discussed in Sect.~\ref{sec:background:CSP}, compiling a RoboChart model into CSP$_\mathrm{M}$ produces both an untimed and a timed semantics.  
When verifying assertions in FDR, the intended semantics can be selected by labelling the assertion as \texttt{untimed} or \texttt{timed}; if no label is provided, both variants are checked by default.  
Accordingly, the template provides three functions for FDR:~ 
\texttt{deadlock\_} \texttt{free\_untimed} checks use the untimed semantics,  
\texttt{dead-} \texttt{lock\_free\_timed} the timed semantics,  
and \texttt{deadlock\_} \texttt{free} leaves the semantics unspecified. %, thereby triggering FDR’s default behaviour of evaluating both.  

In addition, \texttt{deadlock\_free\_isa} indicates verification in Isabelle/HOL, using the Z-Machine mechanisation of RoboChart state machines~(Sect.~\ref{sec:background:z-machine}).  
This method~\cite{yan2023automated} automatically generates a deadlock-freedom condition and discharges it using proof tactics, providing an alternative that mitigates state explosion.  

% \noindent Example (R-DDLK):  
% \emph{The Movement state machine shall be deadlock-free.}  

Requirement R3 from Table~\ref{table:mvmt_requirements} can be encoded as follows. The placeholder \emph{\color{blue}scope} is instantiated to the state machine \RC{Movement}.  The chosen function is \texttt{dead-} \texttt{lock\_free}, and the Boolean statement is \texttt{always}.  
%Instantiated using the RT-DDLK template, the requirement becomes:  

\begin{tcolorbox}[colback=white, boxrule=0.5pt, arc=0pt, halign=flush left, 
  left=2pt, right=2pt, top=2pt, bottom=2pt, boxsep=2pt]
\begin{footnotesize}
The property \texttt{deadlock\_free}(\texttt{Movement}) 
shall \texttt{always} be true.
\end{footnotesize}
\end{tcolorbox}

\paragraph{Reward-based requirements}
As said, they capture quantitative objectives over execution paths, and are verified using PRISM.  
They typically state that a reward target must be achieved for given constant values and reward specifications.  
We have designed the reward-based template (RT-RWD) using the Kapture \texttt{when} template as follows.

\begin{tcolorbox}[colback=white, boxrule=0.5pt, arc=0pt, halign=flush left, 
  left=2pt, right=2pt, top=2pt, bottom=2pt, boxsep=2pt]
\begin{footnotesize}
\textbf{RT-RWD} \label{RT-RWD}\\[4pt]
When \textcolor{blue}{\emph{constantsConfigs}} and \textcolor{blue}{\emph{rewardEvent}} and \textcolor{blue}{\emph{rewardValue}} is true,\\\vspace{0.5ex}
Until \textcolor{blue}{\emph{pathFormula}} becomes true,\\\vspace{0.5ex}
\textcolor{blue}{\emph{rewardTarget}} shall also be true.\\
\vspace{1.5ex}

$\ast$ \emph{pathFormula} ~\,::= \texttt{pathFormula\_}\textcolor{blue}{\emph{RPathFormula}}\\\vspace{0.5ex}
$\ast$ \emph{RPathFormula} ::=  (\texttt{Reachable} ~|~ \texttt{LTL} ~|~ \texttt{Cumul} ~|~ \\\vspace{0.5ex}~~~~~~~~~~~~~~~~~~~~~~~~~~~~~~~~~~~\texttt{Total}) \textcolor{blue}{\emph{pExpr}}\footnote{The full syntax of \emph{pExpr} is given in~\cite{Miyazawa2016}. One of its forms, \emph{QualifiedNameToElement}, denotes a RoboChart element.}\\\vspace{0.5ex}
$\ast$ \emph{rewardTarget} \,~::= \texttt{reward\_target\_}\textcolor{blue}{\emph{op}}~\textcolor{blue}{\emph{Expr}}\\\vspace{0.5ex}
$\ast$ \emph{op} ~~~~~~~~~~~~~~~~::= \texttt{>} ~|~ \texttt{>=} ~|~ \texttt{<} ~|~ \texttt{<=} ~|~ \texttt{==}\\\vspace{0.5ex}
$\ast$ \emph{rewardEvent} ~~::= \texttt{reward\_event\_}\textcolor{blue}{\emph{fqEvent}}\\\vspace{0.5ex}
$\ast$ \emph{rewardValue}~~ \,::= \texttt{reward\_value\_}\textcolor{blue}{\emph{Expr}}\\
\end{footnotesize}
\end{tcolorbox}

\noindent%\textit{Remark.}  
Although RT-RWD is written using the \texttt{when} template skeleton, the 
semantics of its three sections differ from those of functional requirements, where the \texttt{When}, \texttt{Until}, and \texttt{Required} sections express a guard-scope-obligation pattern.
For reward-based requirements, we adopt the \texttt{when} skeleton, interpreting its three sections in a specialised way:
the \texttt{When} section sets constant configurations and reward specifications
(\emph{constantsConfigs}, \emph{rewardEvent}, \emph{rewardValue}); the
\texttt{Until} section encodes the \emph{pathFormula}; and the \texttt{Required}
section gives the quantitative bound (\emph{rewardTarget}) to be satisfied.

Since this requirement is to be formalised as an assertion in the PRISM reactive-modules language, to facilitate the assertion generation, %in Sect.~\ref{model checking assertion generation}, 
the placeholders %used in RT-RWD intentionally
follow the terminology of PRISM. %~(Sect.~\ref{sec:background:PRISM}).  
We describe each of the sections next.

\paragraph{When section}  
This corresponds to the line of the template that starts with the keyword {When}. It sets constant values for PRISM verification and introduces reward specifications. Instantiation of the placeholder \emph{constantsConfigs} specifies one or more RoboChart constants with assigned values~(single assignments or ranges). %, which parameterise the generated PRISM model.  
Its full grammar is given in Appendix~\ref{app:constants-configs}. Instantiation of the placeholder  
\emph{\color{blue}rewardEvent} identifies the event contributing to the reward and its name must be prefixed with ``\texttt{reward\_event\_}''. %serving as a label during assertion generation to identify the type of the assertion to be created.
Instantiation of  \emph{rewardValue} gives the amount added per occurrence of \emph{\color{blue}rewardEvent}. That expression must be labelled with the prefix ``\texttt{reward\_value\_}''.

\paragraph{Until section.} 
This corresponds to the line of the template starting with Until. 
It specifies the path condition~(described by a \emph{\color{blue}pathFormula}) under which the reward is accumulated.  
Here, \emph{pathFormula} is labelled with the prefix ``\texttt{pathFormula\_}'' and given by a term of the syntactic category \emph{RPathFormula} from the Robo\-Chart assertion language~ (see Sect.~\ref{sec:background:PRISM} and\cite{Miyazawa2016}).  
%We include the definition of \emph{RPathFormula} here so that the subsequent example can be followed without requiring prior knowledge of PRISM.

\paragraph{Required section.} This corresponds to the line of the template starting with a quantitative bound (\emph{rewardTarget}) to be satisfied by the accumulated reward, whose definition must be labelled with the prefix ``\texttt{re-} \texttt{ward\_target\_}''.

\bigskip
\noindent%
We illustrate the use of this template in Section~\ref{sec:cases}.

\subsection{Requirement formalisation\label{Claim Structuring}}

% This section corresponds to Step~1 in Fig.~\ref{fig:evidence_generation_method}.
% The requirement templates introduced in Sect.~\ref{subsec:Claim Template Design} are used as input to this step, guiding the translation of natural language requirements into a structured form in Kapture.

\rc{1.3(a)}{
Requirement formalisation in Kapture remains a manual activity and requires familiarity with several aspects of the RoboChart model under verification. 
First, the software engineers need to understand the structure of the RoboChart model, including module, controller, and state machine names, which is used to construct the fully qualified event names such as \texttt{sys::ctrl::Movement::flag.out}. 
This is easily achieved by understanding the hierarchical structure of the RoboChart models; with graphical RoboChart models, this structure is straightforward to grasp. 
Second, the engineers should be able to distinguish between input and output events (e.g., \texttt{.in} vs.\ \texttt{.out}) and identify whether an event carries data. 
This can be facilitated by checking the CSP$_{\mathrm{M}}$ models derived from RoboChart in Robo\-Tool.
Third, the engineers shall choose the appropriate requirement template for each requirement, for instance, distinguishing between untimed and timed safety requirements, or between probabilistic and reward-based properties, which requires an understanding of the classification in Table~\ref{table:req_coverage}.
Since the currently available templates are distinct from one another, this selection process is not difficult.

Common errors may lie in: (i)~resolving the correct fully qualified name for an event, especially in models with multiple controllers or nested state machines;
(ii)~determining the appropriate scope for reachability or deadlock-freedom properties; 
and (iii)~an incorrect instantiation of the requirement template. These difficulties are relatively manageable, but they represent a non-trivial skill requirement that should be considered when applying the approach. 
But none of these requires the engineers to have a deep understanding of the CSP and PRISM languages. 
Therefore, the templates do shield the engineers from the underlying CSP and PRISM syntax, which would otherwise require formal methods expertise.}

\rc{1.3}{\sout{Writing a requirement in Kapture is a manual activity.
The user is expected to know the RoboChart model under consideration and to follow the naming conventions of RoboChart events and variables, % in their CSP and PRISM semantics, 
and also be familiar with the syntax of the RoboChart assertion languages. Importantly, however, no expertise in the underlying formal languages~(CSP or PRISM) is required, as the templates shield the user from these details.}}

For illustration, we consider the requirement template R-UNTIMED again. % from Sect.~\ref{subsec:Claim Template Design}.
The process of obtaining the encoding of a requirement using this template begins by declaring \emph{\color{blue}guardEvent} and \emph{\color{blue}requiredEvent} in Kapture’s \emph{data dictionary}, using RoboChart events as their identifiers.
These declared events are then referenced in the \emph{requirements} section, where the \texttt{when} template is instantiated.
%For requirement R-UNTIMED, the placeholders \emph{guardEvent} and \emph{requiredEvent} are filled with the previously defined entries from the Data Dictionary, while the \emph{Until} and \emph{Otherwise} parts of the template are not required.

% Kapture allows each requirement to carry traceability metadata. 
% % This connects the software requirement to both its source~(backward traceability to the originating claim) and its verification target~(forward traceability to the supporting claim). 
% Backward traceability links the requirement to its corresponding claim in the AC, 
% while forward traceability links it to the lower-level claim that specifies the verification method or tool to be used. 
\rc{2.62}{Kapture allows requirements to carry traceability information that is essential for automated evidence integration. Two types of links are used. \textit{Backward traceability} links the structured requirement to the AC claim that it supports~(e.g., the requirement-level claim ``Requirement R1 is implemented'').
This link identifies which claim the generated evidence should ultimately support. 
\textit{Forward traceability} links the requirement to a lower-level verification-method claim that specifies how the requirement is to be verified~(e.g., by FDR model checking).
This lower-level claim is directly supported by the verification evidence generated from the requirement.
The overall structure of such an AC is illustrated later in Fig.~\ref{fig:Mail_robot GSN before PRISM}~(Section~\ref{sibsubsec: mail-robot}), where each software-level requirement is specified as a verification-method claim.
These traceability links enable the automated integration of verification evidence into the SACM AC model, as described in Sect.~\ref{evidence model gen for model checking}.}
% This lower-level claim is the one that will be directly supported by our verification evidence.
% In our work, we use such links to integrate verification evidence into the SACM AC model automatically.

Kapture supports multiple output formats, including XML, HTML, and CSV. In our process, the XML export is used.
The result of Step~1 is therefore a semi-formal XML file that encodes: (i)~the requirements instantiated from templates, and (ii)~their associated traceability links.
This XML file serves as the structured input for the assertion generation of Step~2~(see Fig.~\ref{fig:evidence_generation_method}).

%% file: Assertion_generation.tex
%%%%%%%%%%%%%%%%%%%%%%%%%%%%%%%%%%%%%%%
%%%%%%%%%%%%%%%%%%%%%%%%%%%%%%%%%%%%%%%

\section{Assertion generation\label{model checking assertion generation}}

This section presents our approach in Step~2 of Fig.~\ref{fig:evidence_generation_method}, the generation of assertions. 
The outputs of this step are assertions in the RoboChart assertion language, for model checking, and in \rc{2.63}{Isabelle's formal proof language}, Isar, for theorem proving. 
They are obtained by instantiating assertion templates with the structured requirements produced in Step~1.

For model checking, the generated RoboChart assertions are translated by RoboTool into CSP{$_\mathrm{M}$} or PRISM assertions, and verified against the RoboChart models using FDR and PRISM. 
For theorem proving, RoboChart models are transformed into Z-Machines using the approach described in our previous work~\cite{yan2023automated}. These Z-Machines are then verified against Isar assertions in Isabelle/HOL.

The remainder of this section is structured as follows.  
Sect.~\ref{Assertion Template design} introduces the assertion templates defined for different requirement categories.  
Section~\ref{Assertion Generation} describes our assertion generation process, including template selection and template instantiation.

\subsection{Assertion-template design
\label{Assertion Template  design}}

We define a set of assertion templates~(ATs) that are derived from, and aligned with, the requirement templates introduced in Sect.~\ref{subsec:Claim Template Design}. In total, eleven assertion templates have been designed, covering all the requirement categories listed in Table~\ref{table:req_coverage}. Some requirement templates map to more than one assertion template, so the number of ATs is larger than the number of requirement templates.  
Table~\ref{Assertion Template map to claim table} summarises the mapping between requirement templates and assertion templates, and also indicates the verification tool used in each case. This section presents five representative templates in detail, while the remaining ones are provided in Appendix~\ref{sec:appendix:ATs}.

\begin{table*} [htbp]
\caption{Mapping between requirement templates and the assertion requirements}
\label{Assertion Template map to claim table}
\centering
\begin{small}
  \begin{tabular}{lllll}
\toprule
 Requirement Template     &  Section     & Assertion Template        & Section      & Tool             \\

 \hline

 \multirow{2}{*}{RT-UNTIMED }& \multirow{2}{*}{Sect.~\ref{subsec:Claim Template Design}} & AT-UTG & Sect.~\ref{Assertion Template  design}      &  \multirow{2}{*}{FDR}                                    \\

&  &  AT-UTL           &  Appendix~\ref{sec:appendix:ATs}  &                                      \\

 % & & AT-UTLB    &   Appendix~\ref{sec:appendix:ATs}     &                                           \\

 \hline

  RT-TIMED    &   Sect.~\ref{subsec:Claim Template Design}&  AT-DLINE                            &Sect.~\ref{Assertion Template  design}   & FDR              \\
  \hline
 RT-REACH      & Appendix~\ref{sec:appendix:RTs}&  AT-REACH  &Appendix.~\ref{sec:appendix:ATs} & FDR                                                        \\
 \hline
 \multirow{2}{*}{RT-DDLK }    &  \multirow{2}{*}{Sect.~\ref{subsec:Claim Template Design}}&    AT-DDLK-1   &\multirow{2}{*}{Sect.~\ref{Assertion Template  design}}      & FDR                                 \\
 &&AT-DDLK-2&& Isabelle\\
 \hline
RT-DIV  &Appendix~\ref{sec:appendix:RTs}  &  AT-DIV  &Appendix~\ref{sec:appendix:ATs}  & FDR  \\
 \hline
 RT-TERM   & Appendix~\ref{sec:appendix:RTs}&  AT-TERM    & Appendix~\ref{sec:appendix:ATs}&FDR \\
\hline
RT-PROB    & Appendix~\ref{sec:appendix:RTs}&  AT-PROB                          &Appendix~\ref{sec:appendix:ATs}& PRISM                    \\
 \hline
 RT-RWD       & Sect.~\ref{subsec:Claim Template Design}& AT-RWD &Sect.~\ref{Assertion Template  design}   & PRISM    \\
 \hline
  RT-TEMP & Appendix~\ref{sec:appendix:RTs} & AT-TEMP        & Appendix~\ref{sec:appendix:ATs}& PRISM    \\

 \bottomrule
\end{tabular}
\end{small}
\end{table*}

As explained in Sect.~\ref{sec:background:CSP} and Sect.~\ref{sec:background:PRISM}, RoboChart has a domain-specific assertion language~\cite{Miyazawa2016} that is designed to simplify the construction of model checking assertions for FDR and PRISM. This language is the basis of our assertion templates:~each template is expressed in the RoboChart assertion language. %, and the generated assertions can be automatically translated into CSP$_\mathrm{M}$/PRISM within RoboTool for verification.

\rc{1.2a}{A methodological clarification is necessary regarding the relationship between requirement templates (RTs) and assertion templates (ATs). Since Kapture requirement templates do not possess a pre-existing formal semantics, the ATs do not translate an existing formal meaning of the requirement templates. Instead, each AT \emph{provides} the formal semantics for the corresponding RT. Thus, the AT is the formalisation of the RT, not a translation of it.

For each requirement pattern (e.g., untimed response, timed deadline, reward accumulation), we adopt the well-established semantics from the literature~\cite{murray2022safety,yan2023automated,pls} and encode it directly in the CSP/PRISM/Isar constructs of the AT. The operational semantics of these constructs, as explained below for the representative templates, is therefore the formal meaning of the RT. The question of whether this semantics is ``correct'' is not a matter of proving equivalence to an absent reference, but a matter of whether the adopted pattern is appropriate for the intended requirement, which is already validated by the literature and further supported by our case studies (Sect.~\ref{subsec: mail robot}).}

\paragraph{Untimed refinement assertions}

For untimed requirements specified by the RT-UNTIMED template, we define two assertion templates:~AT-UTG and AT-UTL. AT-UTG handles requirements that apply globally, without any scoping restrictions.  
By contrast, AT-UTL is used when the requirement applies only when a state machine is in a particular state.
%Both AT-UTL and AT-UTLB are given in Appendix~\ref{sec:appendix:ATs}. 
%The specification captures the common form that whenever a \textit{guardEvent} occurs, the corresponding \textit{requiredEvent} must also occur. A typical instance could be a requirement that a variable always maintains a fixed target value or matches the value of another variable.
The assertions written using these templates are verified in the FDR model checker by refinement checking. 
Therefore, each assertion template has three parts:~
(1)~the specification process expressing the required behaviour; 
(2)~the implementation process, corresponding to the RoboChart model; and 
(3)~the refinement clause, written in the RoboChart assertion language, 
which states that the implementation refines the specification.
We check refinement in the \emph{traces} model, as our focus is on the safety aspects of the properties.
% Any violation of a safety property is witnessed by a finite trace, which can be reported as a counterexample.

The specification and implementation sections are expressed as CSP$_\mathrm{M}$ blocks delimited by the keywords \textbf{csp-begin} and \textbf{csp-end}. 
The requirement itself provides the content for the specification section.
% In this work, we consider the safety-related requirements, so that we use the CSP$_\mathrm{M}$ blocks to capture the safety aspect of the specifications and therefore check refinement in the \emph{traces} model.
The AT-UTG template is shown below. Keywords of the RoboChart assertion language are displayed in normal font. 
Placeholders are written in black italics. %these will later be instantiated with values extracted from the XML requirements. 
Fixed parts of a template. %, which can not be changed, 
are written in teletype black font. 
\begin{tcolorbox}[colback=white, boxrule=0.5pt, arc=0pt, halign=flush left,
  left=2pt, right=2pt, top=2pt, bottom=2pt, boxsep=2pt]
 \renewcommand{\baselinestretch}{1.1}\selectfont
 \begin{footnotesize}
 \textbf{AT-UTG}:\\[4pt]

\textbf{\color{eclipsePurple}{untimed csp}} Spec\\
csp-begin\\
\texttt{Spec} = \texttt{CHAOS(Events) [\!|\{\!|} \textit{guardEvent} \texttt{|\!\}|>}
\texttt{(RUN(\{\!|} \textit{guardEvent}\texttt{|\!\}) /\textbackslash (} \textit{requiredEvent} \texttt{ -> } \textit{Spec} \texttt{))}\\
csp-end\\[6pt]

\textbf{\color{eclipsePurple}{untimed csp}} Spec\_impl 
\textbf{\color{eclipsePurple}{associated to}}
\textit{module\_name}\\
csp-begin\\
\texttt{spec\_impl} = \textit{module\_name}%::O\_\_(0) 
% ~|\textbackslash \{|~\textit{guardEvent}, \\\quad \textit{requiredEvent\_without\_value}~|\!\}
\\
csp-end\\[6pt]

%%%%%%%%%%%%%%%%%%%%%%%%%%%%%%%%
%NOTE: the projection operator is removed from spec_impl, because the property is to check there is no other events between guard and required. 
% With projection, if there are other events between the guard and the required event, which should make the property unsatisfied, they would be hidden, and the property would incorrectly appear to be satisfied.
%%%%%%%%%%%%%%%%%%%%%%%%%%%%%%%%

\textbf{\color{eclipsePurple}{untimed assertion}} A\_\textit{Claim\_ID} :\\
Spec\_impl \textbf{\color{eclipsePurple}{refines}} Spec 
\textbf{\color{eclipsePurple}{in the traces model}}
\end{footnotesize}
\end{tcolorbox}

\noindent
The identifier of the assertion, that is, \textit{A\_Claim\_ID}, is derived from the claim identifier in the AC models, providing traceability between the assertion and the evidence.

In its specification section, AT-UTG uses the CSP$_\mathrm{M}$ operators 
\texttt{[\!||\!>} (Exception) and \texttt{/\textbackslash} (Interrupt)~\cite{fdr4.2}, 
together with the processes \texttt{CHAOS(A)} and \texttt{RUN(A)}. 
The process defined as \texttt{P [\!|A|\!> Q} behaves as the process \texttt{P}, but if \texttt{P} performs an event from \texttt{A} (the set of exception events), it then behaves as the process \texttt{Q}. 
The process defined as \texttt{P/\textbackslash Q} behaves as the process \texttt{P}, but if any action of \texttt{Q} is performed, \texttt{P} is abandoned and the execution proceeds with \texttt{Q}. 
\texttt{CHAOS(A)} offers a nondeterministic choice over all events in \texttt{A} but may deadlock at any point, while \texttt{RUN(A)} offers a perpetual choice over all events in \texttt{A}.

So, in AT-UTG, the specification \textit{Spec} is the process that may behave nondeterministically over all events in scope, which are those in the set ``Events''. 
If the \textit{guardEvent} occurs, however, then further occurrences of \textit{guardEvent} may still take place until \textit{requiredEvent} occurs, and then $Spec$ recurses.
%
%, and further occurrences of \textit{guardEvent} may still take place. 
% A more formal account of how this CSP specification corresponds to the intended semantics of RT-UNTIMED is given in the later paragraphs.
%
%The pattern built using CSP{$_\mathrm{M}$} operators can be used for the common requirement where when \textit{guardEvent} occurs, then \textit{requiredEvent} shall occur.
%
The implementation part of AT-UTG refers to the CSP$_\mathrm{M}$ semantics of the RoboChart model under verification.
For this reason, it directly uses the identifiers of the corresponding components. %as generated from the RoboChart semantics.
% The operator \texttt{|\!\textbackslash} (Project) is applied to restrict the system to the events relevant for this assertion and to filter out all unrelated events.
% Since the purpose of the assertion is to check refinement against the specification, the exact value the \textit{requiredEvent} carries in the implementation must not be fixed.
% Instead, the template employs \textit{requiredEvent\_without\_value}, which denotes the same event name but without its associated input/output value.
%
The third part of AT-UTG is the assertion section itself, written in the RoboChart assertion language for CSP. 
% The keyword used for refinement checking is \textbf{refines}. 

% \paragraph{Semantic correspondence of RT-UNTIMED (trace semantics).}
% The intended semantics of RT-UNTIMED is that whenever the guard event $g$ occurs, the required event $r$ shall afterwards eventually be seen. 
% When checking by CSP refinement in the \emph{traces} model, this is formulated as a safety condition:~after any $g$, the first event that is not $g$ (if any) must be $r$.
% Formally, for any finite trace $tr \in Events^{*}$ and any decomposition $tr = u \cdot g \cdot v$ (where $\cdot$ denotes concatenation, 
% $u$ is a prefix trace, $g$ is the guard event, 
% and $v$ is the remaining suffix), the suffix $v$ must either consist of further occurrences of $g$, or have the form $g^{n} \cdot r \cdot v'$ for some $n \ge 0$.

% This safety clause is captured by the AT-UTG template:
% \[
% \texttt{CHAOS(Events)[\!|\{\!|}g\texttt{|\!\}|>(RUN(\{\!|}g\texttt{|\!\})/\textbackslash (} r\texttt{->Spec))},
% \]
% Its CSP specification uses \texttt{interrupt} to ensure that, after each occurrence of the guard event $g$, only further occurrences of $g$ may appear until $r$ occurs, after which the local behaviour recurs.
% We adopt AT-UTG to capture this safety aspect and therefore check refinement in the traces model\footnote{Liveness obligations (e.g., that every $g$ is eventually followed by $r$), when required, can be verified separately (e.g., via a temporal assertion).%using \textbf{AT-TEMP}).
% }.

\paragraph{Timed refinement assertion}
As mentioned, Robo\-Chart's timed semantics is given using \textit{tock}-CSP~\cite{Roscoe1998}.
This variant of CSP models discrete time using an event   \texttt{tock} to represent the passage of one unit of time, which is an abstraction for an arbitrary amount of time:~1 ms or 1 s, for instance. CSP$_\mathrm{M}$ provides a \textbf{Timed} section for \textit{tock}-CSP processes, enabling verification in FDR.

AT-DLINE is the assertion template for timed refinement. It captures %requirements of the form:~if a \emph{triggerEvent} occurs, then the \emph{targetEvent} must take place within a given time bound (the \emph{deadline}, measured in \texttt{tock}s).
%This template corresponds to the 
requirements expressed using RT-TIMED. The specification \textit{Spec} is defined inside a \textbf{Timed} section. (The OneStep parameter defines that events are instantaneous, like in RoboChart.) %\footnote{\url{https://cocotec.io/fdr/manual/cspm/definitions.html\#csp-timed-section}}
After the \emph{triggerEvent} occurs, $Spec$ enters a timed monitoring phase in which any events may occur, but time is constrained by an interrupting waiting process \texttt{WAIT} (\emph{deadline}); \texttt{STOP\textsubscript{U}}. \texttt{WAIT(t)} is a process that engages in \texttt{t} occurrences of \texttt{tock} before terminating. \texttt{STOP\textsubscript{U}} represents a \emph{timelock} state in which no event is offered and time cannot progress.
The interrupt \texttt{/\textbackslash} in $Spec$ therefore ensures that if the waiting period expires, time cannot advance. So, the exception takes over \texttt{[\!|\{\!|\!}\textit{targetEvent}\texttt{\!|\!\}|>\,SKIP}, and so  \emph{targetEvent} has to happen and then the monitoring phase terminates (\texttt{SKIP}). 
The recursion ensures that subsequent \emph{triggerEvent}s are handled in the same way.

Finally, the wrapper \texttt{timed\_priority} is needed due to a technicality of FDR:~it enforces the \emph{maximal progress principle}. 
%By default, time in CSP advances through the event \texttt{tock}, which may occur even when internal events are still enabled. 
%Without \texttt{timed\_priority}, this could lead to premature time advancement, allowing the process to reach the deadline (\texttt{WAIT(deadline)}) and raise a timelock even though a \emph{targetEvent} might still have been reachable. 
In other words, \texttt{timed\_priority} gives precedence to internal events over \texttt{tock}, ensuring that time passes only when no further internal actions are possible.
\begin{tcolorbox}[breakable, colback=white, boxrule=0.5pt, arc=0pt, halign=flush left,
  left=2pt, right=2pt, top=2pt, bottom=2pt, boxsep=2pt]
 \renewcommand{\baselinestretch}{1.1}\selectfont
 \begin{footnotesize}
 \textbf{AT-DLINE}:\\[4pt]

\textbf{\color{eclipsePurple}{timed csp}} Spec \\
csp-begin\\
\texttt{Timed(OneStep) \{}\\
\quad \texttt{Spec} = \texttt{timed\_priority}(\\
\quad\quad \texttt{CHAOS(Events) [\!| \{\!|} \textit{triggerEvent} \texttt{|\!\} |> SKIP;}\\
\quad\quad \texttt{((CHAOS(Events) /\textbackslash (WAIT(} \textit{deadline} \texttt{); STOPU))}\\
\quad\quad \texttt{[\!| \{\!|} \textit{targetEvent} \texttt{|\!\} |> SKIP);} \texttt{Spec);}\\
\quad\texttt{\}}\\
csp-end\\[6pt]

\textbf{\color{eclipsePurple}{timed csp}} Spec\_impl 
\textbf{\color{eclipsePurple}{associated to}} \textit{module\_name}\\
csp-begin\\
\texttt{Timed(OneStep) \{}\\
\quad \texttt{spec\_impl = timed\_priority(}\\
\quad\quad \textit{module\_name} \\
\quad\quad \texttt{|\!\textbackslash \{\!|} \textit{triggerEvent}\texttt{, }\textit{targetEvent}\texttt{, tock |\!\});}\\
\quad\texttt{\}}\\
csp-end\\[6pt]

\textbf{\color{eclipsePurple}{timed assertion}} A\_\textit{Claim\_ID} :\\
Spec\_impl \textbf{\color{eclipsePurple}{refines}} Spec 
\textbf{\color{eclipsePurple}{in the traces model}}
 \end{footnotesize}
\end{tcolorbox}

The implementation section of AT-DLINE uses the \emph{Project} operator~(\texttt{$|\backslash$}) to hide, in the process under verification~(that, the process that defines the RoboChart semantics) all events, except $triggerEvent$ and $targetEvent$, as well as \texttt{tock}, so that passage of time can be observed.

% \paragraph{Semantic correspondence of AT-DLINE (timed trace semantics).}

% The intended semantics of RT-TIMED is that whenever the trigger event $g$
% occurs, the required event $r$ must occur within $d$ time units.
% Formally, in tock-CSP timed traces 
% $tr \in (Events \cup \{\texttt{tock}\})^{*}$, 
% for any factorisation $tr = u \cdot g \cdot v$ 
% (where $u$ and $v$ are traces, and $g$ is an event), 
% the suffix $v$ must contain an occurrence of $r$ such that 
% the number of \texttt{tock} events between $g$ and $r$ 
% is at most $d$.

% The AT-DLINE specification realises this semantics by combining
% a timing constraint process with an interrupt that enforces the
% deadline. The process \texttt{WAIT($d$); STOP\textsubscript{U}} 
% acts as a timer that prevents any further time advancement once 
% $d$ units have elapsed, while the interrupt ensures that the 
% \textit{targetEvent} must occur before the timer expires.
% Because refinement in the \emph{traces} model rules out behaviours
% that exhibit $d{+}1$ or more \texttt{tock}s between $g$ and $r$,
% the CSP encoding in AT-DLINE is semantically equivalent to the 
% intended meaning of RT-TIMED.

\paragraph{General assertions}

Five assertion templates are provided for general software properties:~AT-REACH (for reachability), AT-DDLK-1 and AT-DDLK-2~(for deadlock freedom), AT-DIV~(for divergence freedom), and AT-TERM~(for termination). AT-REACH, AT-DIV, and AT-TERM are expressed in the RoboChart assertion DSL for FDR. AT-DDLK-1 is for model checking in FDR and AT-DDLK-2 for theorem proving in Isabelle/HOL.

Each template is explicitly labelled as \texttt{untimed} or \texttt{timed} to indicate which semantics it uses.  
If no label is given, the assertion is interpreted against both timed and untimed semantics.  
AT-DDLK-1 and AT-DDLK-2 are shown below; the others are provided in Appendix~\ref{sec:appendix:ATs}.  
%Assertions using template AT-DDLK-1 are verified using FDR, and assertions using AT-DDLK-2 are discharged by theorem proving in Isabelle/HOL. 
\begin{tcolorbox}[colback=white, boxrule=0.5pt, arc=0pt, halign=flush left, left=2pt, right=2pt, top=2pt, bottom=2pt, boxsep=2pt]
 \renewcommand{\baselinestretch}{1.1}\selectfont
 \begin{footnotesize}
 \textbf{AT-DDLK-1 (Deadlock-freedom, FDR)}:\\[4pt]

[\textbf{\color{eclipsePurple}{untimed}} \,|\, \textbf{\color{eclipsePurple}{timed}}]\;
\textbf{\color{eclipsePurple}{assertion}}\; A\_\textit{Claim\_ID} : \\
\quad \textit{scope} \, \textit{boolTerm} \, \textbf{\color{eclipsePurple}{deadlock-free}}.\\

$\ast$~\textit{boolTerm} ::= \textbf{\color{eclipsePurple}{is}}\;|\; \textbf{\color{eclipsePurple}{is not}}
 \end{footnotesize}
\end{tcolorbox}
%This template uses the same EBNF notation introduced in Sect.~\ref{subsec:Claim Template Design}, so the symbols $[\,\cdot\,]$, $\{\,\cdot\,\}$, and `\,$|$\,' are meta-notation in EBNF, used to indicate optional parts, repetition, and alternatives, and are not part of the concrete assertion syntax.
%
%The template for deadlock-freedom assertions in Isabelle is shown below:
%
\noindent%
AT-DDLK-1 uses CSP$_{M}$, and AT-DDLK-2 uses Isar syntax to pose a lemma that claims deadlock-freedom. 

\begin{tcolorbox}[colback=white, boxrule=0.5pt, arc=0pt, halign=flush left, left=2pt, right=2pt, top=0pt, bottom=0pt, width=\columnwidth,boxsep=2pt]
 \renewcommand{\baselinestretch}{1.1}\selectfont
 \lstset{style=IsabelleStyle}{\small\begin{lstlisting}
AT-DDLK-2 (Deadlock-freedom, Isabelle)
lemma ~\textit{Claim\_ID}~_deadlock_free: 
"deadlock_free ~\textit{scope}~"
  apply deadlock_free
  \end{lstlisting}
  }
\end{tcolorbox}
\noindent% 
In both AT-DDLK-1 and AT-DDLK-2, the placeholder \textit{scope} is as defined in the specification of RT-DDLK.
%AT-DDLK-1 assertion follows the syntax of RoboChart assertion language for CSP.
%AT-DDLK-2 assertion uses the proof method \lstinline{deadlock_free} defined in~\cite{yan2023automated} to automate the proof.

\paragraph{Reward assertion} Three assertion templates, AT-PROB, AT-RWD, and AT-TEMP, are provided for probabilistic and temporal requirements.
These templates are expressed in the RoboChart assertion language for PRISM.
%During verification, each template is instantiated with structured requirements exported in XML (using RT-PROB, RT-RWD, or RT-TEMP), producing concrete assertions.
%The instantiated assertions are then automatically translated into PRISM assertions and verified in PRISM through RoboTool.
%
AT-RWD is presented below; the others in Appendix~\ref{sec:appendix:ATs}. %“Reward assertion template” 
AT-RWD is used to generate assertions for requirements written with the RT-RWD template.
%This template uses the EBNF notation.

\begin{tcolorbox}[colback=white, boxrule=0.5pt, arc=0pt, halign=flush left, 
  left=2pt, right=2pt, top=2pt, bottom=2pt, boxsep=2pt]
 \renewcommand{\baselinestretch}{1.1}\selectfont
 \begin{footnotesize}
 \textbf{AT-RWD}:\\[4pt]

\textbf{\color{eclipsePurple}{rewards}}\; reward\_\textit{Claim\_ID} \\
\quad = [\,\textit{rewardEvent}\,]\; (true\;|\; false) : \textit{rewardValue} \\
\textbf{\color{eclipsePurple}{endrewards}}\\[6pt]

\textbf{\color{eclipsePurple}{prob property}}\; R\_\textit{Claim\_ID}: \\
 \textbf{\color{eclipsePurple}{Reward}}\, reward\_\textit{Claim\_ID}\,
\textit{rewardTarget}\,
\textbf{\color{eclipsePurple}{of}}\, \par [\,\textit{pathFormula}\,] \\[4pt]

[\,\textbf{\color{eclipsePurple}{with constant}}\;
   \{ \textit{constant}\}\, \{ \textit{multiConstant} \} \,] 
 \end{footnotesize}
\end{tcolorbox}

\noindent
%Unlike the CSP-based templates such as AT-UTG, which follow a three-part  structure (\textit{Spec}, \textit{Spec\_impl}, and the refinement assertion), 
AT-RWD is written in the RoboChart assertion language for PRISM.  Accordingly, its structure follows PRISM’s reward and property declarations. %, rather than defining a refinement checking.

AT-RWD has three sections. The first section, introduced by \textbf{\color{eclipsePurple}{rewards}}, specifies the event that generates a reward, together with its corresponding reward value. The second section, introduced by \textbf{\color{eclipsePurple}{prob property}}, defines the actual property to be checked. Here, the identifier \textit{R\_Claim\_ID} is derived from the AC claim identification, ensuring traceability. The property asserts that the accumulated reward, as recorded by \textit{reward\_Claim\_ID}, satisfies the required bound (\textit{rewardTarget}) along the paths described by \textit{pathFormula}. The third section introduces optional constant values. %configurations, which parameterise the PRISM model. These are expressed through 
The syntax for the placeholders \textit{constant} and \textit{multiConstant} is given in Appendix~\ref{app:constants-configs}.

\rc{1.3(b)}{For AT-RWD, the situation differs from the CSP refinement templates above. The placeholders of RT-RWD are already structured expressions in the Robo\-Chart assertion language for PRISM: \emph{pathFormula} is a term of the syntactic category \emph{RPathFormula}, \emph{rewardTarget} is a comparison expression, and \emph{rewardEvent} and \emph{rewardValue} identify a reward specification. The assertion template arranges these expressions into the corresponding positions of a \texttt{rewards} declaration and a \texttt{prob property} query, without constructing any additional formal structure around them.

The three sections of RT-RWD map directly to three blocks in AT-RWD: 
the \textsc{When} section provides the content of the \texttt{rewards} block (reward event and reward value); the \textsc{Until} section provides the \emph{pathFormula}; and the \textsc{Required} section provides the \emph{rewardTarget}. No additional encoding is performed beyond this arrangement. The semantics of AT-RWD is therefore given directly by PRISM's reward semantics, as defined in the RoboChart manual~\cite{Miyazawa2016}. The generated assertion denotes the reward property the engineer supplied through the placeholders. (Whether the controlled English rendering of RT-RWD conveys that property clearly to the engineer is a question of usability; the knowledge required to use the templates is discussed in Sect.~\ref{Claim Structuring}.)}
%AT-RWD directly corresponds to the requirement template RT-RWD (Sect.~\ref{subsec:Claim Template Design}). In particular, the placeholders \textit{rewardEvent}, \textit{rewardValue}, \textit{rewardTarget}, and \textit{pathFormula} in AT-RWD match those of RT-RWD, so that instantiation from structured requirements is systematic. 

% \paragraph{Structural correspondence of AT-RWD.}

% Unlike templates such as AT-UTG, where the CSP specification encodes the semantics of the requirement, AT-RWD directly relies on the PRISM assertion language. The requirement template RT-RWD already uses placeholders that follow the terminology of PRISM assertion language (for rewards, path formulas, and targets). 

% The correspondence therefore lies in structure rather than in translation of semantics. Each section of RT-RWD maps directly to a block in AT-RWD:~the \texttt{WHEN} section instantiates the \textbf{rewards} block, the \texttt{UNTIL} section provides the \textit{pathFormula}, and the \texttt{REQUIRED} section yields the \textit{rewardTarget}. As a result, every piece of information required to construct a valid PRISM property is explicitly captured in RT-RWD and transferred unchanged into AT-RWD. 

% This one-to-one mapping ensures that the generated assertion is both complete and faithful to the structured requirement.%, without the need for an additional semantic justification as in the CSP-based templates.

\subsection{Assertion generation}\label{Assertion Generation}

An assertion is produced in two steps:~(1)~selection of the appropriate assertion template for each requirement, and  
(2)~instantiation of that template with concrete values taken from the requirement. Algorithm~\ref{Al: integrated process of assertion generation} describes both steps, and is implemented using the Epsilon model-transformation engine, and in particular its domain-specific language EGL~\cite{kolovos2008epsilon}.
The input of Algorithm~\ref{Al: integrated process of assertion generation} is an XML document \textit{kapture\_xml} exported from Kapture containing requirements, and the output is a set of well-formed assertion instances, one for each requirement.

%%%%%%%%%%%%%%%%%%  Template selection %%%%%%%%%%%%%
\paragraph{Template selection.}
Before an assertion can be created, the correct assertion template must be identified.
%Algorithm~\ref{Al: integrated process of assertion generation} defines this selection mechanism, which is encoded directly in the EGL program.
The XML file exported from Kapture may contain multiple requirements, each represented as an entry $r$ (line~\ref{alg:loop-req}). Each entry includes its claim identifier (\textit{backwardsTr}, line~\ref{alg:get-claimid}) and requirement type (\textit{templ}, line~\ref{alg:get-template}).
For requirements defined using the \textit{when} Kapture template (line~\ref{alg:if-when}), an entry $r$ also has a \textit{requiredCondition}~(line~\ref{alg:rcond}) and a \textit{guardCondition}~(line~\ref{alg:gcond}), which record the instantiation of the placeholders of the Kapture template discussed in Section~\ref{Kapture Template Design}.
\begin{algorithm}[tbp]
\small
\caption{Assertion generation }
\label{Al: integrated process of assertion generation}
\SetAlgoLined
\DontPrintSemicolon
\KwIn{\textit{kapture\_xml}}
\KwOut{A set of instantiated assertions}
\BlankLine
% \hrule

\For{$r \in \mathit{kapture\_xml.reqs}$}{                                   \label{alg:loop-req}
    $claim\_ID \leftarrow r.backwardsTr$;                                  \label{alg:get-claimid}\\
    $templ \leftarrow r.template$;                                         \label{alg:get-template}\\
    
    \If{$templ = \text{``when''}$}{                                        \label{alg:if-when}
        \textit{rCond} $\leftarrow r.requiredCondition$;\label{alg:rcond}\\
        \textit{gCond} $\leftarrow r.guardCondition$;\label{alg:gcond}\\
        \textit{prefix} $\leftarrow \texttt{head}(rCond)$;            \label{alg:get-prefix}

        \uIf{\textit{prefix} $ = \text{``prob\_target''}$}{                          \label{alg:prob-branch}
            \textsc{GenerateProbAsst}($r$, $claim\_ID$);                   \label{alg:gen-prob}
        }
        \ElseIf{\textit{prefix} $ = \text{``reward\_target''}$}{                     \label{alg:reward-branch}
            \textsc{GenerateRewardAsst}($r$, $claim\_ID$);         \tcp*{reward branch}        \label{alg:gen-reward}
        }
        \ElseIf{\textit{prefix} $ = \text{``term''}$}{                               \label{alg:term-branch}
            \textsc{GenerateTempAsst}($r$, $claim\_ID$);                   \label{alg:gen-temp}
        }
        $C \leftarrow \{\text{``constant''},\ \text{``multi\_constants''}\}$;  \label{alg:set-constants}\\
        \For{$cond \in guardCond.conds$}{                           \label{alg:loop-elems}
            \If{$cond.prefix \in C$}{                                      \label{alg:if-constant}
                \textsc{GenConstant}(\dots);                               \label{alg:gen-constant}
            }
        }\label{alg:gen-cons-endfor}

        \If{\textit{prefix} $ \notin \{\text{``prob\_target''},\ \text{``reward\_target''},\ \text{``term''}\}$}{ 
            \label{alg:untimed-cond}
            \textsc{GenerateUntimedAsst}($r$, $claim\_ID$);                \label{alg:gen-untimed}
        }\label{alg:prefix-finish}
    }

    \ElseIf{$templ = \text{``trigger on event''}$}{                        \label{alg:trigger}
        \textsc{GenerateTimedAsst}($r$, $claim\_ID$);                      \label{alg:gen-timed}
    }

    \ElseIf{$templ = \text{``every''}$}{                                   \label{alg:every}
        \textsc{GenerateGeneralAsst}($r$, $claim\_ID$);                    \label{alg:gen-general}
    }
}
\end{algorithm}

We recall that probabilistic, reward-based, temporal, and untimed requirements use the \textit{when} Kapture template. As discussed in Section~\ref{subsec:Claim Template Design}, we add prefixes to their \textit{requiredCondition} placeholder to explicitly identify the requirement type defined. 
Algorithm~\ref{Al: integrated process of assertion generation} uses this \textit{prefix}~(line~\ref{alg:get-prefix}) to determine whether the requirement matches a probabilistic~(line~\ref{alg:prob-branch}), reward-based~(line~\ref{alg:reward-branch}), temporal~(line~\ref{alg:term-branch}), or untimed assertion if there is no specific prefix~(line~\ref{alg:untimed-cond}).  
For example, if the placeholder  \textit{requiredCondition} is prefixed with ``\textit{reward\_target\_}''~(line~\ref{alg:reward-branch}), then this is a reward-based requirement, and we select AT-RWD.

The assertions using AT-PROB, AT-RWD, and AT-TEMP may have a section for setting the constant values, which are recorded in the\textit{guardCondition} placeholder with prefix ``constant'' or ``multi\_constants''.
Each constant appears as a separate element in \textit{guardCondition.conds}.
Algorithm~\ref{Al: integrated process of assertion generation} uses a local variable $C$ (line~\ref{alg:set-constants}) to store these two prefixes.
Algorithm~\ref{Al: integrated process of assertion generation} iterates over all the elements of \textit{guardCondition.conds} (lines~\ref{alg:loop-elems}–\ref{alg:gen-cons-endfor}) to identify the elements whose prefixes fall into \textit{C} and to generate constant configurations for the chosen assertion template accordingly. This is achieved by a call to a procedure \textsc{GenConstant}(\dots).  We omit the details of the call here, which can be found, together with the definition of \textsc{GenConstant}, in Appendix~\ref{sec:appendix:algs}.

The requirements defined using the \textit{when} template that have no prefix are untimed requirements. Requirements defined using the templates \textit{trigger on event} (line~26) and \textit{every}~(line 29) also do not require a prefix analysis. 

For each kind of template, we have a \textsc{Gen} procedure, further described below, that defines 
an assertion generator. %entirely and dispatch directly to the timed (line~\ref{alg:gen-timed}) and general (line~\ref{alg:gen-general}) assertion generators.
The definitions for each of the assertion generators called in Algorithm~\ref{Al: integrated process of assertion generation} are provided in Appendix~\ref{sec:appendix:algs}.

As an example, we consider a requirement defined using the Kapture template RT-UNTIMED.
Its XML entry records \textit{templ = when}, and its \textit{requiredCondition} begins with the prefix ``requiredEvent''.
Since this prefix does not match the specialised branches for probabilistic, reward, or temporal assertions~(line~\ref{alg:untimed-cond}), Algorithm~\ref{Al: integrated process of assertion generation} follows the untimed branch~(line~\ref{alg:gen-untimed}), and calls the function \textsc{GenerateUntimedAsst} to generate the assertion by instantiating the AT-UTG assertion template.

%%%%%%%%%%%%% Instantiation %%%%%%%%%%%%%
\paragraph{Instantiation.}
Once a template is selected, its placeholders are instantiated with concrete values extracted from the XML requirement.  
This is the role of the \textsc{Gen} procedures called in Algorithm~\ref{Al: integrated process of assertion generation}. Our assertion templates~(as described in Sect.~\ref{Assertion Template design}) are encoded in EGL, and the \textsc{Gen} procedures use values extracted from the Kapture templates to instantiate the assertion templates automatically.
%The instantiation itself is carried out by the generator functions invoked in Algorithm~\ref{Al: integrated process of assertion generation}; each such function implements the instantiation logic for its corresponding template.
%Their definitions are provided in Appendix~\ref{sec:appendix:algs}.
For illustration, we consider the following instance of our Kapture template  RT-UNTIMED. 
\begin{tcolorbox}[colback=white, boxrule=0.5pt, arc=0pt, halign=flush left, 
  left=2pt, right=2pt, top=2pt, bottom=2pt, boxsep=2pt]
\begin{footnotesize}
When \texttt{sys::stop.in} occurs,\\
\texttt{sys::flag.out} shall also be seen.
\end{footnotesize}
\end{tcolorbox}
\noindent%
Here the placeholders \textit{guardEvent} and \textit{requiredEvent} are instantiated with \texttt{sys::stop.in} and \texttt{sys::flag.out}.  
Since AT-UTG was selected, the instantiation yields the following CSP$_\mathrm{M}$ specification and refinement assertion.

\begin{tcolorbox}[colback=white, boxrule=0.5pt, arc=0pt, halign=flush left, left=2pt, right=2pt, top=2pt, bottom=2pt, boxsep=2pt]
 \renewcommand{\baselinestretch}{1.1}\selectfont
 \begin{footnotesize}
% ---- keep your original instantiated CSPM template here ----
\textbf{\color{eclipsePurple}{untimed csp}} Spec\\
csp-begin\\
\texttt{Spec} =
\texttt{CHAOS(Events) [\!|\,\{\!|} \texttt{sys::stop.in}\texttt{ |\!\} |>}
~~~~\texttt{(RUN(\{\!|} \texttt{sys::stop.in}\texttt{ |\!\})\,/\textbackslash}
\texttt{sys::flag.out -> Spec)}\\
csp-end\\[1ex]

\textbf{\color{eclipsePurple}{untimed csp}} Spec\_impl
\textbf{\color{eclipsePurple}{associated to}} \textit{module\_name}\\
csp-begin\\
\texttt{spec\_impl} = \texttt{sys::O\_\_(0)}\\
csp-end\\[1ex]

\textbf{\color{eclipsePurple}{untimed assertion}} A\_1 :\\
Spec\_impl \textbf{\color{eclipsePurple}{refines}} Spec
\textbf{\color{eclipsePurple}{in the traces model}}
\end{footnotesize}
\end{tcolorbox}
\noindent% 
The assertion A\_1 can be translated to CSP$_\mathrm{M}$ by RoboTool and verified using FDR. The result contributes to AC evidence generation, as discussed in Sect.~\ref{evidence model gen for model checking}.

\paragraph{Complexity and correctness.}
Let $R$ be the number of requirements and $E$ the total number of guard elements.
The algorithm performs one pass over all requirements and, for each requirement, a linear scan of its guard elements.
Thus, the running time is $O(R+E)$, and the auxiliary space consumption is $O(1)$, since only a constant number of temporary variables~(namely, \textit{claim\_ID}, \textit{templ}, \textit{rCond}, \textit{gCond}, \textit{prefix}, and \textit{C}) are stored.

With well-formed XML entries, each Kapture requirement matches exactly one case of the selection logic, ensuring that one assertion template is chosen unambiguously.  
The chosen template’s placeholders are instantiated using the structured fields of the XML, guaranteeing that the generated assertions remain well-formed, consistent with Table~\ref{Assertion Template map to claim table}, and traceable via the claim identifier.

%% file: Evidence_model_generation_and_integration.tex
\section{Evidence model generation and integration}
\label{evidence model gen for model checking}

We recall from Sect.~\ref{sec:approach} that we assume the existence of a preliminary model-based AC created with predefined SACM patterns, where the evidence elements either remain as placeholders, since verification has not yet been performed, or need updating.  
Building on this assumption, our process~(Fig.~\ref{fig:evidence_generation_method}) generates assertions for the claims~(Steps~1--2), obtains the corresponding verification results~(Step~3), and then automatically generates and integrates evidence models into the AC~(Step~4).  
Each evidence model carries explicit traceability information linking the verification outcome to its supported claim. Any earlier evidence models are replaced with new ones derived from updated verification results.  
This section focuses on Steps~3 and~4.  

\begin{figure}[t]
    \centering
    \begin{adjustbox}{frame=0.1pt 1pt}
        \includegraphics[width= 1.0 \linewidth]{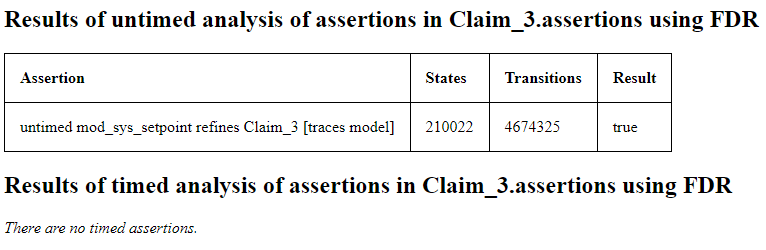}
    \end{adjustbox}
    \caption{Excerpt of a RoboTool FDR model-checking report, showing untimed and timed analysis results.}
    \label{fig: html_result_example}
\end{figure}

\begin{figure}[t]
    \centering
    \begin{adjustbox}{frame=0.1pt 1pt}
        \includegraphics[width=1.0 \linewidth]{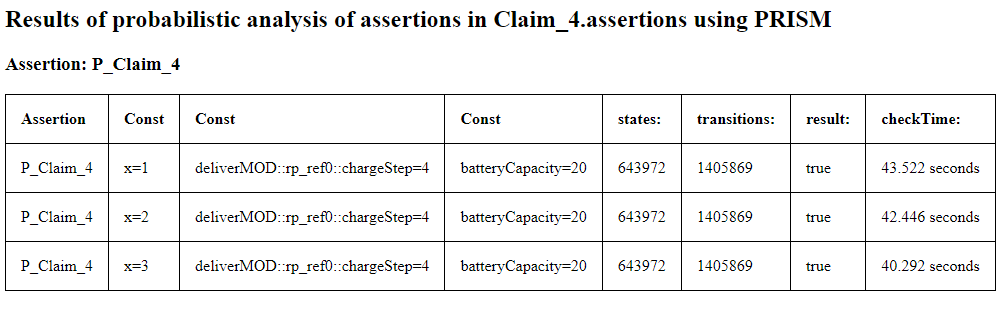}
    \end{adjustbox}
    \caption{Excerpt of a RoboTool PRISM model-checking report, showing probabilistic analysis results.}
    \label{fig: html_result_example_PRISM}
\end{figure}

With the assertions generated from Step~2 and the formal semantics of RoboChart generated by RoboTool, we can conduct the verification using FDR, PRISM or Isabelle within RoboTool.
During model checking, RoboTool outputs verification results as HTML reports, where each report corresponds to the evaluation of a single assertion file.
These reports record whether the checked assertions hold. Examples of reports of FDR and PRISM results are shown in Figs.~\ref{fig: html_result_example} and \ref{fig: html_result_example_PRISM}.  
During theorem proving, Isabelle can be invoked in RoboTool by communicating with the Isabelle server to discharge the proof, and the resulting proof log, containing the requirement identification and the result, can be reported.

% To provide an overview, Fig.~\ref{fig:overview_evidence_pipeline} presents the workflow for evidence generation and integration as a high-level pipeline, highlighting the main inputs, processing steps, and outputs.  

\begin{algorithm}[h]
\small
\caption{AC evidence generation and integration}
\label{Al:model checking evidence_gen}
\SetAlgoLined
\DontPrintSemicolon

\KwIn{\textit{report}, \textit{kapture\_xml}, \textit{ac}} \label{alg2:mceg-input}
\KwOut{updated \textit{ac} with integrated evidence} \label{alg2:mceg-output}
\BlankLine

$reqId      \gets \textsc{getClaimId}(report)$          \label{alg2:mceg-get-claimid}\;
$reqEntry   \gets \textsc{lookupBackTr}(kapture\_xml, reqId)$   \label{alg2:mceg-select-xml}\;
$acClaimId  \gets reqEntry.forwardTr$                    \label{alg2:mceg-vclaimid}\;

$tool \gets \textsc{getToolType}(report)$ \label{alg2:mceg-get-tool}\;
$result \gets true$ \label{alg2:mceg-init-result}\;

\If{$tool =$ PRISM}{ \label{alg2:mceg-if-prism}
    \For{$r \in report.tables.first.rows$}{ \label{alg2:mceg-for-prism}
        $result \gets result \wedge \textsc{getResult}(r)$ \label{alg2:mceg-prism-update}\;
    }
}\label{alg2:mceg-if-prism-end}
\If{$tool =$ FDR}{ \label{alg2:mceg-if-fdr}
    \For{$t \in report.tables$}{ \label{alg2:mceg-for-fdr}
        $result \gets result \wedge \textsc{getResult}(t)$ \label{alg2:mceg-fdr-update}\;
    }
}\label{alg2:mceg-if-fdr-end}
\If{$tool =$ Isabelle}{ \label{alg2:if-isabelle}
 $result \gets \textsc{getResult}(report)$\label{alg2:if-isabelle-result}\;
}
$evidence \gets \textsc{genEvidence}(result, acClaimId, tool)$ \label{alg2:mceg-generate-evidence}\;

$evidenceLink \gets \textsc{findByTgt}(acClaimId)$\label{alg2:find-evidence-link}\;
$\textsc{clear}(evidenceLink.source)$ \label{alg2:mceg-clear-source}\;
        $evidenceLink.source \gets \{evidence\}$ \label{alg2:mceg-update-source}\;
        
% \For{$assertedEv \in ac.argElm.\textsc{select}(c\:|\:c.type = AssertedEvidence)$}{
%     \label{alg2:mceg-for-assertedev}
%     \If{$assertedEv.target = acClaimId$}{ \label{alg2:mceg-if-target-match}
%         $assertedEv.source.\textsc{clear}()$ \label{alg2:mceg-clear-source}\;
%         $assertedEv.source \gets \{evidence\}$ \label{alg2:mceg-update-source}\;
%     }
% } \label{alg2:mceg-for-assertedev-end}
\end{algorithm}

With the verification results available, Algorithm~\ref{Al:model checking evidence_gen} formalises the workflow for generating evidence from those results and integrating it into the assurance case.
The algorithm takes as input the verification report \textit{report}, the requirements \textit{kapture\_xml} exported from Kapture, and the current AC model~\textit{ac}.
Its output is an updated version of \textit{ac} with integrated evidence.

Algorithm~\ref{Al:model checking evidence_gen} is defined based on the structure of the verification report; therefore, we briefly outline this structure.
It differs slightly depending on which tool is used to generate the report:~for FDR, a report may contain one table~(untimed or timed) or two tables if no label for a model is specified. In each table, there is a row per assertion.  Verification succeeds when the Result column entries for all rows are ``true'' in the single-table case, and only if both tables record ``true'' otherwise. For PRISM, a report always contains a single table with multiple rows and again verification succeeds only if all rows record ``true''.  
As shown in Figs.~\ref{fig: html_result_example} and \ref{fig: html_result_example_PRISM}, the title of each table encodes both the identifier of the requirement %~(line~\ref{alg2:mceg-get-claimid}) 
and the verification tool %~(line~\ref{alg2:mceg-get-tool}) 
used, which facilitates the evidence generation. For example, a title for a report can be as follows.
\begin{quote}
\textit{Results of probabilistic analysis of assertions in SR1\_1\_1.assertions using PRISM}
\end{quote}
Here, the tool is PRISM and the requirement is \textit{SR1\_1\_1}. %This identifier is extracted as the \textit{claim\_id} for traceability. %The claim in the XML specification is retrieved using its backward trace~(\textit{backwardsTr}), while the directly supported lower-level claim is identified through the forward trace~(\textit{forwardTr}). This mapping ensures that evidence is correctly attached to the appropriate level of the AC pattern.  
For Isabelle, the proof log returned can be converted into an EMF report, which contains the requirement identification, the verification tool, and the result.

%%%%%%%%%%%%%%%%%%%%%%%%%%%

%Algorithm~\ref{Al:model checking evidence_gen} takes three inputs: the verification \textit{report}, the Kapture requirement model $kapture\_xml$, and the assurance case model $ac$.
The $ac$ input of Algorithm~\ref{Al:model checking evidence_gen} is assumed to adopt the AC pattern defined in~\cite{yan2022model}, where each software requirement corresponds to a requirement-level claim that is further refined into one or more verification-method claims.
These sub-level claims describe how the requirement is demonstrated~(for instance, by model checking, theorem proving, or testing), and that they are the claims that the evidence provided directly supports. It is this evidence that our work generates and integrates or updates automatically.

Consistent with this structure, the Kapture template records two traceability links:~the backwards traceability link % $BackwardsTr$ 
refers to the requirement, and the forward traceability link $forwardsTr$ 
identifies the verification-method claim.
Algorithm~\ref{Al:model checking evidence_gen} first uses the backwards trace to locate the requirement and then the forward trace to obtain the identifier of the AC claim supported by the evidence.

Algorithm~\ref{Al:model checking evidence_gen} operates in three stages.
First, it identifies the claim to be supported by the evidence~(lines~\ref{alg2:mceg-get-claimid}--\ref{alg2:mceg-vclaimid}). Second, it generates the evidence~(lines~\ref{alg2:mceg-get-tool}--\ref{alg2:mceg-generate-evidence}). 
Finally, it updates the AC model~~(lines~\ref{alg2:find-evidence-link}--\ref{alg2:mceg-update-source})
\rc{2.64}{by replacing the existing evidence for the identified claim with the newly generated evidence; this is described in detail in the following paragraph.}
In the first stage, Algorithm~\ref{Al:model checking evidence_gen} extracts the requirement identifier \textit{reqId} from the \textit{report}~(line~\ref{alg2:mceg-get-claimid}). For that, we use a simple projection function~\textsc{getClaimId}; its definition and that of other simple projection functions are omitted. Next, Algorithm~\ref{Al:model checking evidence_gen}  retrieves the matching requirement entry \textit{reqEntry} from $kapture\_xml$ using the backwards trace~(line~\ref{alg2:mceg-select-xml}), and obtains the corresponding verification-method claim identifier \textit{acClaimId} via the forward trace~(line~\ref{alg2:mceg-vclaimid}).

To generate the evidence artefact, Algorithm~\ref{Al:model checking evidence_gen} first determines the verification tool~(PRISM, FDR or Isabelle) from the report metadata~(line~\ref{alg2:mceg-get-tool}). Afterwards, the aggregated verification result is recorded in a local variable $result$~(lines~\ref{alg2:mceg-init-result}--\ref{alg2:if-isabelle-result}).
For PRISM, all rows must evaluate to \emph{true}~(lines~\ref{alg2:mceg-if-prism}--\ref{alg2:mceg-if-prism-end}); for FDR, every table must yield \emph{true}~(lines~\ref{alg2:mceg-if-fdr}--\ref{alg2:mceg-if-fdr-end}). As we only check deadlock freedom using Isabelle, its verification outcome is already a single Boolean result, and therefore, no aggregation is required~(line~\ref{alg2:if-isabelle-result}).
The result is then used to construct an evidence artefact using the function \textsc{GenEvidence}~(line~\ref{alg2:mceg-generate-evidence}).

%In line~\ref{alg2:mceg-generate-evidence}, the function \textsc{GenEvidence} abstracts the construction of an evidence artefact from a verification outcome. 
Given a Boolean \textit{result}, a claim identifier \textit{acClaimId}, and a verification \textit{tool} used, \textsc{GenEvidence} produces an evidence element that records the verification method, the associated claim, and the outcome. Evidence is generated for both successful and failed verification outcomes, allowing the AC to faithfully reflect the current verification status of each claim.
A result \emph{true} indicates that the verification-method claim is supported, whereas a result of \emph{false} records a verification failure and indicates that the claim is not satisfied. We assume that the resulting AC is reviewed by an engineer prior to submission to the certification authority. In our implementation, this evidence artefact is realised as an EMF model conforming to SACM, but this representation choice is not essential. % to the approach.  

\rc{1.5}{We discussed the generation and integration of verification outcomes; it is also important to consider the practical workflow when verification fails. When a verification attempt fails, the HTML report produced by RoboTool indicates which assertions failed, but does not include diagnostic information such as counterexamples. To diagnose the cause of failure, the engineer must manually inspect the diagnostic output provided by the underlying verification tools. For FDR, this involves using FDR's capability to identify the counterexample traces; for PRISM, the simulator or counterexample analysis can be used; for Isabelle, ``quickcheck'' and ``nitpick'' tools can be used for counterexamples. The current implementation does not automatically obtain or integrate this diagnostic information into the AC evidence model.}

\rc{1.5}{The engineer uses this diagnostic information to support an iterative update: they may update the RoboChart model, correct the requirement structuring in Kapture, or adjust the assertion template parameters. After updating, the workflow is re-executed, and Algorithm~\ref{Al:model checking evidence_gen} automatically replaces the previous evidence in the AC via lines~\ref{alg2:find-evidence-link}--\ref{alg2:mceg-update-source}, ensuring that the AC remains consistent with the updated model or requirements. Automating the extraction and presentation of diagnostic feedback to support iterative updates is future work.}
 
The final stage updates the AC model so that the verification-method claim identified by \textit{acClaimId} is supported by the newly generated $evidence$.
In the SACM metamodel, the relationship between a claim and the evidence that supports it is represented by an \texttt{Asserted} \texttt{Evidence} link.
\rc{2.65}{This link has two key attributes: a \texttt{target} (the claim being supported) and a \texttt{source} (the evidence artefact providing the support). In our AC structure, we allow one evidence artefact to support a single claim, and evidence can be replaced when verification is re-executed.}
So, Algorithm~\ref{Al:model checking evidence_gen} first locates the $evidenceLink$ whose target matches \textit{acClaimId} by invoking \textsc{findByTgt} (line~\ref{alg2:find-evidence-link}).
Once that link has been identified, its existing source is cleared by removing either a placeholder evidence element or outdated verification results (line~\ref{alg2:mceg-clear-source}).
Finally, the new $evidence$ artefact is assigned as the source of the link (line~\ref{alg2:mceg-update-source}), thereby updating the AC so that the verification-method claim is supported by the most recent verification result.

%For theorem proving, since we currently only verify deadlock freedom in Isabelle, the Z-Machine semantics and the proof obligation for deadlock freedom are generated together within a single Isabelle theory file, using the assertion template \textsc{AT-DDLK-2}.The identification of \textit{acClaimId}, generation and integration of the evidence are implemented manually following the procedure in Algorithm~\ref{Al:model checking evidence_gen}. Future work will focus on its full automation.

The correctness of Algorithm~\ref{Al:model checking evidence_gen} is straightforward, as it propagates verification outcomes into evidence models while preserving traceability links, and this behaviour is validated in our case studies described next.

%% file: case_study.tex
%%%%%%%%%%%%%%%%%%%%%%%%%%%%%%%%%%%%%%%%%%
%%%%%%%%%%%%%%%%%%%%%%%%%%%%%%%%%%%%%%%%%%

% \begin{table*}[t]
% \centering
% \caption{Summary of our case studies.
% The model size is given in terms of the number of states and transitions in the RoboChart model.
% The automated generation time are reported as averages over 10 runs of the total time to check all assertions and to generate and integrate evidence models for each case study using the indicated back-end tool.}

% \label{tab:efficiency}
% \begin{tabular}{lcccc}
% \hline
% \textbf{Case study} & \makecell{\textbf{\#}\\\textbf{Assertions}} & \makecell{\textbf{Model size}\\\textbf{(states/transitions)}} & \makecell{\textbf{Verification}\\\textbf{runtime}} & \makecell{\textbf{Automated}\\\textbf{generation}\\(assert+evidence)} \\
% \hline
% Mail robot~\cite{pls}           & 3  & 8 / 11 & 450s (PRISM)     & 39ms + 56ms \\
% HV controller~\cite{murray2022safety}       & 4  & 7 / 10 & 405s (FDR)      & 277ms + 87ms \\
% Underwater vehicle~\cite{yan2023automated}  & 1  & 5 / 18 & 20s (Isabelle) & 580ms + 92ms \\
% Maintenance robot       & 4  & 25 / 35 & \textcolor{red}{???s} (FDR)      & 67ms + 112ms \\
% \hline
% \end{tabular}
% \end{table*}

\begin{table*}[t]
\centering
\caption{Summary of the case studies and automation overhead.
Model size is given in terms of the number of states and transitions of the RoboChart model. Times are averages over 10 runs in a 1.6 GHz Core i5 computer with 8GB of RAM.}
\label{tab:case_summary_perf}

\begin{tabular}{lcccc}
\hline
\textbf{Case study} 
& \makecell{\textbf{\#}\\\textbf{Assertions}} 
& \makecell{\textbf{Model size}\\\textbf{(states / transitions)}} 
& \makecell{\textbf{Assertion}\\\textbf{generation (ms)}} 
& \makecell{\textbf{Evidence}\\\textbf{generation (ms)}} \\
\hline
Mail robot~\cite{pls}          
& 3  
& 8 / 11  
& 39  
& 56 \\

Painting robot~\cite{murray2022safety}       
& 4  
& 7 / 10  
& 277  
& 87 \\

Underwater vehicle~\cite{yan2023automated}  
& 1  
& 5 / 18  
& 580  
& 92 \\

Maintenance robot       
& 4  
& 25 / 35  
& 67  
& 112 \\
\hline
\end{tabular}
\end{table*}

\begin{table*}[t]
\small
\centering
\caption{Requirement templates (RT) and assertion templates (AT) used in the case studies.}
\label{tab:case_summary_templates}

\begin{tabular}{lcc}
\hline
\textbf{Case study} 
& \textbf{Requirement} \textbf{Templates (RT)}
& \textbf{Assertion} \textbf{Templates (AT)} \\
\hline
Mail robot          
& RT-PROB, RT-RWD, RT-TEMP  
& AT-PROB, AT-RWD, AT-TEMP  \\

Painting robot      
& RT-UNTIMED, RT-REACH  
& AT-UTG, AT-UTL, AT-REACH  \\

Underwater vehicle
& RT-DDLK  
& AT-DDLK-2  \\

Maintenance robot       
& RT-UNTIMED, RT-REACH, RT-DDLK, RT-DIV
& AT-UTG, AT-REACH, AT-DDLK-1, AT-DIV  \\
\hline
\end{tabular}
\end{table*}

\section{Evaluation}
\label{sec:cases}

In this section, four case studies are presented to demonstrate the application of our algorithms for evidence model generation through FDR, PRISM, and Isabelle (Sect.~\ref{subsec: mail robot}). % to \ref{subsec:unscrew_robot}). The evaluation presented in 
Sect.~\ref{subsec: contr2_evaluation} uses these case studies to answer the research questions described in Sect.~\ref{subsec:RQ}. % based on the outcomes of these case studies. 
In Sect.~\ref{subsec: contr2_evaluation_threat}, we discuss threats to the validity of our results.

%%%%%%%%%%%%%%%%%%%%%%%%%%%%%%%%%%%%%%%%%
% OLD RQs
% \textbf{RQ1 Generality}:
% Can the approach be used for AC evidence generation in different domains and with various system development techniques?

% \textbf{RQ2 Coverage}:
% Does the classification of requirements sufficiently address the diverse needs of safety-critical RAS systems?

% \textbf{RQ3 Completeness}:
% Are the ACs constructed using our approach complete? Specifically, does the AC structure comply with the AC template, and does the AC module conclude with evidence referencing a formal verification result?

% \textbf{RQ4 Efficiency}:
% Does our approach generate formal evidence with acceptable computational and time overhead?
%%%%%%%%%%%%%%%%%%%%%%%%%%%%%%%%%%%%

\subsection{Research Questions}\label{subsec:RQ}

We have used four case studies to answer the following four research questions and evaluate our process.

\begin{description}
  \item[\textbf{RQ1~(Generality):}] 
To what extent can our automated evidence-generation approach be applied across different robotic domains, and how dependent is it on the choice of modelling language and verification tools?

  \item[\textbf{RQ2~(Coverage):}] 
  Do the requirement classification and the associated templates provide sufficient coverage for the kinds of software requirements that typically arise in robotic and autonomous systems, and can the scheme be extended when new requirement categories are encountered?

  \item[\textbf{RQ3~(Completeness of integration):}] 
  Does the approach ensure that all formalised requirements are systematically transformed into verification results and that these results are consistently integrated into the assurance-case argument structure?

  \item[\textbf{RQ4~(Scalability and efficiency):}] 
  Is the approach scalable and efficient in practice, both in terms of handling larger models and requirement sets, and in reducing the manual effort required for maintaining evolving assurance cases?
\end{description}

\noindent
One assumption has been made for the approach:~the software's modelling language has a formal semantics that enables proper formal verification. Therefore, the formalisation of the software models is outside the scope of our approach and is not evaluated.

The four selected case studies provide a diverse and representative evaluation of the proposed approach across different robotic domains, verification back-ends, and classes of software requirements. A mail delivery robot and a High Voltage Controller demonstrate evidence generation via PRISM and FDR model checking for probabilistic and discrete-event control systems, respectively. 
An Autonomous Underwater Vehicle case study focuses on deadlock-freedom verification using Isabelle theorem proving, illustrating applicability beyond model checking and to systems with real-valued parameters. 
Finally, a maintenance robot involves a larger and more structurally complex RoboChart model.
Together, these case studies provide a strong basis to answer our research questions. %enable assessment of the approach’s generality, coverage, completeness of integration, scalability and efficiency.

\subsection{Case studies description}\label{subsec: mail robot}

We describe each of our case studies below.
\rc{2.4}{The implementation of the workflow and the case study artefacts presented in this paper are publicly available online\footnote{\url{https://github.com/laila-fangyan/formal-evidence-gen-for-ac}.}}

\subsubsection{Mail delivery robot}
\label{sibsubsec: mail-robot}
A case study of a mail-delivering robot~\cite{pls} has been conducted to demonstrate the application of our work for generating evidence models using PRISM model checking.

The robot delivers mail to eight offices~(1 to 8), with Office~0 serving as the charging station. %~(see Fig.~\ref{fig:mail}). 
An operator can request mail delivery between offices, excluding the charging station. The robot handles one task at a time:~it picks up mail from the source office and delivers it to the destination. After completing a task, the robot becomes available for the next request. 
The robot may either remain in its current office or move to one of the offices that are adjacent according to a fixed connectivity relation; all available options are chosen with equal probability. For instance, from Office 3, it can stay, move to Office 2, or move to Office 5, each with a 1/3 chance.
The robot has a finite battery capacity, which is recharged at the charging station in discrete steps with a fixed charging rate.
% \begin{figure}[t]
%     \centering
%  	\includegraphics[width= 0.4 \linewidth]{images/mail robot.png}
% 	\caption{Map of the workplace~\cite{pls}}
% 	\label{fig:mail}
% \end{figure}

The RoboChart models for the mail delivery robot can be found online\footnote{\url{https://robostar.cs.york.ac.uk/prob\_case\_studies/mail\_delivery\_robot/index.html}}.
A top-level system requirement considered in this case study is as follows.

\smallskip

\textbf{TR1} The robot shall deliver the mail to the destination.

\smallskip 

\noindent
One possible cause for violation of this requirement is the robot running out of battery while away from the charging station. 
The robot starts with a battery capacity of 20 units, each movement consumes one unit of battery charge, and the battery is recharged in discrete steps of 4 units per update.
This scenario leads to a set of software-level requirements as follows.
\begin{enumerate}
    \item SR1\_1\_1: The probability of the robot running out of power at each non-charging position shall be less than 0.5 when the starting battery capacity and the charge step are 20 and 4.
    \item SR1\_1\_2: The average number of moves before running out of power shall be greater than 8 when the starting battery capacity and the charge step are 20 and 4.
    \item SR1\_1\_3: There is a path through which the robot will not get stuck.
\end{enumerate}
It is these requirements that we handle. 

\paragraph{Initial AC generation}

\begin{figure}[t]
    \centering
 	\includegraphics[width= 0.95 \linewidth]{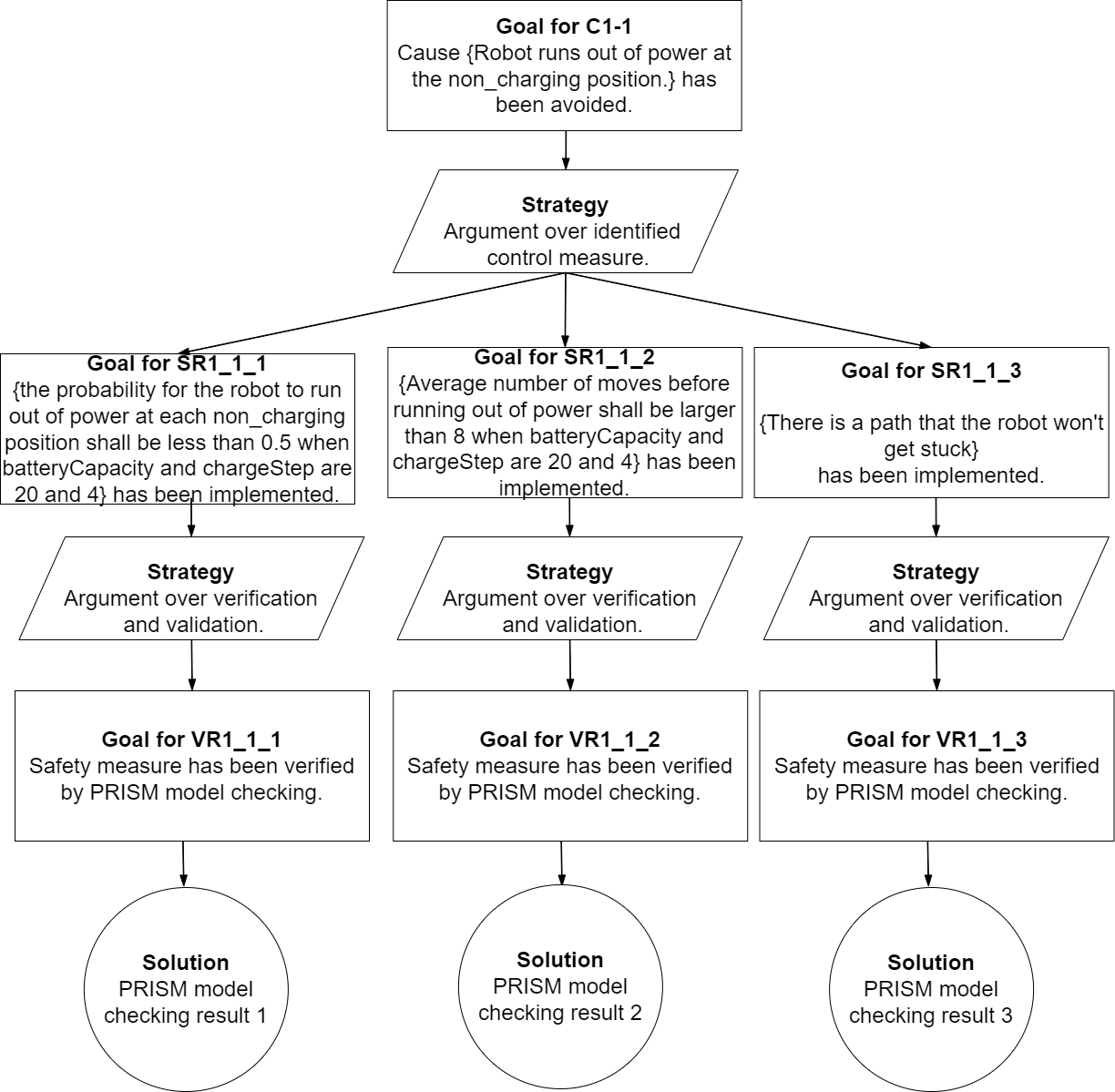}
	\caption{AC in GSN notation for the mail-delivering robot.}
	\label{fig:Mail_robot GSN before PRISM}
\end{figure}

For the system-level mission requirement TR1, an initial AC argumentation structure is constructed.
The resulting AC is shown in Fig.~\ref{fig:Mail_robot GSN before PRISM} using GSN notation\rc{2.67}{, with context elements omitted for simplicity.}
In GSN, AC claims are referred to as goals.

In this AC, the three software-level goals ``Goal for SR1\_1\_1'', ``Goal for SR1\_1\_2'', and ``Goal for SR1\_1\_3'' are supported by the corresponding ver\-i\-fi\-ca\-tion-level goals ``Goal for VR1\_1\_1'', ``Goal for VR1\_1\_2'', and ``Goal for VR1\_1\_3'',
all of which are intended to be discharged using PRISM model checking of a RoboChart model.

We assume that, at the time this initial AC is developed, the verification activities have not been performed.
As a result, no concrete verification results are yet available to be referenced as evidence.
Instead, the corresponding evidence nodes in the AC are represented by placeholders (``result~1'', ``result~2'', and ``result~3'').
So, the next step is to prepare for evidence generation as described below. %by %structuring the software requirements referenced in the software-level claims. This is achieved  using the Kapture tool together with the requirement templates introduced in Sect.~\ref{subsec:Claim Template Design}.

\paragraph{Requirement structuring}

\begin{figure}[t]
    \centering
 	\frame{\includegraphics[width= 0.8\linewidth]{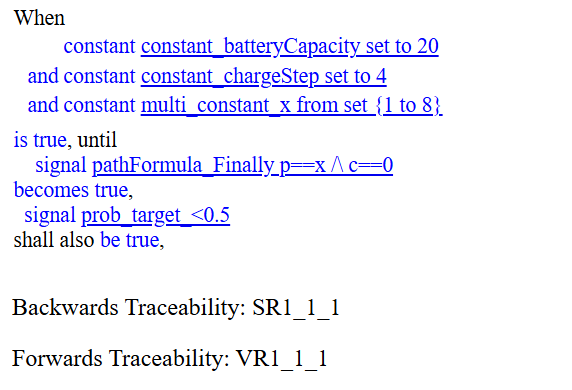}}
	\caption{SR1\_1\_1 exported from Kapture.}
	\label{fig:claim1} 
\end{figure}

\begin{figure}[t]
    \centering
 	\frame{\includegraphics[width= 0.8 \linewidth]{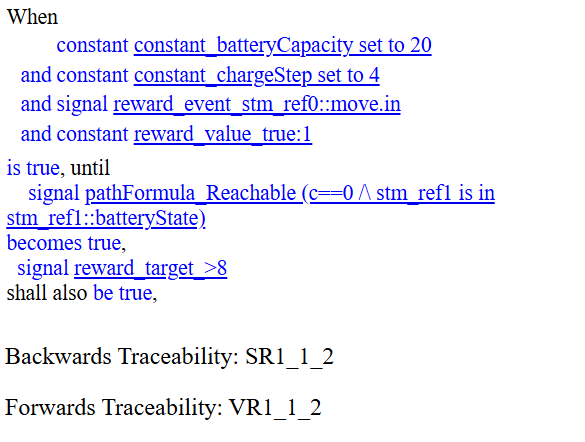}}
	\caption{SR1\_1\_2 exported from Kapture.}
	\label{fig:claim2} 
\end{figure}

\begin{figure}[t]
    \centering
 	\frame{\includegraphics[width=0.8 \linewidth]{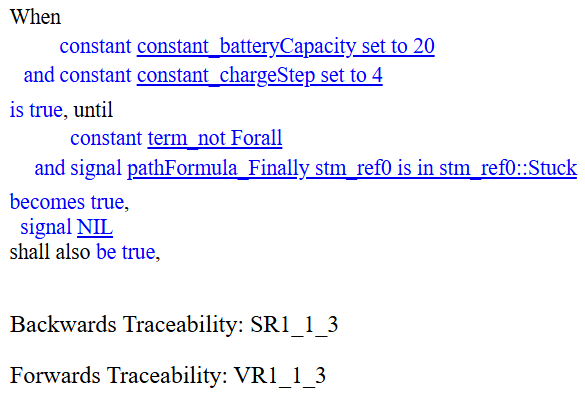}}
	\caption{SR1\_1\_3 exported from Kapture.}
	\label{fig:claim3} 
\end{figure}

SR1\_1\_1 is a probabilistic requirement, SR1\_1\_2 is a reward-based requirement, and SR1\_1\_3 is a temporal requirement. Accordingly, the requirement templates RT-PROB, RT-RWD, and RT-TEMP have been used to describe them. Based on the RoboChart model, the requirements referenced by the AC claims are structured in Kapture as illustrated in Figs.~\ref{fig:claim1}--\ref{fig:claim3}. For clarity, we do not use fully qualified names here.

The software behaviour of the robot is modelled in a RoboChart module \RC{deliverMOD}. 
The robotic platform \RC{rp\_ref0} provides the software controller with two variables, recording the robot’s position \RC{p} and the current battery level \RC{c}, as well as two constants, recording the starting battery capacity \RC{batteryCapacity} and the amount of battery charged per update \RC{chargeStep}. 

The movement control behaviour of the robot software is modelled by the state machine \RC{stm\_ref0}. 
This machine includes a state \RC{Stuck}, which denotes that the robot becomes stuck and is unable to move further. 
The event \RC{move} is used to trigger movement when the current position differs from the desired goal. Battery management is modelled by the state machine \RC{stm\_ref1}, which includes a state \RC{batteryState} representing battery charging behaviour. 
%The definitions of these state machine models can be found online\footnote{\url{https://robostar.cs.york.ac.uk/prob\_case\_studies/mail\_delivery\_robot/index.html}}.

For SR1\_1\_1~(see Fig.~\ref{fig:claim1}), the \texttt{When} clause defines values for the RoboChart constants in accordance with the state of the requirement:~\texttt{batteryCapacity} and \texttt{chargeStep} are set as 20 and 4.   The \texttt{When} clause also defines a variable \texttt{x}~(prefixed with ``multi\_constant\_'' in Kapture), ranging over values 1 to 8, needed for the specification of the path formula. The scenario in which the robot runs out of power is characterised by the battery capacity \texttt{c} being zero (\texttt{c == 0}). The robot being located at a non-charging position is represented by \texttt{p == x}, where, as said above, \texttt{x} ranges from 1 to 8.
Accordingly, the placeholder \textit{pathFormula} is instantiated as the conjunction of \texttt{p == x} and \texttt{c == 0}, using the temporal operator \texttt{Finally}. 
The probability target is less than 0.5.

For each requirement~(Figs.~\ref{fig:claim1}--\ref{fig:claim3}), Kapture provides two traceability labels: a \texttt{Backwards Traceability} label that identifies the corresponding software requirement, and a \texttt{Forwards Traceability} label that identifies the verification-method claim to which the generated evidence will be attached. 
This traceability information is used to integrate the verification evidence back into the AC model.

SR1\_1\_2 characterises the number of movements the robot can perform before battery exhaustion. The \texttt{When} clause gives values to RoboChart constants and defines the reward event as \texttt{move} with a reward value of~1, indicating that each movement consumes one unit of battery capacity. 
% The battery exhaustion condition is characterised by \textit{c == 0}. 
%% note: move event is in the state machine\RC{movingSTM}, but we need to use the name \RC{stm\_ref0} because at the module level, the state machine is referenced by stm\_ref0
The resulting reward expression therefore measures the expected number of movements performed before the battery is depleted~(\texttt{c == 0}) and the battery is ready to be charged~(defined in RoboChart as \texttt{stem\_ref1\allowbreak\ is\allowbreak\ in\allowbreak\ stm\_ref1\allowbreak::\allowbreak bat\-tery} \texttt{State}).

SR1\_1\_3 requires the robot to avoid entering a stuck condition, which is represented by the state machine \RC{stm\_ref0} not being in the \RC{Stuck} state. The \texttt{When} clause just defines the values of the constants \texttt{batteryCapacity} and \texttt{chargeStep}. The placeholder \textit{pathFormula} is instantiated as  \texttt{stm\_ref0\allowbreak\ is\allowbreak\ in\allowbreak\ stm\_\allowbreak ref0\allowbreak::\allowbreak Stu\-ck}, combined with the temporal operator \texttt{not Forall}, requiring that there exists at least one execution path along which the robot does not become stuck.

With the Kapture templates, we can now generate assertions automatically as explained next. 

\paragraph{Assertion and evidence generation}

Assertion generation for this case study follows the generic procedure defined in Algorithm~\ref{Al: integrated process of assertion generation}, which takes as input the XML file exported from Kapture and produces a set of instantiated RoboChart assertions. 
For the mail delivery robot, the Kapture XML contains three requirement entries corresponding to SR1\_1\_1, SR1\_1\_2, and SR1\_1\_3.
Each entry records the claim identifier to be supported, the Kapture template used, and the instantiated place- hold\-ers of the requirement.

Taking SR1\_1\_1 as an example, its Kapture entry uses the \texttt{when} template, and its \textit{requiredCondition} is prefixed with \texttt{prob\_target}.
Accordingly, Algorithm~\ref{Al: integrated process of assertion generation} follows the probabilistic branch and invokes \textsc{Gen\-er\-ate-} \textsc{Prob\-Asst}, which instantiates the AT-PROB assertion template.
The constant configurations specified in the \textit{guardCondition} placeholder (e.g.\ battery capacity and charge step) are then processed to complete the assertion instantiation.
The same mechanism is applied to SR1\_1 \_2 and SR1\_1\_3, which are identified as reward-based and temporal requirements, respectively, and handled using the AT-RWD and AT-TEMP templates.

\begin{figure}[t]
    \centering
 	\frame{\includegraphics[width= 1.0 \linewidth]{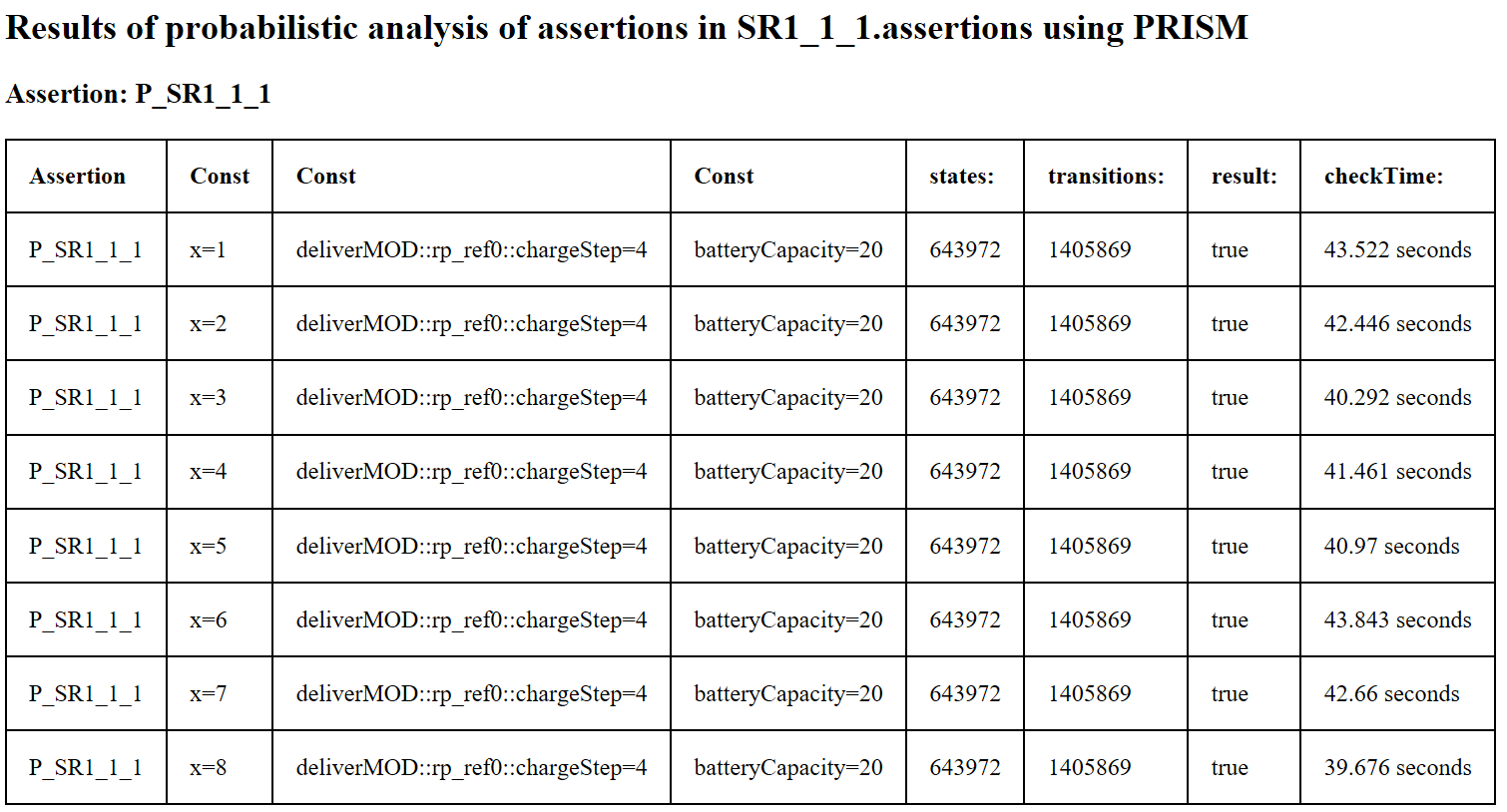}}
	\caption{Model checking result for SR1\_1\_1.}
	\label{fig:SR1 result}
\end{figure}

As a result, three RoboChart DSL assertions are automatically generated for this case study using the templates AT-PROB, AT-RWD, and AT-TEMP.
Robo- Tool then automatically translates the RoboChart models into PRISM models.
By invoking the PRISM model checker within RoboTool, the generated assertions are translated into PRISM specifications and verified.
The verification results are reported in HTML format, using fully qualified names.
An example verification result for SR1\_1\_1 is shown in Fig.~\ref{fig:SR1 result}.

% Finally, evidence models of the AC are created using the verification results.

% \begin{figure}[t]
%     \centering
%  	\frame{\includegraphics[width= 1.0\linewidth]{images/mail_prob_asst1.png}}
% 	\caption{Probability assertion for SR1\_1\_1.}
% 	\label{fig:prob asst} 
% \end{figure}

% \vspace{-10pt}
% \begin{figure}[t]
%     \centering
%  	\frame{\includegraphics[width= 1.0\linewidth]{images/mail_reward_asst1.png}}
% 	\caption{Reward assertion for SR1\_1\_2.}
% 	\label{fig:reward asst} 
% \end{figure}

% \begin{figure}[t]
%     \centering
%  	\frame{\includegraphics[width= 1.0\linewidth]{images/mail_temporal_asst1.png}}
% 	\caption{Temporal assertion for SR1\_1\_3.}
% 	\label{fig:temporal asst} 
% \end{figure}

% \subsubsection{AC update with new evidence models}
% \label{subsubsec:mail_robot_AC_update}After verification, evidence generation and AC update are performed automatically following Algorithm~\ref{Al:model checking evidence_gen}.
Using the verification result produced by PRISM together with the traceability information recorded in the Kapture requirement entries, the algorithm identifies the verification-method claim to be supported and generates the corresponding evidence artefact.
Using the verification results produced by PRISM together with the traceability information recorded in the Kapture requirement entries, Algorithm~\ref{Al:model checking evidence_gen} is applied to generate the corresponding evidence artefact.
For example, for SR1\_1\_1, the verification report shown in Fig.~\ref{fig:SR1 result} encodes both the requirement identifier and the verification tool used.
Algorithm~\ref{Al:model checking evidence_gen} extracts this identifier, resolves the associated verification-method claim via the forward traceability link, and generates an evidence model from the aggregated verification result.
This evidence is then integrated into the AC by replacing the placeholder solution element supporting the verification-method claim VR1\_1\_1\_1.

\begin{figure}[t]
    \centering
 	\includegraphics[width= 0.95 \linewidth]{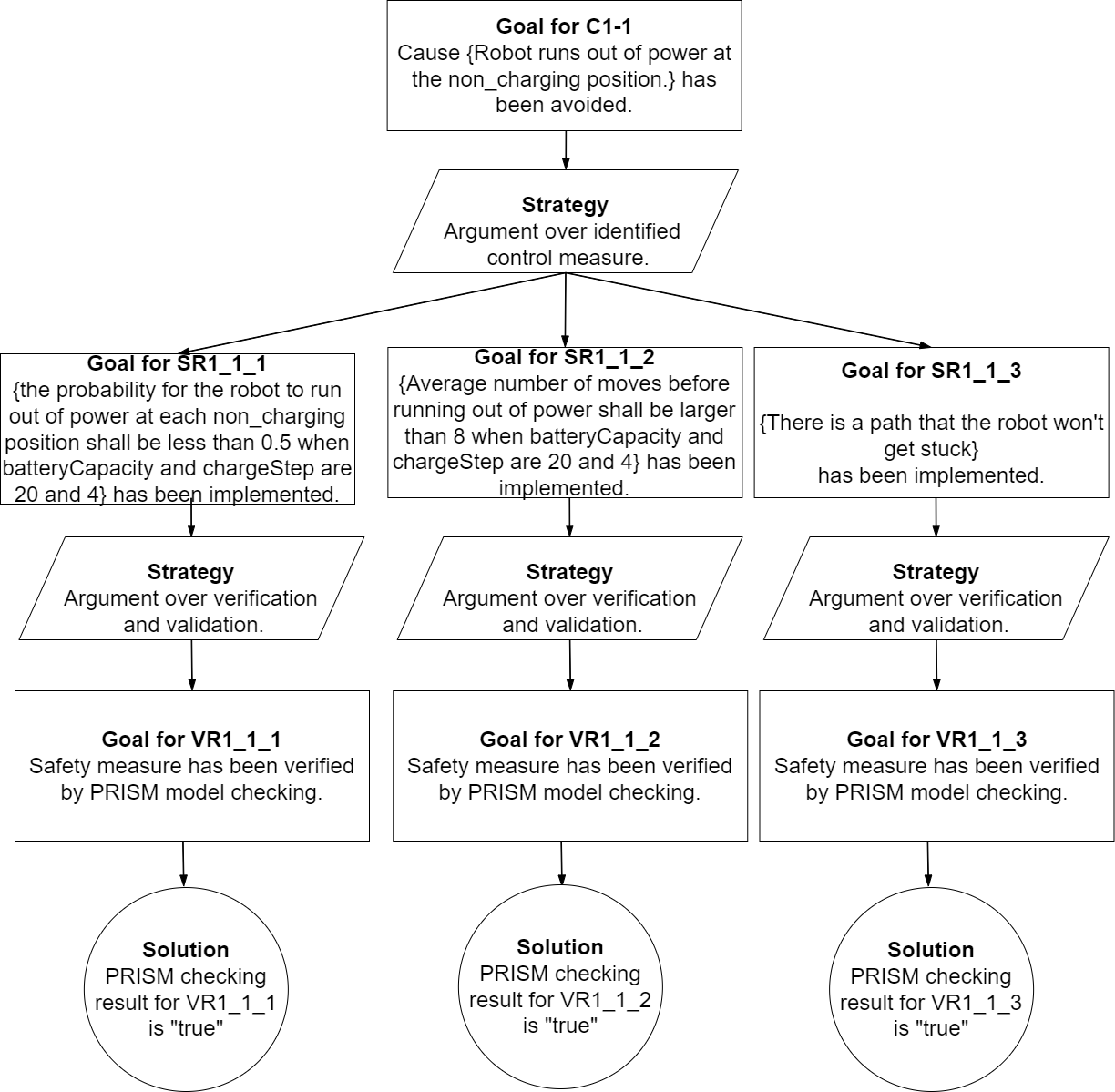}
	\caption{AC module for mail delivery robot updated with formal evidence.}
	\label{fig:Mail_robot GSN after PRISM}
\end{figure}

As a result, the preliminary AC shown in Fig.~\ref{fig:Mail_robot GSN before PRISM} is updated with concrete evidence derived from model checking, yielding the AC shown in Fig.~\ref{fig:Mail_robot GSN after PRISM}.

With this example, we demonstrate the application of our workflow and tools from start to end.  In the next section, we describe our additional examples, covering different tools~(FDR and Isabelle) and focusing on the extra lessons learned in those case studies.  

\subsubsection{Painting robot\label{subsec: hvc}}
%The painting robot is that presented in ~\cite{murray2022safety}. % is conducted to demonstrate the application of the proposed approach for evidence model generation using FDR model checking.

%\subsubsection{System description}

The %his case study considers an 
industrial painting robot ~\cite{murray2022safety} uses high voltage to charge paint particles, improving efficiency and quality. 
However, this poses safety risks, as high voltage can cause discharge sparks, potentially igniting paint particles in the air and leading to a fire. 
To ensure safety, a High Voltage Controller~(HVC) has been developed.
 
The system is powered by a 24V power signal, which supplies electricity to the HVC. The HVC uses the desired voltage setting %\textit{setPoint} 
to calculate a Pulse Width Modulation~(PWM) value%, referred to as \textit{PWM\_Output}
. This PWM value, ranging from 0\% to 100\%, corresponds to an analogue voltage signal between 0 and 10V, which is amplified by a transformer. The amplified signal serves as the input to the HVC, which adjusts the actual high voltage output.
The system continuously measures the actual voltage %\textit{HV.Actual} 
and the current %\textit{IM} 
in real time, providing feedback to the HVC. These measurements are used to dynamically adjust the PWM output %\textit{PWM\_Output} 
to ensure that the output voltage matches the desired set point.%\textit{HV.setPoint}.
%The RoboChart models for the painting robot can be found online\footnote{\url{https://robostar.cs.york.ac.uk/case_studies/hvc/index.html}}.

%A key safety concern is the risk of fire in the painting cell caused by sparks from unintended high-voltage output. 
At the software level, three scenarios are identified that could lead to hazardous high-voltage behaviour:
\begin{itemize}
    \item the HVC does not respond to changes of the voltage set point, resulting in a constant voltage output;
    \item the voltage set point is not disabled when the 24V power supply fails, causing the HVC to continue outputting high voltage;
    \item the PWM output is not disabled when the 24V power supply is off, leading to continued high-voltage output.
\end{itemize}
%
%
% The software behaviour of the paint robot is modelled as a RoboChart module \RC{mod\_sys}. We introduce the components necessary for understanding the example requirements. The complete RoboChart models can be found online\footnote{\url{https://robostar.cs.york.ac.uk/case_studies/hvc/index.html}}.
% The event \RC{ext\_pow24VStatus} is of an enumerate type \RC{Power} with two field \RC{ON} and \RC{OFF} to denote the status of the 24V power supply.
% The event \RC{ext\_setPoint} communicate on the value of the voltage set point.
% The behaviour of the HVC is modelled in the state machine \RC{State\_machine} which has multiple states, among which the \RC{ClosedLoop} state is the  normal state for operation, and is where the controller is regulating the
% voltage in relation to the set-point. In case the voltage is breaching the upper or
% lower limits, the state machine \RC{State\_machine} moves from \RC{ClosedLoop} to \RC{ErrorMode}.
%
The software behaviour of the painting robot is modelled as a RoboChart module \RC{mod\_sys}. 
Here, we describe only the elements of that module necessary to understand the example requirements; the complete RoboChart model is available online\footnote{\url{https://robostar.cs.york.ac.uk/case_studies/hvc/index.html}}.
The event \RC{ext\_} \RC{pow24VStatus} has an enumerated type \RC{Power} with two values, \RC{ON} and \RC{OFF}, representing the status of the external 24V power supply.
The event \RC{ext\_setPoint} is used to communicate the desired voltage set point.
The behaviour of the HVC is modelled by the state machine called \RC{State\_machine}. Its \RC{ClosedLoop} state specifies the normal operation, where the controller regulates the output voltage with respect to the set point.
If the voltage exceeds predefined upper or lower bounds, the state machine transitions from \RC{ClosedLoop} to another state \RC{ErrorMode}.

%To address these software-level causes, 
The following software requirements are identified.
\begin{enumerate}
    \item SR1\_1\_1: The actual voltage shall follow the set-point during normal operation.
    \item SR1\_2\_1: The voltage set point shall be set to 0 when the 24V power supply is switched off.
    \item SR1\_3\_1: The PWM output shall be set to 0 when the 24V power supply is off.
    \item SR1\_1\_1\_VA: The \RC{ClosedLoop} state shall be reachable in the state machine \RC{State\_machine}.
\end{enumerate}
%
\begin{comment}
\begin{figure}[ht!]
    \centering
 	\frame{\includegraphics[width= 0.8 \linewidth]{images/HVC_hazardlog_emf.png}}
	\caption{HVC hazard analysis EMF models.}
	\label{fig:HVC hazard log}
\end{figure}    
\end{comment}
%
%\subsubsection{Initial AC generation}
An initial AC has been constructed to argue that the software-level causes of unintended high-voltage behaviour are mitigated by the HVC software. The AC structure focuses on the satisfaction of the software requirements identified above. For brevity, the AC diagram is omitted. In this AC, the four goals corresponding to SR1\_1\_1, SR1\_1\_2, SR1\_1\_3, and SR1\_1\_1\-\_VA are each supported by a verification-method goal. 
All verification activities for these goals are performed using FDR.

\paragraph{Requirement structuring}

\begin{figure}[t]
    \centering
 	 \fbox{ \includegraphics[width= 0.95 \linewidth]{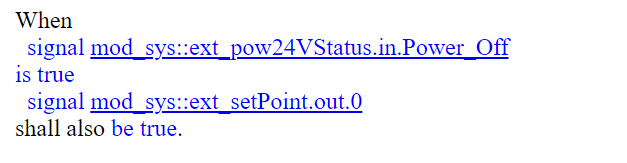}}
	\caption{HVC Claim for SR1\_2\_1.}
	\label{fig:HVC Kapure claim 2}
\end{figure}

\begin{figure}[t]
    \centering
 	 \fbox{ \includegraphics[width= 0.95 \linewidth]{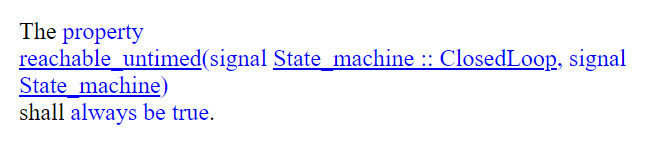}}
	\caption{HVC Claim for SR1\_1\_1 validation.}
	\label{fig:HVC Kapure claim 4}
\end{figure}

To structure the requirements in Kapture, for the requirements SR1\_1\_1, SR1\_2\_1, and SR1\_3\_1 we have used the RT-UNTIMED template, while for SR1\_1\_1\_VA we have used RT-REACH. We use SR1\_2\_1 and SR1\_1\_1\_VA as examples. 
Figs.~\ref{fig:HVC Kapure claim 2} and~\ref{fig:HVC Kapure claim 4} show their requirement specifications in Kapture.
The complete set of structured requirements and generated assertions is available in~\cite{yan2023assurance}.

% SR1\_1\_1 requires the actual voltage to follow the set-point during normal operation~(\RC{ClosedLoop} state). 
% To meet this, the placeholder \textit{guard\_event} in RT-UNTIMED uses two RoboChart events: \RC{currentState} set to \RC{State\_} \RC{ClosedLoop}, indicating that the HVC is in the normal operation mode, and \RC{ext\_setPoint} to receive any set-point value.
% The \textit{required\_event} is instantiated with \RC{int\_ActualHV}, which should output the same value as the \RC{setPoint}.
% Since \RC{int\_ActualHV} is asynchronous, it is prefixed with \textit{buffered} to facilitate assertion generation in the next step.

% \begin{figure}[ht]
%     \centering
%  	 \fbox{ \includegraphics[width= 0.95 \linewidth]{images/HVC claim 1 no trace.png}}
% 	\caption{HVC Claim for SR1\_1\_1.}
% 	\label{fig:HVC Kapure claim 1}
% \end{figure}

For SR1\_2\_1, the placeholder \texttt{guard\_event} is instantiated with the event \RC{ext\_pow24Vstatus} (identified by its fully qualified name) carrying the value \RC{Power\_Off}, and the \texttt{required\_event} is instantiated with the event \RC{ext\_setPoint} producing the output value 0.

For SR1\_1\_1\_VA, the \texttt{function} placeholder is instantiated with the predefined \texttt{reachable\_untimed} function, and constraining the analysis to the state machine \RC{State\_machine} via the \texttt{scope} placeholder.

\paragraph{Assertion generation and evidence generation}

% \begin{figure}[ht]
%     \centering
%  	 \fbox{ \includegraphics[width= 0.95 \linewidth]{images/hvc assertion 1 untimed buffered.png}}
% 	\caption{HVC assertion for SR1\_1\_1 verification.}
% 	\label{fig:HVC Kapure asst 1}
% \end{figure}

% SR1\_1\_1 involves communication between the platform \RC{RP1} and the state machine in the controller \RC{ctrl0} via events \RC{ext\_setPoint} and \RC{int\_ActualHV}. Hence, a buffer needs to be created in the CSP{$_\mathrm{M}$}  module of the assertion.
% Since the property is constrained within the \RC{ClosedLoop} state, it is considered a local property.
% Thus, the assertion for SR1\_1\_1 is generated using AT-UTLB, as shown in Fig.~\ref{fig:HVC Kapure asst 1}.

\begin{figure}[t]
    \centering
 	 \fbox{ \includegraphics[width= 0.95 \linewidth]{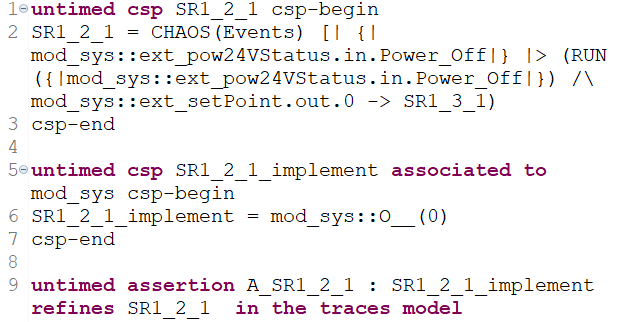}}
	\caption{HVC assertion for SR1\_2\_1 verification.}
	\label{fig:HVC Kapure asst 2}
\end{figure}

The assertion for SR1\_2\_1, shown in Fig.~\ref{fig:HVC Kapure asst 2}, is generated using the assertion template AT-UTG. We refer to the explanation of AT-UTG in Section~\ref{model checking assertion generation} for a detailed explanation of how the CSP definitions in Fig.~\ref{fig:HVC Kapure asst 2} capture our requirement.  We highlight that its definition requires considerable expertise in CSP, which is obviated by the use of our approach and tool. 
%In the CSP$_\mathrm{M}$ specification, the behaviour initially allows any event (\texttt{CHAOS}). If the event \texttt{mod\_sys::ext\_pow24VStatus.in.} \texttt{Power\_Off} occurs, this event may occur repeatedly until the event \texttt{mod\_sys::ext\_setPoint.out.0} is observed, after which the behaviour recurses to \texttt{CHAOS}. The CSP$_\mathrm{M}$ implementation module is obtained from the corresponding Robo\-Chart model under its CSP semantics.

% The assertion for SR1\_1\_1\_VA, shown in Fig.~\ref{fig:HVC Kapure asst validation}, is generated using the AT-REACH template.

% \begin{figure}[t]
%     \centering
%  	 \fbox{ \includegraphics[width= 0.95 \linewidth]{images/hvc assertion 4 reachability.png}}
% 	\caption{HVC assertion for SR1\_1\_1\_VA validation.}
% 	\label{fig:HVC Kapure asst validation}
% \end{figure}

In Robo\-Tool, the RoboChart models are automatically translated into CSP models for verification.
The FDR model checker then checks the generated assertions against these models under the CSP$_\mathrm{M}$ semantics.
% For brevity, the detailed verification results are omitted here.

% The subsequent steps of evidence model generation and integration into the AC follow the same process as described in Sect.~\ref{subsubsec:mail_robot_AC_update}, and are therefore not repeated.

\subsubsection{Autonomous Underwater Vehicle}\label{subsec:LRE}

The Autonomous Underwater Vehicle~(AUV) can operate under human supervision or autonomously to perform underwater maintenance and intervention tasks. 
A key safety issue for such systems is the risk of collision with subsea infrastructure, which may arise from both operator actions and autonomous behaviour.

The core safety component of the AUV is the Last Response Engine~(LRE), which monitors system behaviour and enforces safety constraints. 
In this case study, we apply our approach to verify deadlock-free\-dom of the LRE, which is captured as a verification-method claim in the AC. The property is verified using Isabelle.
We use the requirement template RT-DDLK and the assertion template AT-DDLK-2 to generate the assertion.
The generated assertion and the proof are as follows.

\smallskip

 \lstset{style=IsabelleStyle}{\small\begin{lstlisting}
lemma LRE_deadlock_free: "deadlock_free LREMachine"
    apply deadlock_free
    by (metis St.exhaust_disc)
  \end{lstlisting}
  }

\smallskip

\noindent
The formalised assertion is \lstinline{deadlock_free LREMachine}, where \lstinline{LREMachine} denotes the identifier of the LRE model in the Z-Machine notation.
The property is automatically verified in Isabelle by applying the \texttt{dead\-lock\_free} proof method and invoking the sledgehammer~\cite{bohme2010sledgehammer} Isabelle tool, which generates the proof line \lstinline{by (metis St.exhaust_disc)} automatically.
Details of the Robo\-Chart formalisation using Z-Machine and the proof method definition are in~\cite{yan2023automated}.

Because the LRE model includes real number parameters, it cannot be directly verified using the FDR model checker.
This case study therefore illustrates the complementary role of theorem proving within our approach, enabling the use of verification of properties beyond the scope of model checking in AC development.
The generation and integration of the corresponding evidence into the AC follow the same process as in the previous case studies. % and are therefore not repeated here.

\subsubsection{Maintenance robot}\label{subsec:unscrew_robot}

The fourth case study is a maintenance robot that unscrews laptop screens by identifying the screws on the laptop and applying the driver force to the screws automatically. 
The way used to model the robot in RoboChart is in two steps via the method provided in~\cite{baxter2025formal}.
The first step defined a RoboArch~\cite{BCM22} model for the robot; then, that model was translated automatically to RoboChart.
This RoboChart model represents the architecture of the robot software, and application details were subsequently added to describe the complete behaviour of the robot.

The RoboChart model of this robot adopts a variant of the MAPE-K~\cite{kephart2003vision} design structure. The model contains an \RC{Adaptation} controller that specifies the main MAPE-K loop.
The controller has six state machines to represent the behaviour of each MAPE-K component, among which we discuss two of them:~the state machine \RC{Adaptation\_Knowledge} and the state machine \RC{Adaptation\_Plan}.
\RC{Adaptation\_Knowledge} stores all the variables shared between components of the \RC{Adaptation} controller. 
The \RC{image} variable records the image received from the managed system. %, which is sent to the \RC{Adaptation\_Knowledge} state machine from the \RC{Adaptation\_Monitor} state machine.
The robot is equipped with a camera that takes images of the laptop screws. Every time \RC{Adaptation\_Knowledge} receives an image request through the event \RC{get\_image}, it should send out the stored image through the event \RC{image}.
The state machine \RC{Adaptation\_Plan} creates a plan to adapt to an identified anomaly and the key state of the machine is the \RC{MakePlan} state, where the plan is created.

An AC has been built based on the structural Robo\-Chart model to justify the safety goal of the robot. A set of software requirements referenced in the AC claims that require model checking is given below.
\begin{itemize}
    \item SR1: The \RC{Adaptation\_Knowledge} state machine shall always respond via the event \RC{image} when the event \RC{get\_image} is received.
    \item SR2: The \RC{MakePlan} step of the plan phase shall be reachable in the \RC{Adaptation\_Plan} state machine.
    \item SR3: The \RC{Adaptation\_Plan} state machine shall be deadlock-free.
    \item SR4: The \RC{Adaptation\_Plan} state machine shall be divergence-free.
\end{itemize}
We first used the Kapture templates RT-UNTIMED, RT-REACH, RT-DDLK, and RT-DIV to create requirements.
% SR1 is an untimed requirement that fits in the template RT-UNTIMED. The corresponding requirement in Kapture is shown in Fig.~\ref{fig:kap_DTI_R1}.
% The requirements in Kapture are provided in Appendix~\ref{sec:appendix:unscrew_robot_assts}.
Secondly, we have used the assertion templates AT-UTG, AT-REACH/DDLK/DIV to generate assertions.
Both the requirements exported from Kapture and the CSP assertions generated are in Appendix~\ref{sec:appendix:unscrew_robot_assts} and~\cite{robosapiensD12}.
%The process of evidence model generation and integration is not repeated here.

\subsection{Evaluation\label{subsec: contr2_evaluation}}

% To evaluate the effectiveness and generality of the formal evidence generation approach using model checking, the approach was applied to generate the AC modules for two case studies: the mail robot case study in Section~\ref{subsec: mail robot} and the HVC in Section~\ref{subsec: hvc}. 

This section answers our research questions proposed in Sect.~\ref{subsec:RQ}.

\noindent\textbf{RQ1 (Generality)} 
Our approach generates formal evidence through formal verification in support of AC claims. 
\rc{1.4}{Its applicability depends on the availability of a modelling language with formal semantics and associated verification tools. 
In this paper, the approach is instantiated for RoboChart, and the current implementation is tightly coupled with RoboChart's EMF metamodel, its CSP/PRISM/Z-Machine semantics, and the Robo\-Tool infrastructure. 
The approach is therefore currently demonstrated within the RoboChart ecosystem.}

\rc{1.4}{However, the workflow itself is not inherently Robo\-Chart-specific: it consists of (i)~structuring requirements using templates, (ii)~generating formal assertions via template instantiation, (iii)~orchestrating multiple verification backends, and (iv)~integrating results into an AC model. 
These steps are reusable and rely only on the existence of a suitable metamodel, formal semantics, and verification infrastructure. 
Applying the approach to another modelling language would require redesigning the requirement and assertion templates for that language's syntactic categories and semantics, but the overall workflow structure, including traceability management, verification orchestration, and evidence integration, would remain unchanged.}

The workflow's independence from the application domain is illustrated by our case studies: the mail delivery robot, the high-voltage controller, the underwater vehicle, and the unscrew robot originate from different domains of application, yet those domains do not affect the evidence model generation process. 
What is essential to this independence, however, is the use of RoboChart, whose formal semantics in CSP and PRISM for model checking and in Z-Machine for theorem proving enable the automatic generation of verification models. 
\rc{1.4}{\sout{In principle, other modelling languages with comparable formal semantics could also be adopted in place of RoboChart.}}

\rc{1.4}{Our approach currently incorporates the FDR and PRISM model checkers and the Isabelle theorem prover. 
The assertion templates are defined in terms of CSP{$_\mathrm{M}$}, PRISM, and Isar, reflecting the specific input languages of these tools. 
Supporting additional verification backends would require developing new assertion templates aligned with their formal semantics, which is part of our future work.}

\medskip\noindent\textbf{RQ2 (Coverage)} 
The coverage of our approach stems from the support for multiple formal semantic models of RoboChart and their associated verification tools. 
Unlike single-method frameworks that rely on one formalism or verification tool,  our workflow combines model checking and theorem proving across the CSP, Z-Machine, and PRISM semantics. 
This multi-seman\-tics design enables different classes of properties to be verified within a unified framework, 
thereby mitigating the limitations of individual tools and techniques. 
Table~\ref{table:req_coverage} summarises the requirement types currently supported across these domains.

Within this framework, the CSP semantics is used to address behavioural properties, such as untimed and timed safety and general behavioural correctness (reachability, divergence-freedom, deadlock-freedom, and termination). 
The Z-Machine semantics is used to deal with scalable state-based reasoning, supporting theorem proving of deadlock-freedom as a complementary form of verification. 
Finally, with the PRISM semantics, we extend the coverage to both probabilistic (quantitative) and non-probabilistic (qualitative temporal) properties, 
allowing the analysis of reliability, performance, and liveness.

Each requirement category is realised by one or more requirement templates, 
which define the syntactic and semantic patterns used to generate corresponding assertion templates. 
The current templates deliberately focus on common and well-defined verification patterns within each semantic domain. 
In the CSP domain, the templates are event-based and trace-oriented, each designed for a specific verification pattern such as the occurrence of a guard event followed by either the same guard or a required event. 
These templates address the most common forms of safety requirements over event traces, but they do not yet cover all possible behavioural combinations. For example, cases where a guard must be immediately followed by a required event would require a separate template. 
Furthermore, because the CSP semantics focuses on event interactions rather than explicit states, they are less suited to capturing data-dependent or condition-based requirements directly. 
Such cases can instead be expressed through the PRISM domain’s qualitative temporal templates, which provide higher-level expressiveness via temporal operators over state predicates.

For timed requirements, the framework currently supports deadline constraints, where a trigger event must be followed by a target event within a bounded time interval. 
This represents a common real-time software requirement pattern in embedded and robotic systems~\cite{akesson2022comprehensive}. 
However, time-related properties are inherently more complex and may require additional specialised templates to address case-specific needs. 
Our approach is designed to accommodate such extensions by allowing new requirements templates and their corresponding assertion templates to be incorporated into the workflow and tool. 

In the PRISM domain, the templates for probabilistic and temporal requirements adopt a deliberately general structure, 
with their expressive power concentrated in the \texttt{pathFormula} placeholder. 
This design follows the semantics of PRISM, where temporal and probabilistic behaviours are captured by logic formulas over state predicates. 
As a result, even though the template structure itself is simple, 
it can accommodate properties of varying complexity, ranging from basic reachability or safety checks to nested temporal and reward-based expressions, without requiring multiple specialised templates.

Overall, the current set of templates achieves practical coverage across the behavioural, state-based, and probabilistic aspects of RoboChart, 
and has proved sufficient for the range of requirements encountered in our case studies. 
While the expressive scope is intentionally conservative, 
the framework is designed to support incremental extension of the template library to accommodate additional behavioural and timing patterns when required.

\medskip\noindent\textbf{Limitations of template coverage.}
While the current templates cover common requirement patterns encountered in our case studies, certain patterns are not yet supported and would require additional templates in future, and we briefly discuss here.
\begin{itemize}[leftmargin=*,nosep]
\item \textbf{Immediate-next relations:} Requirements of the form ``event A must be \emph{immediately} followed by event B'' (without any intervening events) cannot be expressed by the current CSP templates, which focus on \emph{eventual} occurrence patterns.

\item \textbf{Complex data-dependent conditions:} Requirements involving arithmetic constraints or predicates over multiple variables that cannot be naturally encoded as RoboChart events are better suited to the PRISM domain's state-based templates.

\item \textbf{Nested probabilistic properties:} Properties requiring multiple reward structures or nested probabilistic operators are not covered by the current PRISM templates, though they could be added if needed.
\end{itemize}
The framework is designed to accommodate such extensions: new requirement and assertion templates can be defined following the patterns established in Sections~\ref{subsec:Claim Template Design} 
and~\ref{Assertion Template  design}, and integrated into the approach with minimal modification to the core workflow.

\noindent\textbf{RQ3 (Completeness of Integration)} 
To answer this research question, we evaluate the completeness of integration in terms of the structure of the resulting ACs. 
Specifically, we consider two aspects of structural completeness. 
First, end-to-end traceability is maintained throughout the workflow: from the claim to be supported, to the structured requirement referenced in the claim, to the assertion automatically generated from the requirement, and finally to the verification result produced by the verification tool. 
This traceability ensures that every claim linked to a formal requirement is systematically discharged with corresponding evidence, and that the evidence is correctly integrated into the AC modules. 

Second, the instantiated evidence models follow a predefined AC pattern introduced in~\cite{yan2022model}, implemented by Algorithm~\ref{Al:model checking evidence_gen}. 
This pattern explicitly defines the expected structure of an AC fragment, including the relationships between claims, verification-method claims, and evidence links. 
As evidence integration is performed by construction according to this pattern, the resulting AC fragments are structurally complete by design.
Across the case studies, we inspected the generated AC fragments and confirmed that they conform to the intended pattern, with each verification-method claim being associated with a corresponding evidence link. 
This provides confidence that the approach consistently produces AC fragments that are structurally complete with respect to their claims and evidence links. 

While the overall completeness of an AC also depends on the adequacy of the claims and requirements themselves, 
our workflow guarantees structural completeness of integration: once the links between claims and requirements are provided, 
every linked requirement is (automatically) formalised, verified, and integrated into the AC with its corresponding evidence.

\noindent\textbf{RQ4 (Scalability and Practical Efficiency)} 
To evaluate efficiency, we focus on the steps that are automated by our approach: the generation of assertions from structured requirements, and the instantiation and integration of evidence models from verification results. 

All case studies were executed on a 1.6 GHz Core i5 computer with 8GB of RAM. Table \ref{tab:case_summary_perf} reports the average execution time for the case studies, calculated over 10 runs. 
\rc{1.1}{The times reported in Table 5 correspond only to the automated transformation steps (assertion generation and evidence integration), not to the manual requirement structuring activity in Step 1.}
The automated steps for assertion generation and evidence integration completed within milliseconds, with the largest example~(the underwater vehicle) requiring less than 0.7 seconds in total. 
% In contrast, verification runtimes ranged from tens of seconds to several minutes, depending on the model and tool. 
These results confirm that the scalability of our workflow is determined by the verification back-ends, and that our work does not introduce any significant additional overhead.
% Verification runtimes in Table~\ref{tab:efficiency} are reported as single representative runs, since repeated executions produced very similar results for each case study.

This low computational cost is because both assertion generation and evidence instantiation are light\-weight transformations. 
Their complexity grows linearly with the number of requirements: assertion generation has complexity $O(R+E)$ 
where $R$ is the number of requirements and $E$ is the total 
number of guard elements (Algorithm~\ref{Al: integrated process of assertion generation}), 
and evidence integration has complexity $O(R)$ 
(Algorithm~\ref{Al:model checking evidence_gen}), where $R$ is the number of requirements. The true scalability bottleneck lies 
in the verification backends (FDR, PRISM, Isabelle), which are well-studied and widely used in industry. Our contribution of assertion generation and evidence integration operates independently of model size, with verification time dominating the workflow based on model complexity and tool performance rather than our approach.

In engineering practice, formal verification is typically applied to a focused set of safety-critical requirements, usually in the order of tens. This aligns with the scale of our case studies and 
indicates that the automation overhead remains negligible even for realistic verification scenarios with larger models.

%In terms of efficiency, the case studies show that the automated steps complete quickly (each run was under one second in our four cases). 
\rc{1.1, 1.3(b)}{The automated steps of our workflow (assertion generation and
evidence integration) complete within milliseconds per assertion
(Table~\ref{tab:case_summary_perf}). However, these measurements do not
include the manual requirement structuring step~(Step~1), which remains the
dominant human effort in the end-to-end workflow. In our experience across
the four case studies, structuring a requirement in Kapture typically takes
between 5 and 10 minutes per requirement, depending on the complexity of the RoboChart model and the engineer's familiarity with the naming conventions and template options. This is significantly larger than the automated transformation times reported in Table~\ref{tab:case_summary_perf},
which are on the order of milliseconds. The practical benefit of our approach, therefore, is not that it eliminates all manual effort, but that it eliminates the subsequent manual steps of writing CSP or PRISM assertions
(which would otherwise take 3--10 minutes per assertion) and manually updating AC evidence models (which would take 1--2 minutes per claim).
For a typical system with tens of requirements, the accumulated time saving is substantial, even though the initial requirement structuring remains a manual activity.}

\rc{1.3(c)}{A natural question is whether the overhead of learning the Kapture template conventions is justified compared to writing CSP or PRISM assertions directly, particularly for engineers already familiar with these languages.
However, this assumes they are already familiar with the RoboChart $\text{CSP}_{\mathrm{M}}$ models that serve as the implementation to be verified via refinement checking. 
The specifications or assertions they write must match the naming conventions of the $\text{CSP}_{\mathrm{M}}$ models generated from RoboChart. 
Therefore, both domain experts writing formal specifications directly and non-experts using Kapture templates face the same prerequisite: they must understand the structure of the RoboChart models and distinguish between input and output events.

Admittedly, expert users can bypass learning the Kapture templates if they write specification/assertions directly.
Nevertheless, our templates offer two practical benefits even for expert users. First, the templates encode common property patterns that are correct by construction, reducing the risk of subtle errors in assertion formulation (e.g., the incorrect use of CSP exception or interrupt operators). Second, the templates enforce a uniform structure across requirements, facilitating traceability from requirements to generated assertions, and ultimately to the acceptance evidence. 
For engineers without formal methods expertise, the benefit is even more pronounced: the templates make assertion generation accessible without requiring a deep understanding of CSP or PRISM semantics, acting as a key enabler for broader adoption in engineering practice.}

In contrast, a manual workflow without this automation would require
engineers to write each assertion and link each verification result to an
AC module by hand. For case studies with tens of assertions, this
accumulates to hours of manual effort, whereas our automation reduces human
effort to triggering the workflow; the runtime overhead then grows roughly
linearly with the number of assertions and, based on our measurements
reported in Table~\ref{tab:case_summary_perf}, is expected to remain in the
order of seconds. In addition, the automation also eliminates many of the
manual editing steps, thereby reducing the likelihood of human error in
evidence generation and integration.

Our evaluation based on the four case studies involves relatively small RoboChart models and only a few assertions. 
These are sufficient to demonstrate feasibility and illustrate different requirement categories, but they do not yet represent industrial-scale applications. 
As future work, we plan to apply the workflow to larger models with more complex requirements, in order to further validate its scalability and practical efficiency.

\subsection{Threats to validity\label{subsec: contr2_evaluation_threat}} 

Construct validity threats may arise from the assumptions about the validity of the software requirements and design models. 
To mitigate this, the safety requirements and RoboChart models used in the case studies have been mainly taken from previously validated examples~\cite{ye2022probabilistic, murray2022safety, foster2020formal}. 
This increases our confidence in their suitability.

External validity threats concern the restricted use of RoboChart as the modelling language and FDR, PRISM, and Isabelle as the verification back-ends. 
\rc{1.4}{\sout{Although other languages and tools have not been considered, the core concept of our workflow is not limited to RoboChart:~it can be applied to any modelling notation with formal semantics and associated verification support. 
Extending the evaluation to additional notations and tools will further improve the validity and generalisability of the findings.}}
\rc{1.4}{The current implementation is coupled with RoboChart's
metamodel, its semantics, and Robo\-Tool's infrastructure. The core
concept, including embedding formal verification into the assurance workflow via
template-based assertion generation, could in principle be adapted to other modelling languages, though such adaptation would require redesigning the requirement and assertion templates for the target language's syntactic categories and semantics.
Extending the evaluation to additional notations and tools is part of our
future work and would further improve the generalisability of the findings.}

Internal validity threats may arise from the implementation of the approach, in particular, the accuracy of generating assertions from requirements. 
To address this, we relied on three complementary measures.

\rc{1.2c}{\sout{First, the correctness of several key assertion templates has been analysed during their design, as discussed in Sect.~\ref{Assertion Template  design}, providing assurance that the generated assertions reflect the intended semantics. 

Second, formalised assertions from the literature have been used for comparison, ensuring that the automatically generated assertions are consistent with prior work.}}
\rc{1.2(c)}{First, the operational interpretations of the assertion templates, provided in Sect.~\ref{Assertion Template  design}, make explicit what semantics each AT assigns to its corresponding RT. These interpretations provide a reasoned basis for evaluating the encoding, without presupposing a formal semantics for Kapture.
Second, we validated the implementation of the instantiation procedures through oracle comparison. For requirements whose formal assertions had been independently written in prior work~\cite{murray2022safety,yan2023automated,pls}, we compared the automatically generated assertions against these manually written oracles on the same RoboChart models. In all cases, the generated assertions produced the same verification outcomes (pass/fail) in FDR, PRISM, and Isabelle as the oracles. This provides confidence that the EGL implementation faithfully realises the template designs.}

Third, extensive testing has been carried out on the evidence model generation, comparing the produced outputs with expected results, which increases confidence in the reliability of the generated evidence models.

\rc{1.2c}{We acknowledge that empirical validation is inherently finite; it only covers the patterns and models represented in the case studies. The semantic adequacy of the templates themselves is grounded in the operational interpretations provided in Sect.~\ref{Assertion Template  design}, together with the established pattern semantics from the literature. New template patterns, if added in future work, should be validated by the same combination of operational interpretation and case-study testing.}

%% file: relwork.tex
\section{Related work}\label{sec:relwork}

A common approach to applying formal verification in the AC process is referencing existing verification results within the AC evidence component~\cite{Calinescu2018b, jee2010assurance, Prokhorova2015}. 
Thus, the process of obtaining verification results to support claims is usually not addressed. Furthermore, previous research typically focuses on generating evidence from formal design models, whereas RAS systems are more commonly developed using MDE techniques and semi-formal modelling languages. 
A comprehensive solution is therefore needed to effectively incorporate formal verification into ACs for evidence generation. 
Our work proposes an integrated approach to derive formal evidence from argument claims for RAS systems modelled in RoboChart, thereby streamlining the AC process with formal verification support.

Early efforts towards integration include the work of Denney and Pai~\cite{Denney2018}, who integrated formal verification into the Assurance Case Automation Toolset (AdvoCATE).
AdvoCATE invokes AutoCert, a static code analysis tool that verifies source code generated from system models against formally specified properties and produces machine-checkable certificates, which are subsequently integrated into the assurance case.
% More recent work~\cite{denney2026advocate} by Denney and Pai further extends AdvoCATE beyond argument construction to encompass broader safety risk management activities, including hazard analysis, risk modelling, and evidence organisation.
% These extensions focus on assurance case engineering and management, rather than on the automated generation of formal verification evidence, which remains the focus of our work.
%Their approach primarily focuses on structuring and integrating verification results into assurance arguments, rather than on automating requirement formalisation or assertion generation.
Prokhorova~et~al.~\cite{Prokhorova2015} demonstrated how different requirement types can be addressed by combining multiple verification methods, but their workflow relied on manual steps and did not achieve automation. 
% Their contribution is primarily methodological, showing how results from different formal methods can be used to support assurance arguments, rather than providing tool-supported automation for evidence generation or traceability.
Bourbouh~et~al.~\cite{Bourbouh2021a, bourbouh2020integration} investigated the integration of heterogeneous formal methods into ACs, combining the contracts with Simulink-to-Lustre translation and Event-B refinement proofs. 
The contracts are written in the restricted
natural language of the Formal Requirements Elicitation Tool (FRET)~\cite{giannakopoulou2020formal}, referred to as FRETISH.
However, at the component verification level, the Lustre and Event-B models were created from Simulink models manually, so the consistency between these models and the design artefacts required expert review.

For requirement formalisation, several approaches translate natural language or semi-structured requirements into temporal logic. 
The FORM-L approach~\cite{bouskela2022formal} employed ontology and templates to resolve ambiguities, producing semi-formal requirements for Modelica models, but its expressiveness is limited. 
Bolton~et~al.~\cite{bolton2014automatically} generated LTL assertions from predefined templates to verify Human-Automation Interaction behaviour, but only for a restricted set of properties. 
ARSENAL~\cite{ghosh2016arsenal} applied parsing and domain ontologies to translate natural language into LTL, providing more flexibility at the cost of manual ontology development. 

% A prominent representative of structured requirement formalisation is FRET~\cite{giannakopoulou2020formal}, an open-source framework developed by NASA.
% FRET allows requirements to be written in FRETISH, which requires users to manually populate six fields, and automatically translated into future-time and past-time LTL.%, enabling model checking of Simulink-based designs via tools such as CoCoSim and Kind2.
% Recent work by Mavridou~et~al.~\cite{mavridou2025automated} extends FRET with explicit support for probabilistic requirements by enriching the FRETISH language and providing an automated translation into PCTL$^*$, together with validation of the generated formulas.

\rc{1.6}{% ======================================================================
% Expanded FRET/FRETISH comparison for Section 9 (Related Work).
% Replaces the existing FRET discussion of the original manuscript.
%
% Addresses RC1.6 points (a), (b), and (c).
% ======================================================================

A prominent representative of structured requirement formalisation is FRET~\cite{giannakopoulou2020formal}, an open-source framework developed by NASA.
FRET allows requirements to be written in FRETISH, a restricted natural language whose sentences are parsed into six fields: scope (the time interval in which the requirement is enforced), condition (a Boolean trigger), component (the system element the requirement applies to), shall (the obligation keyword), timing (when the response must occur), and response (a Boolean expression the component must satisfy). From these fields, FRET automatically
generates future-time and past-time LTL formulas. Recent work by Mavridou et al.~\cite{mavridou2025automated} extends FRETISH with a seventh field, probability, and provides an automated translation into PCTL$^{*}$, together with validation of the generated formulas.

Since both FRET and our approach aim to bridge the gap between
informal requirements and formal verification, a detailed comparison is warranted.}

\rc{1.6(a)}{\paragraph{Structural comparison.}
FRETISH and our \texttt{Kapture-\allowbreak based} templates adopt different
decompositions. FRETISH uses a fixed sequence of six or seven
fields that are domain-independent: the response field is a
non-temporal Boolean expression, and the scope and timing fields
capture temporal structure. By contrast, our requirement templates
are built on three Kapture base templates (\emph{when},
\emph{trigger on event}, \emph{every}), each with placeholders
that refer directly to RoboChart model elements such as fully
qualified events, states, and clocks. This design reflects a
fundamental difference in target formalisms: FRETISH targets
propositional temporal logics (LTL, PCTL$^{*}$), whereas our
templates additionally target CSP refinement assertions, which
encode behavioural properties using process-algebraic constructs
(exception, interrupt, and parallel composition) that cannot be directly expressed in temporal logics. On the other hand, FRETISH's
domain-independent design makes it applicable beyond robotics,
whereas our templates are specific to RoboChart models.}

\rc{1.6(b)}{\paragraph{Probabilistic requirements.}
The probabilistic extension of FRET~\cite{mavridou2025automated} adds a probability field to FRETISH and translates the resulting sentence
into PCTL$^{*}$, covering probability-threshold properties (e.g., ``with probability $> 0.95$ satisfy $\phi$''). Our RT-PROB template
serves a similar purpose. 
However, FRET does not currently support reward-based properties (described as future work in~\cite{mavridou2025automated}), such as expected cumulative rewards or reward bounds over paths. These are handled by our RT-RWD template,
which generates PRISM reward declarations and reward-based property
queries. Reward properties are important in robotic applications for reasoning about resource consumption (e.g., expected battery usage
before depletion, as in our mail delivery robot case study).}

\rc{1.6(c)}{\paragraph{Relative advantages.}
FRET offers several strengths that our approach does not share:
domain independence, a broader coverage of temporal logic patterns,
built-in realizability checking, interactive semantic diagrams that
help users validate their intent, and a large, NASA-backed user
community. Our approach, on the other hand, offers three
capabilities that FRET does not provide. First, it covers CSP
refinement properties, which are essential for verifying
event-based safety requirements of the kind commonly encountered
in RoboChart models; the CSP refinement assertion templates
(AT-UTG, AT-UTL, AT-DLINE) encapsulate the most technically
demanding aspect of formalisation, where deep knowledge of CSP
syntax and semantics would otherwise be required. Second, it
supports reward-based properties via RT-RWD. Third, it integrates
requirement formalisation, assertion generation, verification
execution, and evidence integration into a single workflow for
assurance case development, whereas FRET focuses on the
requirement-to-formula translation step without addressing
assurance case integration.}

\rc{1.6}{The two approaches are complementary rather than competing:
FRET excels at domain-independent temporal and probabilistic
requirement formalisation, while our approach targets the
RoboChart-specific, process-algebraic, and assurance-case
dimensions that FRET does not cover. Exploring the use of
FRET as an alternative front-end for the temporal and
probabilistic subset of our requirement templates is a
promising direction for future work.}

\rc{2.1}{To further position the novelty of our templates with respect to the broader literature on property specification, we compare our approach with the existing pattern catalogues. The pattern catalogues of Dwyer et al.~\cite{dwyer1999patterns}, Konrad and Cheng~\cite{konrad2005real}, Grunske~\cite{grunske2008specification}, Menghi et al.~\cite{menghi2019specification,menghi2022mission}, Autili et al.~\cite{autili2015aligning}, and Etumi et al.~\cite{etumi2025towards} share a common characteristic: they provide classifications of commonly occurring property types (e.g., response, universality, absence, real-time deadlines, probabilistic thresholds) and express them as abstract temporal logic schemas. They define \emph{what} kinds of properties exist, but do not provide concrete, tool-executable assertions for specific verification backends such as CSP{$_\mathrm{M}$}, PRISM, or Isar, nor do they address integration with assurance workflows.}

\rc{2.1}{By contrast, our templates define \emph{how} such properties are expressed as executable assertions for particular verification tools, and \emph{how} the resulting evidence is integrated into an assurance case. Specifically: (i)~our untimed and timed templates (RT-UNTIMED/AT-UTG, RT-TIMED/AT-DLINE) encode response and real-time patterns as CSP{$_\mathrm{M}$} refinement assertions checkable by FDR, using process-algebraic constructs~(exception, interrupt, timed priority) specific to CSP; (ii)~our probabilistic template~(RT-PROB) follows Grunske's probabilistic patterns, generating PCTL formulas for PRISM; and (iii)~our reward-based template~(RT-RWD) extends existing catalogues by supporting reward properties~(e.g., expected cumulative rewards), which are not covered by Grunske's ProProST catalogue or the other cited works.}

\rc{2.1}{Among the templates we present, the CSP refinement templates~(AT-UTG, AT-UTL, AT-DLINE) and their integration with AC evidence generation are novel. The reward-based template (RT-RWD, AT-RWD) extends the state of the art by covering property types not present in existing catalogues. The probabilistic and temporal templates correspond to existing pattern classes but are distinguished by their integration into the AC evidence workflow and their generation of PRISM-ready assertions.}

More recently, large language models~(LLMs) were explored as a means of automating requirement formalisation. 
Zhao~et~al.\ proposed NL2CTL~\cite{zhao2024nl2ctl}, which translates requirements into CTL/LTL specifications; Cosler~et~al.\ presented nl2spec~\cite{cosler2023nl2spec}, which supports interactive correction of sub-formula translations; Li~et~al.\ introduced a self-supervised framework for safety-compliant LTL~\cite{li2025automatic}; and Xu~et~al.\ improved robustness by learning from failed translations~\cite{xu2024learning}. 
These works showed the promise of LLMs but remain limited to propositional temporal logics~(LTL/CTL) and achieve only partial accuracy as reported in the literature. By contrast, our approach covers CSP refinement, and CTL and PCTL properties.%, domains not covered by existing LLM-based methods.

The work of Denney and Pai~\cite{Denney2018} also addressed the formalisation of properties for formal verification within assurance cases. Among its practical functions for AC construction, AdvoCATE provides a claim formalisation (CF) pattern, through which users interactively supply the values of the pattern parameters based on informal claims and the language and logic adopted for formalisation. 
One of the parameters is the claim itself, directly expressed as a logical formula.
In this way, informal claims are formalised and represented as structured sub-claims within the instantiated pattern.
The formula is placed into an atomic slot of the pattern. As a result, the internal structure of the formula is not constrained by the pattern itself. This design offers greater expressive freedom but requires users to possess substantial expertise in the underlying formal languages and logics.
In this respect, the CF pattern is conceptually analogous to our RT templates in that both aim to bridge informal statements and formal verification artefacts. However, while CF supports the interactive formalisation of assurance case claims by allowing users to directly provide logical formulae, our RT templates focus on the systematic structuring of software requirements and constrain the form of the resulting assertions to reduce the need for user-level knowledge of specific formal languages.

Domain-specific solutions have also been proposed for RoboChart.
Lindoso~et~al.~\cite{Lindoso2021a} have developed a diagrammatic notation mixing UML activity elements with RoboChart constructs, from which CSP specifications can be derived and verified with FDR. 
Their diagrams must be created manually, in contrast to our template-based automatic assertion generation. Use of Lindoso's diagrams can, however, be an alternative to Kapture templates. 

The Isabelle/SACM framework~\cite{Nemouchi2019,foster2020formal,foster2021integration} formalises ACs by embedding the SACM metamodel into Isabelle, yielding machine-checkable cases that can be verified for logical consistency. 
Since cases are represented in Isabelle/Isar, results from formal proofs, such as deadlock freedom of RoboChart models, can be linked directly as evidence. 
The emphasis of this work is on AC formalisation and consistency checking, with limited automation support.

\rc{2.2}{A separate strand of work focuses on the management and evolution of assurance cases themselves. Viger et al.~\cite{viger2020just} formalise decomposition strategies to ensure that AC structures are deductively valid. 
Di Sandro et al.~\cite{disandro2020mmint}, building on earlier work by Fung et al.~\cite{fung2018mmint}, provide tool support for the creation, analysis, and evolution of safety models through MMINT-A.
Chechik et al.~\cite{chechik2020uncertainty} address the management of uncertainty in assurance cases. Gallina and colleagues~\cite{carlan2022automating} propose semi-automated change impact analysis to identify which parts of an AC are affected by system changes.}

\rc{2.2}{Building on SACM-based foundations, Wei et al.~\cite{wei2023automated} developed the ACME tool for AC management, integrating model validation via Constrained Natural Language rules translated into Epsilon EVL. The ACCESS methodology extends this vision by positioning ACs as central artefacts, integrating Isabelle/SACM for formalised cases and introducing dynamic ACs that evolve at runtime with system data~\cite{wei2024access}.}

\rc{2.2}{These frameworks focus on the AC itself, including its structure, consistency, evolution, and uncertainty, rather than on the production of the evidence that populates it. By contrast, our work focuses on the generation of assurance evidence through formal verification: we automate the production of verification results and their integration into ACs. The two lines of work are complementary: our evidence generation could supply evidence artefacts to AC management frameworks such as MMINT-A, while their capabilities could help maintain the AC structure as the system evolves.}

% Building on SACM-based foundations, Wei~et~al.~\cite{wei2023automated} developed the ACME tool for AC management. 
% One strand of this work integrates model validation into the AC process by allowing claim nodes to be annotated with Constrained Natural Language~(CNL) rules, which are automatically translated into Epsilon EVL and executed on EMF-based RoboChart models, with outcomes recorded as evidence. 
% The ACCESS methodology extends this vision by positioning ACs as central artefacts in system development, integrating Isabelle/SACM for formalised cases, supporting model validation with Epsilon rules, and introducing dynamic ACs that evolve at runtime with system data~\cite{wei2024access}. 
% Together, these approaches focus on formalisation, validation, and lifecycle management of ACs. 
% In contrast, our work does not formalise the AC itself, but integrates formal verification results from FDR, PRISM, and Isabelle into the assurance workflow. 

\rc{2.3}{Beyond the formalisation and management of ACs, another related line of work concerns the design of assurance case templates. For example, Wassyng et al.~\cite{wassyng2015product,chowdhury2017principles} propose product-specific assurance case templates that define the structure of the argument itself. That is, what claims a system must make, what strategies to use, and what kinds of evidence to provide. These templates are developed for product lines and instantiated per product, serving as blueprints for complete ACs.
A related notion appears in our own work: the AC pattern defined in our previous work~\cite{yan2022model} structures the local relationship between a verification-method claim and the evidence that supports it. This pattern specifies, for a given verification activity (model checking, theorem proving, or testing), how the corresponding evidence artefact should be linked to the claim. }

\rc{2.3}{The key difference is one of scope. Wassyng's templates provide a global blueprint for the entire AC structure, whereas our AC pattern provides a local pattern for the evidence-to-claim relationship that can be instantiated repeatedly within a larger AC. The two are complementary: an AC developed from Wassyng's template could incorporate our evidence pattern for individual claims.}

Sorokin~et~al.~\cite{sorokin2024towards} employed the Evidential Tool Bus~(ETB) to orchestrate testing and simulation tools for incremental AC maintenance. 
Their focus, however, is on non-formal verification artefacts, whereas our work integrates formal verification results into the AC workflow.

Sljivo~et~al.~\cite{sljivo2023guided} proposed reusable tool confidence patterns to strengthen the justification of verification results in assurance cases.
These patterns explicitly capture the assumptions, guarantees, and applicability conditions of verification tools, making the use of tool-generated evidence more transparent and credible.
The patterns are realised within the AdvoCATE framework, but the contribution primarily concerns the structuring of arguments about tool trustworthiness rather than the automation of verification or evidence generation.
Our automatically generated verification evidence can complement this line of work by providing formally verified results that can be embedded within such patterns.

\begin{table*}[t]
\centering
\caption{Coverage of assurance case workflow stages across related lines of work.}
\label{tab:rw-matrix}
\renewcommand{\arraystretch}{1.2}
\scriptsize
\setlength{\tabcolsep}{6pt}

\begin{tabular}{p{0.5\textwidth}cccccc}
\toprule
\textbf{Reference (line of work)} 
& \makecell{Req.\\struct.} 
& \makecell{Assert.\\gen.} 
& \makecell{Model\\transf.} 
& \makecell{Verif.\\exec.} 
& \makecell{Evid.\\integ.} 
& \makecell{Trace-\\ability} \\
\midrule

Denney \& Pai -- AutoCert / AdvoCATE~\cite{Denney2018}
& \xmark & \cmark & \cmark & \cmark & \cmark & \cmark \\

Prokhorova et al. -- Multi-tool FV (Event-B-centred)~\cite{Prokhorova2015}
& \xmark & \xmark & \xmark & \cmark & \cmark & \xmark \\

Bourbouh et al. -- Heterogeneous FM (FRET/Lustre/Event-B)~\cite{Bourbouh2021a, giannakopoulou2020formal, mavridou2025automated}
& \cmark & \cmark & \cmark & \cmark & \cmark & \xmark \\

Bouskela et al. -- FORM-L~\cite{bouskela2022formal}
& \cmark & \cmark & \xmark & \xmark & \xmark & \xmark \\

Bolton et al. -- Template-based LTL~\cite{bolton2014automatically}
& \cmark & \cmark & \xmark & \cmark & \xmark & \xmark \\

Ghosh et al. -- ARSENAL~\cite{ghosh2016arsenal}
& \cmark & \cmark & \xmark & \xmark & \xmark & \xmark \\

Zhao et al.; Cosler et al.; Li et al.; Xu et al. -- LLM-based NL2LTL~\cite{zhao2024nl2ctl,cosler2023nl2spec,li2025automatic,xu2024learning}
& \xmark & \cmark & \xmark & \xmark & \xmark & \xmark \\

Lindoso et al. -- Diagrammatic CSP (RoboChart)~\cite{Lindoso2021a}
& \cmark & \cmark & \cmark & \cmark & \xmark & \xmark \\

Nemouchi et al.; Foster et al. -- Isabelle/SACM~\cite{Nemouchi2019,foster2020formal,foster2021integration}
& \xmark & \xmark & \xmark & \cmark & \cmark & \cmark \\

Wei et al. -- CNL / ACME~\cite{wei2023automated}
& \cmark & \cmark & \xmark & \cmark & \cmark & \cmark \\

Wei et al. -- ACCESS~\cite{wei2024access}
& \xmark & \xmark & \xmark & \cmark & \cmark & \cmark \\

Sorokin et al. -- ETB~\cite{sorokin2024towards}
& \xmark & \xmark & \xmark & \xmark & \cmark & \cmark \\

Sljivo et al. -- Tool Confidence Patterns~\cite{sljivo2023guided}
& \xmark & \xmark & \xmark & \xmark & \cmark & \cmark \\

\textbf{This work}
& \cmark & \cmark & \cmark & \cmark & \cmark & \cmark \\

\bottomrule
\end{tabular}

\vspace{0.8ex}
\footnotesize
\textit{Notation:} \cmark~the stage is explicitly addressed in the work; 
\xmark~the stage is not addressed.
\rc{1.7}{The columns are defined as follows:
\textit{Req. struct.}~whether informal requirements are structured or formalised; 
\textit{Assert. gen.}~whether assertions are generated automatically; 
\textit{Model transf.}~whether models are transformed to verification backends; 
\textit{Verif. exec.}~whether verification execution is part of the workflow; 
\textit{Evid. integ.}~whether verification results are integrated as AC evidence; 
\textit{Traceability}~whether links are maintained from claims to evidence.
A \cmark~indicates that the stage is supported by the approach at the level of the framework or tool, not that it is fully automated; the degree of automation 
varies and is discussed in the corresponding text.}

\textit{Note:} Some lines of work (e.g., AdvoCATE~\cite{Denney2018}, FORM-L~\cite{bouskela2022formal}, ARSENAL~\cite{ghosh2016arsenal}, and ETB~\cite{sorokin2024towards}) are explicitly named in the original publications.
Others have no explicit name; we use short descriptive labels (e.g., ``Template-based LTL''~\cite{bolton2014automatically},
``LLM-based L2LTL''~\cite{zhao2024nl2ctl,cosler2023nl2spec,li2025automatic,xu2024learning}, and ``Diagrammatic CSP (RoboChart)''~\cite{Lindoso2021a}) for clarity, while citations always point to the original sources.
CNL/ACME and ACCESS are listed separately as they represent different strands of work by the same authors, addressing distinct stages of the assurance case workflow.

\rc{1.7}{\textit{Regarding specific cases:} AdvoCATE is marked as not addressing requirement structuring because its claim formalisation patterns require users to supply logical formulae directly, and do not provide a systematic means of structuring informal requirements prior to formalisation. This differs from the requirement structuring stage as defined in this paper
(see column definitions above), which covers the systematic capture of requirements in a structured form suitable for automated assertion generation. 
Conversely, our approach is marked as addressing the requirement structuring because the Kapture templates provide such a systematic structure, even though populating the templates remains a manual activity (as discussed in Sect.~5.4).}
\end{table*}

Table~\ref{tab:rw-matrix} summarises which stages of the assurance case workflow
are explicitly addressed by different lines of work. The workflow stages used in the table are defined with respect to the scope of our approach.
As a result, our work naturally addresses all listed stages, while other works cover different subsets according to their respective research focus.
For conciseness, we do not describe each stage individually for every work;
instead, the table supports the discussion of how our approach relates to existing work in terms of workflow focus.
Different strands of work address different aspects of an AC development:~requirement formalisation into temporal logics~(FORM-L, Template-based LTL, ARSENAL, LLM-based NL2LTL), model validation integrated with ACs~(CNL/ACME), AC formalisation and consistency checking~(Isabelle/SACM, ACCESS), and evidence orchestration or tool patterns~(ETB, AdvoCATE). 
AdvoCATE is marked as not addressing requirement structuring but as supporting assertion generation. While AdvoCATE does not provide mechanisms for structuring prior to formalisation, it does support the formalisation of informal statements into verification properties through its claim formalisation patterns.
Our work concentrates on the formal verification dimension:~we automate the generation of CSP and PCTL assertions from structured requirements, verify them with heterogeneous backends~(FDR, PRISM, Isabelle), and inject the results as evidence into ACs. 
We acknowledge that requirement structuring still requires manual input and that traceability is only partially supported via assurance patterns, but our contribution lies in advancing automation for formal verification evidence within AC workflows.

%% file: concl.tex
\section{Conclusions and future work}
\label{sec:concl}

This paper has presented a model-based approach to automating evidence generation for ACs in RAS. By embedding model checking and theorem proving into the assurance workflow, our approach closes the automation gap that left evidence generation outside. Requirements referenced in AC claims are classified and linked to customised templates, enabling the automatic derivation of tool-ready assertions. Verification is then delegated to suitable backends~(FDR, PRISM, Isabelle/HOL), and the results are integrated as assurance evidence. Collectively, these contributions establish a unified framework for AC engineering supported by formal verification.

\rc{1.1}{Our case studies demonstrate that our approach substantially reduces manual effort for assertion generation and evidence integration: the automated steps of assertion generation, verification orchestration, and evidence integration are completed within milliseconds, whereas writing assertions manually would typically take several minutes per assertion. However, requirement structuring (Step 1) remains a manual activity requiring knowledge of the RoboChart model and the Kapture template conventions; the automation benefit lies in eliminating the subsequent manual steps of writing CSP/PRISM assertions and updating AC evidence.}

Among the assertion templates we defined, those for CSP refinement stand out as particularly valuable: they encapsulate some of the most challenging aspects of formalisation, where a deep understanding of CSP syntax and semantics would otherwise be required. By lowering this learning barrier, the templates directly support the wider adoption of formal verification in engineering practice. Other templates, while associated with a less steep learning curve, also contribute to easing the transition from natural language requirements to verification-ready assertions.

Looking ahead, we see several promising directions for future work.
One line of extension is to integrate additional verification and validation techniques, such as testing and simulation, to complement the current focus on formal verification. 
Secondly, we currently check refinement in the traces model to focus on safety; an extension is to broaden this to include richer refinement models, such as the failures model.

Another important direction concerns the automation of theorem-proving-based evidence generation.
At present, deadlock-freedom properties are verified in Isabelle by generating the Z-Machine semantics and proof obligations within a single Isabelle theory file, while the subsequent identification of the supported AC claim and the integration of the resulting evidence follow the same workflow but are carried out manually.
Future work will focus on fully automating this process, aligning theorem-proving-based verification with the level of automation already achieved for model checking.

Another promising direction concerns requirement formalisation. Our current approach reduces the need for formal expertise through structured templates, but it still requires semi-structured input via Kapture. Advances in Natural Language Processing~(NLP) offer opportunities to further automate this step. Traditional NLP pipelines, such as semantic parsing, have been applied (e.g., NL2CTL~\cite{zhao2024nl2ctl}, nl2spec~\cite{cosler2023nl2spec}) to generate temporal logic from natural language, but they rely heavily on domain-specific annotations.
Recent progress in Large Language models~(LLMs) %, which are built on the Transformer architecture,
provides new possibilities. LLMs such as GPT-4 have demonstrated strong capabilities in generating structured representations from free text and could assist in translating natural language requirements into formal assertions. Unlike earlier machine learning approaches, LLMs benefit from large-scale pretraining and therefore require less task-specific data. Exploring the integration of LLMs with our assertion-generation process is a promising avenue of work to improve both automation and usability.

\rc{1.5}{Another direction is to improve support for failed verification outcomes.
Currently, when a verification attempt fails, the evidence model records
the failure, but diagnostic information, such as counterexamples, must be manually inspected by the engineer outside the automated workflow using the corresponding verification tools. We plan to extend our evidence generation framework to automatically extract
and present such diagnostic information, integrating it into the assurance case to provide engineers with actionable feedback on the cause of failure and to support iterative refinement of requirements,
models, and assertions.}

%In addition, the scalability of our method needs to be further explored by applying it to larger and more complex case studies, ideally in industrial contexts. 
Finally, enhancing tool support and integration into AC development environments will be crucial to improve usability and encourage adoption. %Together, these directions will increase the efficiency, scalability, and accessibility of automated AC generation.

%% file: Appendix.tex
\section{Appendix\label{appendix}}

Appendices~\ref{sec:appendix:RTs} and \ref{sec:appendix:ATs} provide the additional requirement and assertion templates that complement those presented in Sect.~\ref{subsec:Claim Template Design} and Sect.~\ref{model checking assertion generation}. 
Appendix~\ref{sec:appendix:algs} presents the algorithms corresponding to the auxiliary procedures invoked in Algorithm~\ref{Al: integrated process of assertion generation} (e.g.\ \textsc{GenerateGeneralAsst}, \textsc{GenerateTimedAsst}, etc.).
%Appendix~\ref{app:proof-sketches} provides proof outlines for the template correctness lemmas stated in Sect.~\ref{sec:template-correctness}.
Appendix~\ref{sec:appendix:unscrew_robot_assts} presents the set of generated assertions for the unscrew robot. 

\subsection{Requirement templates}
\label{sec:appendix:RTs}

This section provides the software requirement templates for the requirement categories of \emph{reachability}, \emph{divergence-freedom}, \emph{termination}, and \emph{probability}, and \emph{temporal}. 
% Each template is derived from the customised Kapture bases (\texttt{when}, \texttt{every}) and presented with a schematic form and a minimal example.

\paragraph{Reachability requirements}
Reachability requirements specify whether a given state shall or shall not be reachable in a RoboChart component. 
We capture this pattern in the reachability template (RT-REACH).

\begin{tcolorbox}[colback=white, boxrule=0.5pt, arc=0pt, halign=flush left,
left=2pt, right=2pt, top=2pt, bottom=2pt, boxsep=2pt]
\begin{footnotesize}
\textbf{RT-REACH} \label{RT-REACH}\\ [4pt]
The property \textcolor{blue}{\emph{function}}(\textcolor{blue}{\emph{state}}, \textcolor{blue}{\emph{scope}}) \
shall \textcolor{blue}{\emph{boolStmt}} be true.\\
\vspace{1.5ex}

$\ast$ \emph{function} \quad::= \texttt{reachable\_untimed} ~|\\
\quad\quad\quad\quad\quad\quad~~~~ \texttt{reachable\_timed} ~|~ \texttt{reachable}\\\vspace{0.5ex}
$\ast$ \emph{state}\quad\quad\quad ::= \textcolor{blue}{\emph{RoboChartStateIdentifier}}\\\vspace{0.5ex}
$\ast$ \emph{scope}\quad\quad\quad ::= \textcolor{blue}{\emph{RoboChartComponentIdentifier}}\\\vspace{0.5ex}
$\ast$ \emph{boolStmt}~~\, ::= \texttt{always} ~|~ \texttt{never}\\
\end{footnotesize}
\end{tcolorbox}

% \noindent Compiling a RoboChart model into CSP$_\mathrm{M}$ produces both an untimed and a timed semantics.
% The template therefore provides three functions:
% \texttt{reachable\_untimed} checks reachability for untimed semantics,
% \texttt{reachable\_timed} for timed semantics,
% and \texttt{reachable} leaves the semantics unspecified, in which case both variants are checked.

% \noindent Example (R-REACH):
% \emph{The state \RC{Wait24VPower} shall be reachable in the voltage controller.}

% \noindent The placeholder \emph{state} is instantiated by \RC{Wait24VPower}, and \emph{scope} by the state machine \RC{vol\_stm}, the chosen function is \texttt{reachable}, and the Boolean statement is \texttt{always}.
% Instantiated using the RT-REACH template, the requirement becomes:

% \begin{tcolorbox}[colback=white, boxrule=0.5pt, arc=0pt, halign=flush left,
% left=2pt, right=2pt, top=2pt, bottom=2pt, boxsep=2pt]
% \begin{footnotesize}
% The property \texttt{reachable}(\texttt{Wait24VPower}, \texttt{vol\_stm}) \
% shall \texttt{always} be true.
% \end{footnotesize}
% \end{tcolorbox}

\paragraph{Divergence-freedom requirements}
The divergence-freedom requirement is captured in the template (RT-DIV).

\begin{tcolorbox}[colback=white, boxrule=0.5pt, arc=0pt, halign=flush left,
left=2pt, right=2pt, top=2pt, bottom=2pt, boxsep=2pt]
\begin{footnotesize}
\textbf{RT-DIV} \label{RT-DIV}\\[4pt]
The property \textcolor{blue}{\emph{function}}(\textcolor{blue}{\emph{scope}}) \
shall \textcolor{blue}{\emph{boolStmt}} be true.\\
\vspace{1.5ex}

$\ast$ \emph{function} ~~\,::= \texttt{divergence\_free\_untimed} ~|~\\\vspace{0.5ex}
\quad\quad\quad\quad\quad\quad\quad\texttt{divergence\_free\_timed} ~|\\\quad\quad\quad\quad\quad\quad\quad
\texttt{divergence\_free}\\\vspace{0.5ex}
$\ast$ \emph{scope} \quad\quad\,::= \textcolor{blue}{\emph{RoboChartComponentIdentifier}}\\\vspace{0.5ex}
$\ast$ \emph{boolStmt}~~ ::= \texttt{always} ~|~ \texttt{never}\\
\end{footnotesize}
\end{tcolorbox}

% \noindent The template allows three function variants depending on whether untimed, timed, or both semantics of the model are considered.

% \noindent Example (R-DIV):
% \emph{The voltage controller shall be diverg-} \emph{ence-free.}

% \noindent Here, the placeholder \emph{scope} corresponds to the state machine \RC{vol\_stm} modelling the voltage controller.
% The chosen function is \texttt{divergence\_free}, and the Boolean statement is \texttt{always}.
% Instantiated using the RT-DIV template, the requirement becomes:

% \begin{tcolorbox}[colback=white, boxrule=0.5pt, arc=0pt, halign=flush left,
% left=2pt, right=2pt, top=2pt, bottom=2pt, boxsep=2pt]
% \begin{footnotesize}
% The property \texttt{divergence\_free}(\texttt{vol\_stm}) \
% shall \texttt{always} be true.
% \end{footnotesize}
% \end{tcolorbox}

\paragraph{Termination requirements}
The template for the termination requirements is as follows.

\begin{tcolorbox}[colback=white, boxrule=0.5pt, arc=0pt, halign=flush left,
left=2pt, right=2pt, top=2pt, bottom=2pt, boxsep=2pt]
\begin{footnotesize}
\textbf{RT-TERM} \label{RT-TERM}\\[4pt]
The property \textcolor{blue}{\emph{function}}(\textcolor{blue}{\emph{scope}}) \
shall \textcolor{blue}{\emph{boolStmt}} be true.\
\vspace{1.5ex}

$\ast$ \emph{function} ~~::= \texttt{terminate\_untimed} ~|\\
\quad\quad\quad\quad~~~~~~~~~~~\texttt{terminate\_timed} ~|\\
\quad\quad\quad\quad\quad\quad~~~\,\texttt{terminate}\\\vspace{0.5ex}
$\ast$ \emph{scope} \quad\quad::= \textcolor{blue}{\emph{RoboChartComponentIdentifier}}\\\vspace{0.5ex}
$\ast$ \emph{boolStmt} ~~::= \texttt{always} ~|~ \texttt{never}\\
\end{footnotesize}
\end{tcolorbox}

% \noindent As with other general templates, the function variants distinguish untimed, timed, or unspecified semantics, with the unspecified case defaulting to both.

% \noindent Example (R-TERM):
% \emph{The voltage controller shall terminate.}

% \noindent Here, the placeholder \emph{scope} corresponds to the state machine \RC{vol\_stm} modelling the voltage controller.
% The chosen function is \texttt{terminate}, and the Boolean statement is \texttt{always}.
% Instantiated using the RT-TERM template, the requirement becomes:

% \begin{tcolorbox}[colback=white, boxrule=0.5pt, arc=0pt, halign=flush left,
% left=2pt, right=2pt, top=2pt, bottom=2pt, boxsep=2pt]
% \begin{footnotesize}
% The property \texttt{terminate}(\texttt{vol\_stm}) \
% shall \texttt{always} be true.
% \end{footnotesize}
% \end{tcolorbox}

\paragraph{Constants configuration for reward/probability/temporal templates.}
\label{app:constants-configs}
The placeholder \emph{constantsConfigs} is used in RT-RWD, RT-PROB, and RT-TEMP. It specifies how constants are assigned for PRISM-based verification. Its grammar is as follows.
\begin{tcolorbox}[colback=white, boxrule=0.5pt, arc=0pt, halign=flush left,
  left=2pt, right=2pt, top=2pt, bottom=2pt, boxsep=2pt]
\begin{footnotesize}
\textbf{Constants configuration (for RT-RWD/RT-PROB/RT-TEMP)} \label{app:constants-configs}\\[4pt]
\emph{constantsConfigs} \,::= [\,\textcolor{blue}{\emph{constants}}\,] [\,\texttt{and}\,] \\
\quad\quad\quad\quad\quad\quad\quad\quad\quad\quad\![\,\textcolor{blue}{\emph{multiConstants}}\,] \\[3pt]

\emph{constants} \quad\quad\quad\quad::= \textcolor{blue}{\emph{constConfig}} \{\,\texttt{and} \textcolor{blue}{\emph{constConfig}}\,\} \\[3pt]

\emph{constConfig} ~~\quad\quad::= \texttt{constant\_}\textcolor{blue}{\emph{RoboChart\_constant}} \\
\quad\quad\quad\quad\quad\quad\quad\quad\quad\quad\texttt\!{ set to }\textcolor{blue}{\emph{Expr}} \\[3pt]

\emph{multiConstants} \quad::= \textcolor{blue}{\emph{constMultiConfig}} \\
\quad\quad\quad\quad\quad\quad\quad\quad\quad\quad\{\,\texttt{and} \textcolor{blue}{\emph{constMultiConfig}}\,\} \\[3pt]

\emph{constMultiConfig} ::= \texttt{multi\_constant\_x from set} \textcolor{blue}{\emph{Expr}}\\
\end{footnotesize}
\end{tcolorbox}

% \noindent\emph{Notation:} square brackets \texttt{[\,\,]} denote optional parts; braces \texttt{\{\,\,\}} denote zero or more repetitions.

% \noindent%
% \texttt{constantsConfigs} declares one or more RoboChart constants and their values (or value sets), which parameterise the quantitative analyses in PRISM.
% \noindent\textbf{Explanation.}  
% \begin{itemize}
%   \item \emph{constantsConfigs} collects one or more constant assignments, optionally combined with ranges.  
%   \item A \emph{constConfig} fixes a RoboChart constant to a specific value, e.g.\  
%   \texttt{constant\_batteryCapacity set to 20}.  
%   \item A \emph{constMultiConfig} introduces a symbolic range, e.g.\  
%   \texttt{multi\_constant\_speed from set \{10,20,30\}}, which expands to multiple PRISM runs.  
%   \item The operator \texttt{and} allows combining several assignments or ranges, so that multiple constants are configured simultaneously.
% \end{itemize}

% Together, these constructs allow requirements to be analysed under fixed parameters or across families of configurations, enabling parametric verification.

\paragraph{Probability requirements.}
These requirements request that the probability of satisfying a path property shall reach a given bound under fixed constants.
The template (RT-PROB) is as follows.

\begin{tcolorbox}[colback=white, boxrule=0.5pt, arc=0pt, halign=flush left, left=2pt, right=2pt, top=2pt, bottom=2pt, boxsep=2pt]
 \renewcommand{\baselinestretch}{1.1}\selectfont
\begin{footnotesize}
\textbf{RT-PROB}\\[4pt]
When \textcolor{blue}{\emph{constantsConfigs}} is true,\\
Until \textcolor{blue}{\emph{pathFormula}} becomes true,\\
\textcolor{blue}{\emph{probTarget}} shall also be true.\\[4pt]

$\ast$ \emph{constantsConfigs}: see above for the grammar\\[2pt]
$\ast$ \emph{pathFormula} ~~\quad::= \texttt{pathFormula\_}\textcolor{blue}{\emph{pExpr}}\\[2pt]

$\ast$ \emph{probTarget} ~~\quad\quad\!::= \texttt{prob\_target\_}\textcolor{blue}{\emph{op}}~\textcolor{blue}{\emph{Expr}}\\

$\ast$ \emph{op}\!\quad\quad\quad\quad\quad\quad\quad ::= \texttt{>} ~|~ \texttt{>=} ~|~ \texttt{<} ~|~ \texttt{<=} ~|~ \texttt{==}\\
\end{footnotesize}
\end{tcolorbox}

\paragraph{Temporal requirements.}
The temporal requirements specify qualitative temporal properties in PRISM, with no numeric bounds. 
The template (RT-TEMP) is as follows.

\begin{tcolorbox}[colback=white, boxrule=0.5pt, arc=0pt, halign=flush left, 
  left=2pt, right=2pt, top=2pt, bottom=2pt, boxsep=2pt]
 \renewcommand{\baselinestretch}{1.1}\selectfont
\begin{footnotesize}
\textbf{RT-TEMP}\\[4pt]
When \textcolor{blue}{\emph{constantsConfigs}} is true,\\
\textcolor{blue}{\emph{termOp}}~\textcolor{blue}{\emph{pathFormula}} shall also be true.\\\vspace{1.5ex}

$\ast$ \emph{constantsConfigs}: see above for the grammar\\\vspace{0.5ex}

$\ast$ \emph{termOp} ~~\quad\quad\quad\,::= \texttt{term\_}[\,\texttt{not}\,](\texttt{forall} ~|~ \texttt{exists})\\

$\ast$ \emph{pathFormula} ~~\quad::= \texttt{pathFormula\_}\textcolor{blue}{\emph{pExpr}}\\
\end{footnotesize}
\end{tcolorbox}
% \noindent Example (R-TEMP):  
% % \emph{Not all paths lead the robot to being stuck.}  
% \emph{The robot can not run out of power capacity when it is not in the charge station 0.} 
% Being stuck is modelled as the state machine entering \texttt{stuck}. Here, \texttt{NIL} marks that no numeric constraint is needed which implies that this is temporal requirement.

% \begin{tcolorbox}[colback=white, boxrule=0.5pt, arc=0pt, halign=flush left, 
%   left=2pt, right=2pt, top=2pt, bottom=2pt, boxsep=2pt]
% \begin{footnotesize}
% When \texttt{constant\_sys::ctrl::stm::batteryCapacity set to 20 } is true,\\
% \texttt{term\_not exists pathFormula\_(ctrl\_stm\_deliverCTRL\_p != 0 \& ctrl\_stm\_deliverCTRL\_c = 0)} shall also be true.\\
% \end{footnotesize}
% \end{tcolorbox}

\subsection{Assertion templates}
\label{sec:appendix:ATs}

This appendix provides the additional templates for the assertion templates not covered in  Sect.~\ref{Assertion Template  design}, namely AT-UTL, AT-REACH, AT-DIV, AT-TERM, AT-PROB, and AT-TEMP. 

\paragraph{Untimed refinement assertion template for local invariants (AT-UTL).}
AT-UTL is used when the requirement applies only when a state machine is in a particular state.
% In the AT-UTL template, keywords of the RoboChart assertion language are shown in purple.
% Placeholders are written in black italics; these will later be instantiated with values extracted from the XML requirements.
% Fixed parts of the template, which cannot be changed, are written in upright black font.
% The identifier of the assertion, written as \textit{A\_Claim\_ID}, is derived from the claim identifier in the AC models, ensuring traceability between the assertion and the evidence structure.

\begin{tcolorbox}[colback=white, boxrule=0.5pt, arc=0pt, halign=flush left,
  left=2pt, right=2pt, top=2pt, bottom=2pt, boxsep=2pt]
 \renewcommand{\baselinestretch}{1.1}\selectfont
 \begin{footnotesize}
 \textbf{AT-UTL}:\\[4pt]

\textbf{\color{eclipsePurple}{untimed csp}} Spec csp-begin\\
\texttt{Spec} = \texttt{CHAOS(Events)} 
\texttt{[\!|\{}\textit{guardEvent\_1}\texttt{\}|>}\\
\quad \texttt{LocalBehaviour} \\[4pt]

\texttt{LocalBehaviour = CHAOS(Events)}\texttt{[\!|\{}\textit{guardEvent\_2}\texttt{\}|>}\\
\quad \texttt{(RUN(\{}\textit{guardEvent\_2}\texttt{\})/\textbackslash (}
\textit{requiredEvent}\texttt{ -> Spec))}\\
csp-end\\[6pt]

\textbf{\color{eclipsePurple}{untimed csp}} Spec\_impl
\textbf{\color{eclipsePurple}{associated to}} \textit{module\_name}\\ csp-begin\\
Spec\_impl = \textit{module\_name}%\texttt{|\!\textbackslash \{} \textit{guardEvent\_1}\texttt{, } \textit{guardEvent\_2}\texttt{, } \textit{requiredEvent\_without\_value}\texttt{\}}
\\
csp-end\\[6pt]

\textbf{\color{eclipsePurple}{untimed assertion}} A\_\textit{Claim\_ID} :\\
Spec\_impl \textbf{\color{eclipsePurple}{refines}} Spec
\textbf{\color{eclipsePurple}{in the traces model}}\\
\end{footnotesize}
\end{tcolorbox}

In the specification section, AT-UTL follows a structure similar to AT-UTG, using the CSP{$_\mathrm{M}$} exception operator \texttt{[\!|\,|\!>} together with the process \texttt{CHAOS(A)} to introduce a scope, and the interrupt operator \texttt{/\textbackslash}  inside the local behaviour.
The principal difference from AT-UTG is the introduction of an explicit scope, denoted by \textit{guardEvent\_1}.
This event acts as a scope delimiter, restricting the requirement to apply when the state machine is in a particular state.
For example, the condition that a state machine is in a particular state $s\_1$ can be modelled using the event \texttt{module::currentState.s1}, where \texttt{currentState} is an enumeration over the states of the machine.
% More generally, \textit{guardEvent\_1} may also represent a regular guard condition, in which case the requirement is enforced only when that guard is satisfied.

Within this scope, the process \textit{LocalBehaviour} enforces the same safety pattern as in AT-UTG, but relative to \textit{guardEvent\_2} (the local guard):
once \textit{guardEvent\_2} occurs, only further \textit{guardEvent\_2} events may appear until the next \textit{requiredEvent}; when \textit{requiredEvent} occurs, the behaviour recurs by returning to \textit{Spec}.
Note that this is a safety property and does not enforce that \textit{requiredEvent} must eventually occur; infinite continuations of \textit{guardEvent\_2} are permitted.
% Also, \textit{requiredEvent} may occur before \textit{guardEvent\_2} within the scope, since the pre-\textit{guardEvent\_2} phase is unconstrained by \texttt{CHAOS}.

\paragraph{Reachability, divergence-free, termination}
Now we introduce the three assertion templates for reachability, divergence-free, and termination.

\begin{tcolorbox}[colback=white, boxrule=0.5pt, arc=0pt, halign=flush left, left=2pt, right=2pt, top=2pt, bottom=2pt, width=\columnwidth,boxsep=2pt]
 \renewcommand{\baselinestretch}{1.1}\selectfont
\begin{footnotesize}
AT-REACH:\\[4pt]
(\textbf{\color{eclipsePurple}{untimed}} ~|~  \textbf{\color{eclipsePurple}{timed}})? \textbf{\color{eclipsePurple}{assertion}} A\_\textit{Claim\_ID}:\\ [2pt]
\textit{state} \textit{boolTerm} \textbf{\color{eclipsePurple}{reachable in}} \textit{scope}. \\[4pt]

%$\ast$  scope ::= \textit{module} $\vert$  \textit{controller} $\vert$  \textit{state\_machine}
$\ast$ \textit{boolTerm} ::= \textbf{\color{eclipsePurple}{is}} ~|~ \textbf{\color{eclipsePurple}{is not}}\\
\end{footnotesize}
\end{tcolorbox}

\begin{tcolorbox}[colback=white, boxrule=0.5pt, arc=0pt, halign=flush left, left=2pt, right=2pt, top=2pt, bottom=2pt, boxsep=2pt]
 \renewcommand{\baselinestretch}{1.1}\selectfont
\begin{footnotesize}
AT-DIV:\\[4pt]
(\textbf{\color{eclipsePurple}{untimed}} ~|~  \textbf{\color{eclipsePurple}{timed}})? \textbf{\color{eclipsePurple}{assertion}} A\_\textit{Claim\_ID}:\\ 
\textit{scope}
\textit{boolTerm}
\textbf{\textbf{\color{eclipsePurple}{divergence-free}}}.\\[4pt]

$\ast$ \textit{boolTermd} ::= \textbf{\color{eclipsePurple}{is}} ~|~ \textbf{\color{eclipsePurple}{is not}}\\
\end{footnotesize}
\end{tcolorbox}

\begin{tcolorbox}[colback=white, boxrule=0.5pt, arc=0pt, halign=flush left, left=2pt, right=2pt, top=2pt, bottom=2pt, boxsep=2pt]
 \renewcommand{\baselinestretch}{1.1}\selectfont
\begin{footnotesize}
AT-TERM:\\[4pt]
(\textbf{\color{eclipsePurple}{untimed}} ~|~  \textbf{\color{eclipsePurple}{timed}})? \textbf{\color{eclipsePurple}{assertion}} A\_\textit{Claim\_ID}:\\
\textit{scope}
\textit{boolTerm}.\\[4pt]

%$\ast$  scope ::= \textit{module} $\vert$  \textit{controller} $\vert$  \textit{state\_machine}
$\ast$ \textit{boolTerm} ::= \textbf{\color{eclipsePurple}{ terminates}} ~|~ \textbf{\color{eclipsePurple}{does not terminate}} \\
 \end{footnotesize}
\end{tcolorbox}

In the above templates, the placeholders \textit{state} and \textit{scope} will be replaced with the elements of the same placeholders from the corresponding requirements, which are structured using RT-REACH, RT-DIV, and RT-TERM.

\paragraph{Probability assertion}

AT-PROB ``Probability assertion template'' is designed as follows:

\begin{tcolorbox}[colback=white, boxrule=0.5pt, arc=0pt, halign=flush left, left=2pt, right=2pt, top=2pt, bottom=2pt, boxsep=2pt]

 \renewcommand{\baselinestretch}{1.1}\selectfont
\begin{footnotesize}
AT-PROB:\\[4pt]
(\textbf{\color{eclipsePurple}{const}} \textit{cname}: \textit{type})$\ast$\\[4pt]
\textbf{\color{eclipsePurple}{prob property}} P\_\textit{Claim\_ID}:\\[2pt]
\textbf{\color{eclipsePurple}{Prob}} \textit{probTarget} \textbf{\color{eclipsePurple}{of}} [\textit{pathFormula}]\\[2pt]

(\textbf{\color{eclipsePurple}{with constant}}\\
(\textit{constant\_i})$\ast$ ,\\
(\textit{multi\_constant\_j})$\ast$ )$\ast$
 \end{footnotesize}
\end{tcolorbox}

The first line of the template declares any constants that the assertion may depend on. The assertion section begins with the keywords \textbf{\color{eclipsePurple}{prob property}}.
The assertion includes the property name and content, which verifies whether the probability of \textit{pathFormula} satisfies the quantitative bound specified in \textit{probTarget}.
The identification of the assertion, \textit{P\_Claim\_ID}, is defined based on the claim identification in AC models, ensuring traceability between the assertions and AC models.

The second section `with constant' is used for the constant configuration. 
\textit{constant\_i} is employed to assign a single value to the constant, while \textit{multi\_constant\_j} is used to assign a set of values to the constant. 

The placeholders within AT-PROB will be substituted with elements of the corresponding type found in requirements written using RT-PROB.

\paragraph{Temporal assertion}

AT-TEMP ``Temporal assertion template'' features two sections similar to AT-PROB. 
These assertions assess whether the property specified in the \textit{pathFormula} is satisfied by either all paths or some paths. The placeholders \textit{temporal\_operator} can have the values of \textit{forall} or \textit{exists}.
Both placeholders will be substituted with the corresponding elements found in the requirements documented in RT-TEMP.

\begin{tcolorbox}[colback=white, boxrule=0.5pt, arc=0pt, halign=flush left, left=2pt, right=2pt, top=2pt, bottom=2pt, width=0.5\textwidth,boxsep=2pt]
 \renewcommand{\baselinestretch}{1.1}\selectfont
\begin{footnotesize}
AT-TEMP:\\[4pt]
\textbf{\color{eclipsePurple}{prob property}} T\_\textit{Claim\_ID}:\\[2pt]
\textit{temporal\_operator} [\textit{pathFormula}]\\[2pt]

(\textbf{\color{eclipsePurple}{with constant}}\\
(\textit{constant\_i})$\ast$ ,\\
(\textit{multi\_constant\_j})$\ast$)$\ast$ \\
\end{footnotesize}
\end{tcolorbox}

The assertion templates and the requirement templates utilise the identical set of placeholders, which simplifies the process of generating assertions by instantiating the assertion placeholders with the content from the requirement placeholders.
% The assertion generation is discussed in Sect.~\ref{Assertion Template Instantiation}.

\subsection{Assertion generation algorithms}
\label{sec:appendix:algs}

This appendix presents six auxiliary algorithms that instantiate assertion templates from structured requirements. 
They implement the mapping rules described in Sect.~\ref{Assertion Template  design} and are invoked as helper functions by the integrated procedure in Algorithm~\ref{Al: integrated process of assertion generation}. 
Since their behaviour is subsumed by the correctness proof of the integrated procedure, their correctness is guaranteed accordingly.

The \texttt{when} template has three instantiable sections \texttt{GuardCondition}, \texttt{UntilCondition}, and \texttt{RequiredCon-} \texttt{dition}.
Depending on the types of claims, these three sections map to different sets of elements.

\begin{algorithm}[h]
\caption{\textsc{GenerateProbAsst}}
\label{Al:GenerateProbAsst}
\SetAlgoLined
\DontPrintSemicolon
\SetKwProg{Proc}{Procedure}{:}{}

\Proc{\textsc{GenerateProbAsst}($r$, $claim\_ID$)}{
    $probTarget \leftarrow r.\texttt{RequiredCondition.removePrefix}()$\;
    $pathFormula \leftarrow r.\texttt{UntilCondition.removePrefix}()$\;
    \texttt{InstProbAsst}($claim\_ID$, $probTarget$, $pathFormula$)\;
}
\end{algorithm}

Algorithm~\ref{Al:GenerateProbAsst} handles probability claims, identified by the prefix ``prob\_target'' in the \textit{RequiredCondition}. The algorithm extracts three elements of \textit{probTarget}, \textit{pathFormula}, and \textit{constant\_i} by de-prefixing the corresponding sections of the requirement. These are then used to instantiate template AT-PROB, partly within this algorithm (lines 2–4) and partly in the main procedure (line~29 of Algorithm~\ref{Al: integrated process of assertion generation}).

% In Algorithm~\ref{Al:GenerateProbAsst}, the assumption is that section ``RequiredCondition'' is prefixed with ``prob\_target", which means that this claim is of type probability. 
% Therefore, the assertion is generated using template AT- 9, which has three elements ``probTarget'', ``pathFormula'', and ``constant\_i'' to be instantiated. 
% ``RequiredCondition'' in the requirement $r$, given as argument, is de-prefixed by the function \textsc{removePrefix()}, and assigned to ``probTarget''. ``UntilCondition'' is assigned to ``pathFormula''.
% The element ``constant\_i'' is assigned a value using the de-prefixed ``GuardCondition''.
% The assertion is generated partially by instantiating the AT-9 template using ``probTarget'', ``pathFormula'' in line~4 of Algorithm~\ref{Al:GenerateProbAsst} and partially by instantiation using ``constant\_i'' in line~29 of Algorithm~\ref{Al: integrated process of assertion generation}.

Algorithm~\ref{Al:GenerateRewardAsst} is similar to Algorithm~\ref{Al:GenerateProbAsst} but has two more elements ``rewardEvent'' and ``rewardValue'' in the assertion template AT-RWD.
These two are mapped back to the ``GuardCondition'' section of the claim.
\begin{algorithm}[h]
\caption{\textsc{GenerateRewardAsst}}
\label{Al:GenerateRewardAsst}
\SetAlgoLined
\DontPrintSemicolon
\SetKwProg{Proc}{Procedure}{:}{}

\Proc{\textsc{GenerateRewardAsst}($r$, $claim\_ID$)}{
    $rewardTarget \leftarrow r.\texttt{requiredCondition.removePrefix}()$\;
    $pathFormula \leftarrow r.\texttt{untilCondition.removePrefix}()$\;
    $rewardEvent \leftarrow$ 
    $r.\texttt{guardCondition.elems.select}(m \mid m.\texttt{prefixed}(\text{``reward\_event''})).\texttt{removePrefix}()$\;
    $rewardValue \leftarrow$ 
    $r.\texttt{guardCondition.all.elems.select}(m \mid m.\texttt{prefixed}(\text{``reward\_value''})).\texttt{removePrefix}()$\;
    \texttt{InstRewardAsst}($claim\_ID$, $rewardTarget$, $pathFormula$, $rewardEvent$, $rewardValue$)\;
}
\end{algorithm}

% \begin{comment}
Algorithm~\ref{Al:GenerateTempAsst} follows the same pattern as the other two presented above to generate assertions using AT-TEMP. 
% \end{comment}

\begin{algorithm}[h]
\caption{\textsc{GenerateTempAsst}}
\label{Al:GenerateTempAsst}
\SetAlgoLined
\DontPrintSemicolon
\SetKwProg{Proc}{Procedure}{:}{}

\Proc{\textsc{GenerateTempAsst}($r$, $claim\_ID$)}{
    $temporal\_operator \leftarrow r.\texttt{requiredCondition.elems.select}(m \mid m.\texttt{prefixed}(\text{``term\_''})).\texttt{removePrefix}()$\;

    $pathFormula \leftarrow r.\texttt{requiredCondition.elems.select}(m \mid m.\texttt{prefixed}(\text{``pathFormula''})).\texttt{removePrefix}()$\;

    \texttt{InstTempAsst}($claim\_ID$, $temporal\_operator$, $pathFormula$)\;
}
\end{algorithm}

Algorithm~\ref{Al:GenerateUntimedAsst} generates the assertions related to untimed refinement, using the assertion template AT-UTG for the global invariant type, AT-UTL for the local invariant type.
The algorithm first decides whether the requirement is of a global or local type.
If there are two elements defined in the ``GuardCondition'' section of the requirement  (line~2), the claim is of type ``local invariant''.
Then, the first element of ``GuardCondition'' is the scope constraint mapped to ``guardEvent\_1'' in the assertion template, the second is the local guard condition mapped to ``guardEvent\_2'', and 
``requiredEvent'' in the assertion is assigned the value of ``requiredCondition'' (lines~3-5).

\begin{algorithm}[t]
\caption{\textsc{GenerateUntimedAsst}}
\label{Al:GenerateUntimedAsst}
\SetAlgoLined
\DontPrintSemicolon
\SetKwProg{Proc}{Procedure}{:}{}

\Proc{\textsc{GenerateUntimedAsst}($r$, $claim\_ID$)}{

  \If{$r.\texttt{guardCondition.elem1Defined()}$}{
  
    $guardEvent\_1 \leftarrow r.\texttt{guardCondition.elems.at(0)}$\;
    
    $guardEvent\_2 \leftarrow r.\texttt{guardCondition.elems.at(1)}$\;
    
    $requiredEvent \leftarrow r.\texttt{requiredCondition}$\;
    
    $requiredEvent\_wo \leftarrow requiredEvent.\texttt{removeValue()}$\;

    % \If{$requiredEvent.\texttt{prefixed("buffered")}$}{
    
    %   \texttt{instLocBufUntimedAsst}($claim\_ID$, $guardEvent\_1$, $guardEvent\_2$, $requiredEvent$, $requiredEvent\_wo$)\;
      
    % }
    % \Else{
    
      \texttt{instLocUntimedAsst}($claim\_ID$, $guardEvent\_1$, $guardEvent\_2$, $requiredEvent$, $requiredEvent\_wo$)\;
      
    % }
  }
  \Else{
  
    $guardEvent \leftarrow r.\texttt{guardCondition}$\;
    
    $requiredEvent \leftarrow r.\texttt{requiredCondition}$\;
    
    $requiredEvent\_wo \leftarrow requiredEvent.\texttt{removeValue()}$\;
    
    \texttt{instGlbUntimedAsst}($claim\_ID$, $guardEvent$, $requiredEvent$, $requiredEvent\_wo$)\;
    
  }
}
\end{algorithm}

% Next, the algorithm decides whether the local claim involves a buffer, i.e., whether the claim involves communication with the RoboChart platform.
% If the $requiredEvent$ is prefixed with `buffered'~(line~7), the requirement is of type local invariant with buffer. The assertion is then generated by instantiating the template AT-UTLB (line~8).
% Otherwise, the requirement is of type  local invariant without a buffer. In this case, 
Next, the assertion is generated by instantiating the template AT-2 (line~10).
If there is one element defined in ``GuardCondition''~(line~12), the requirement is a global type, and the assertion is generated by instantiating AT-1~(line~16).

\textbf{Algorithms}~\ref{Al:GenerateTimedAsst} and \ref{Al:GenerateGeneralAsst} also follow the same pattern:~fetching the elements to instantiate template.
Algorithm~\ref{Al:GenerateGeneralAsst} is for general assertions:~the claim uses the \texttt{every} Kapture template whose sections of ``Condition'' and ``AlwaysOrNever'' are used to identify and generate the elements of the assertion template.

\begin{algorithm}[h]
\caption{\textsc{GenerateTimedAsst}}
\label{Al:GenerateTimedAsst}
\SetAlgoLined
\DontPrintSemicolon
\SetKwProg{Proc}{Procedure}{:}{}

\Proc{\textsc{GenerateTimedAsst}($r$, $claim\_ID$)}{
    $targetEvent \leftarrow r.\texttt{requiredCondition}$\;
    
    $triggerEvent \leftarrow r.\texttt{triggerCondition}$\;
    
    $deadline \leftarrow r.\texttt{duration.getTimeLimit}()$\;
    
    \texttt{instTimedAsst}($claim\_ID$, $targetEvent$, $triggerEvent$, $deadline$)\;
}
\end{algorithm}

\begin{algorithm}[h]
\caption{\textsc{GenerateGeneralAsst}}
\label{Al:GenerateGeneralAsst}
\SetAlgoLined
\DontPrintSemicolon
\SetKwProg{Proc}{Procedure}{:}{}

\Proc{\textsc{GenerateGeneralAsst}($r$, $claim\_ID$)}{

    $asst\_type \leftarrow r.\texttt{condition.getType}()$\;

    $asstTimeFeature \leftarrow asst\_type.\texttt{time}$\;

    \eIf{$asst\_type = \texttt{reachable}$}{          % reachability
        $scope \leftarrow r.\texttt{condition.getScope}()$\;
        
        $state \leftarrow r.\texttt{condition.getState}()$\;
        
        $boolTerm \leftarrow r.\texttt{alwaysOrNever.generateBool}()$\;
        
        \texttt{instReachAsst}($claim\_ID$, $scope$, $state$, $boolTerm$, $asstTimeFeature$)\;
    }{
        \eIf{$asst\_type = \texttt{deadlock\_free}$}{
            $scope \leftarrow r.\texttt{condition.getScope}()$\;
            
            $boolTerm \leftarrow r.\texttt{alwaysOrNever.generateBool}()$\;
            
            \texttt{instDdlkfAsst}($claim\_ID$, $scope$, $boolTerm$, $asstTimeFeature$)\;
        }{
            \eIf{$asst\_type = \texttt{divergence\_free}$}{
                $scope \leftarrow r.\texttt{condition.getScope}()$\;
                
                $boolTerm \leftarrow r.\texttt{alwaysOrNever.}\allowbreak\texttt{generateBool}()$\;
                
                \texttt{instDivfAsst}($claim\_ID$, $scope$, $boolTerm$, $asstTimeFeature$)\;
            }{
                \If{$asst\_type = \texttt{terminate}$}{
                    $scope \leftarrow r.\texttt{condition.getScope}()$\;
                    
                    $boolTerm \leftarrow r.\texttt{alwaysOrNever.}\allowbreak\texttt{generateBool}()$\;
                    
                    \texttt{instTmntAsst}($claim\_ID$, $scope$, $boolTerm$, $asstTimeFeature$)\;
                }
            }
        }
    }
}
\end{algorithm}

\subsection{Assertions generated for unscrew robot}
\label{sec:appendix:unscrew_robot_assts}

\begin{figure}[H]
    \centering
    \fbox{\includegraphics[width=0.45\textwidth]{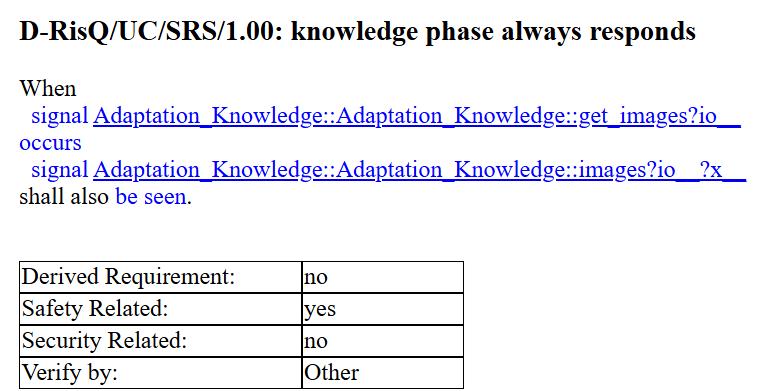}}
    \caption{DTI\_SR1 requirement in Kapture template}
    \label{fig:kap_DTI_R1}
\end{figure}

\begin{figure}[H]
    \centering
    \fbox{\includegraphics[width=0.45\textwidth]{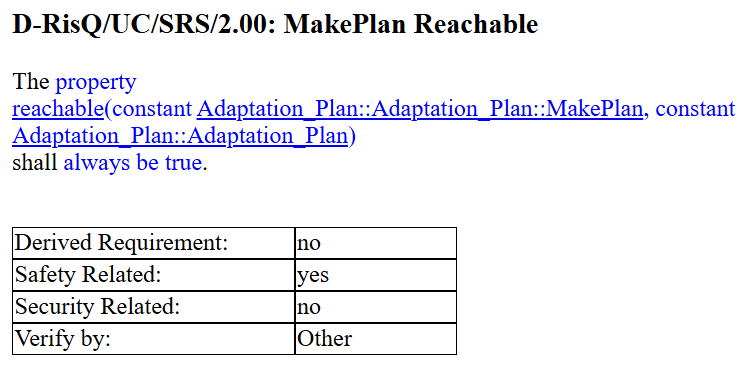}}
    \caption{DTI\_SR2 requirement in Kapture template}
    \label{fig:kap_DTI_R2}
\end{figure}

\begin{figure}[H]
    \centering
    \fbox{\includegraphics[width=0.45\textwidth]{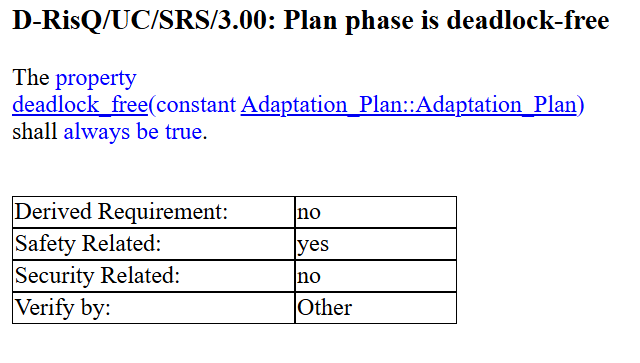}}
    \caption{DTI\_SR3 requirement in Kapture template}
    \label{fig:kap_DTI_R3}
\end{figure}

\begin{figure}[H]
    \centering
    \fbox{\includegraphics[width=0.45\textwidth]{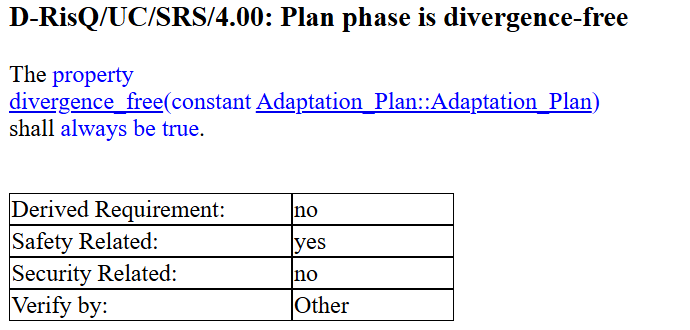}}
    \caption{DTI\_SR4 requirement in Kapture template}
    \label{fig:kap_DTI_R4}
\end{figure}

This appendix shows the Kapture requirements for the unscrew robot case study listed in Sect.~\ref{subsec:unscrew_robot} and the corresponding assertions automatically generated by our toolchain. 
SR1 is an untimed requirement that fits in the template RT-UNTIMED.
The general requirements SR2-SR4 fit in the templates RT-REACH/DDLK/DIV. Their Kapture requirements are shown in Fig.~\ref{fig:kap_DTI_R1}-\ref{fig:kap_DTI_R4}.

The assertions automatically generated from the four Kapture requirements are as follows.

\begin{tcolorbox}[colback=white, boxrule=0.5pt, arc=0pt, halign=flush left, left=2pt, right=2pt, top=2pt, bottom=2pt, boxsep=2pt]
 \renewcommand{\baselinestretch}{1.1}\selectfont
 \begin{footnotesize}
\textbf{\color{eclipsePurple}{untimed csp}} \textit{Spec} csp-begin\\

\textit{Spec} = CHAOS(Events) [|~ \\
    \{|~\textit{Adaptation\_Knowledge::Adaptation\_Knowledge::\\get\_image.in%?io\_\_
    }~|\}~|>\\
    (RUN(\\\{|~\textit{Adaptation\_Knowledge::Adaptation\_Knowledge::\\get\_images.in%?io\_\_ %io means in or out, more general than .in
    }~|\}) /\textbackslash (\textit{Adaptation\_Knowledge::Adaptation\_Knowledge::\\images.out?x\_\_} -> \textit{Spec}))\\
csp-end\\[4pt]
\textbf{\color{eclipsePurple}{untimed csp}} \textit{Spec}\_impl \textbf{\color{eclipsePurple}{associated to}} \\\textit{Adaptation\_Knowledge::Adaptation\_Knowledge} csp-begin\\
\textit{spec}\_impl = \\ \textit{Adaptation\_Knowledge::Adaptation\_Knowledge}::O\_\_(0) \\
csp-end \\[4pt]

\textbf{\color{eclipsePurple}{untimed assertion}} A\_\textit{DTI-1} :\\
\textit{Spec}\_impl \textbf{\color{eclipsePurple}{refines}} \textit{Spec} \textbf{\color{eclipsePurple}{in the traces model}}
\end{footnotesize}
\end{tcolorbox}

\begin{tcolorbox}[colback=white, boxrule=0.5pt, arc=0pt, halign=flush left, left=2pt, right=2pt, top=2pt, bottom=2pt, width=\columnwidth,boxsep=2pt]
 \renewcommand{\baselinestretch}{1.1}\selectfont
\begin{footnotesize}
\textbf{\color{eclipsePurple}{assertion}} A\_\textit{DTI-2}:\\ \textit{Adaptation\_Plan::Adaptation\_Plan::MakePlan} \textit{is} \textbf{\color{eclipsePurple}{reachable in}} \textit{Adaptation\_Plan::Adaptation\_Plan}. \\
\end{footnotesize}
\end{tcolorbox}

\vspace{10pt}

\begin{tcolorbox}[colback=white, boxrule=0.5pt, arc=0pt, halign=flush left, left=2pt, right=2pt, top=2pt, bottom=2pt, boxsep=2pt]
 \renewcommand{\baselinestretch}{1.1}\selectfont
\begin{footnotesize}
\textbf{\color{eclipsePurple}{assertion}} A\_\textit{DTI-3}:\\ 
\textit{Adaptation\_Plan::Adaptation\_Plan}
\textit{is}
\textbf{\color{eclipsePurple}{deadlock-free}}.\\
\end{footnotesize}
\end{tcolorbox}

\vspace{10pt}

\begin{tcolorbox}[colback=white, boxrule=0.5pt, arc=0pt, halign=flush left, left=2pt, right=2pt, top=2pt, bottom=2pt, boxsep=2pt]
 \renewcommand{\baselinestretch}{1.1}\selectfont
\begin{footnotesize}
\textbf{\color{eclipsePurple}{assertion}} A\_\textit{DTI-4}:\\ 
\textit{Adaptation\_Plan::Adaptation\_Plan}
\textit{is}
\textbf{\color{eclipsePurple}{divergence-free}}.\\
\end{footnotesize}
\end{tcolorbox}

%% file: espfor.bib
@MANUAL{RoboChart,
  title = {{RoboChart Reference Manual}},
  organization = {University of York},
  note = {{\url{www.cs.york.ac.uk/circus/RoboCalc/robotool/}}},
}


%% file: publications.bib
@ARTICLE{BCM22,
  author = {W.~Barnett and A.~L.~C.~Cavalcanti and A.~Miyazawa},
  title = {{Architectural Modelling for Robotics: RoboArch and the CorteX example}},
  journal = {Frontiers of Robotics and AI},
  year = {2022},
  doi = {https://doi.org/10.3389/frobt.2022.991637},
}

@ARTICLE{BRC22,
  title = "{Sound reasoning in tock-CSP}",
  journal = "Acta Informatica",
  year = "2022",
  author = "J.~Baxter and P.~Ribeiro and A.~L.~C.~Cavalcanti",
  doi = {10.1007/s00236-020-00394-3},
  volume = "59",
  pages = "125-162",
}

@ARTICLE{LABCGLMOPPTTZ24,
  title = {{Robotic safe adaptation in unprecedented situations: the RoboSAPIENS project}},
  volume = {2},
  doi = {10.1017/cbp.2024.4},
  journal = {Research Directions: Cyber-Physical Systems},
  author = {P.~G.~Larsen and S.~Ali and R.~Behrens and A.~.L.~C.~Cavalcanti and C.~Gomes and G.~Li and P.~Meulenaere and M.~L.~Olsen and N.~Passalis and T.~Peyrucain and J.~Tapia and A.~Tefas and H.~Zhang},
  year = {2024},
}


%% file: reference.bib
@INPROCEEDINGS{WC22,
  author = "M.~Windsor and A.~L.~C.~Cavalcanti",
  editor = "A.~Riesco and M.~Zhang",
  title = "{RoboCert: Property Specification in Robotics}",
  booktitle = "International Conference on Formal Engineering Methods",
  year = "2022",
  publisher = "Springer",
  series = {Lecture Notes in Computer Science},
  volume = {13478},
}

@article{mavridou2025automated,
  title   = {Automated Formalization of Probabilistic Requirements from Structured Natural Language},
  author  = {Mavridou, Anastasia and Farrell, Marie and V{\'a}zquez, Gricel and Pressburger, Tom and Wang, Timothy E and Calinescu, Radu and Fisher, Michael},
  journal = {arXiv preprint arXiv:2512.15788},
  year    = {2025},
  note    = {preprint}
}

@article{gleirscher2019new,
  title={New opportunities for integrated formal methods},
  author={Gleirscher, Mario and Foster, Simon and Woodcock, Jim},
  journal={ACM Computing Surveys (CSUR)},
  volume={52},
  number={6},
  pages={1--36},
  year={2019},
  publisher={ACM New York, NY, USA}
}

@article{murray2022safety,
  title={Safety assurance of an industrial robotic control system using hardware/software co-verification},
  author={Murray, Yvonne and Sirev{\aa}g, Martin and Ribeiro, Pedro and Anisi, David A and Mossige, Morten},
  journal={Science of Computer Programming},
  pages={102766},
  year={2022},
  publisher={Elsevier}
}

@phdthesis{yan2023assurance,
  title={Assurance Case Generation Using Model-based Engineering and Formal Verification},
  author={Yan, Fang},
  year={2023},
  school={University of York}
}

@misc{Miyazawa2016,
author = {Miyazawa, Alvaro and Ribeiro, Pedro and Ye, Kangfeng and Cavalcanti, Ana and Li, Wei and Woodcock, Jim and Timmis, Jon},
title = {{RoboChart Reference Manual}},
year = {2021}
}

@inproceedings{sljivo2023guided,
  title={Guided Integration of Formal Verification in Assurance Cases},
  author={Sljivo, Irfan and Denney, Ewen and Menzies, Jonathan},
  booktitle={International Conference on Formal Engineering Methods},
  pages={172--190},
  year={2023},
  organization={Springer}
}

@inproceedings{baxter2025formal,
  title={Formal Architectural Patterns for Adaptive Robotic Software},
  author={Baxter, James and Van Acker, Bert and Kristensen, Morten and Wright, Thomas and Cavalcanti, Ana and Gomes, Cl{\'a}udio},
  booktitle={International Conference on Fundamental Approaches to Software Engineering},
  pages={145--165},
  year={2025},
  organization={Springer Nature Switzerland Cham}
}

@article{kephart2003vision,
  title={The vision of autonomic computing},
  author={Kephart, Jeffrey O and Chess, David M},
  journal={Computer},
  volume={36},
  number={1},
  pages={41--50},
  year={2003},
  publisher={IEEE}
}

@misc{prism,
title = {{PRISM Manual}},
howpublished = {\url{https://www.prismmodelchecker.org/manual}},
     year = {2017},
note = {Online; accessed 04th May, 2022}
}

@misc{fdr4.2,
title = {{FDR 4.2 Documentation}},
howpublished = {\url{https://cocotec.io/fdr/manual/index.html}},
  year = {2020},
note = {Online; accessed 04th May, 2022}
}

@Book{Roscoe1998,
  title     = {The theory and practice of concurrency},
  publisher = {Prentice Hall},
  year      = {1998},
  author    = {Roscoe, Andrew William},
}

@techreport{robosapiensD12,
  author      = {Baxter, James and Cavalcanti, Ana and Yan, Fang},
  title       = {{Operational Requirements for Case Studies}},
  institution = {University of York},
  type        = {{RoboSapiens Project Deliverable}},
  version     = {1},
  year        = {2025},
  month       = jun,
  note        = {\url{https://drive.google.com/file/d/1TkCEECTpOhxoKph_dM2neyYIyaiydjKa/view}},
  urldate     = {2025-02-17}
}

@article{Bloomfield1998,
author = {Bloomfield, R E and Bishop, P G and Jones, C and Froome, P K D},
journal = {Adelard, 1998. ISBN 0-9533771-0},
title = {{Ascad—adelard safety case development manual}},
note = 
{\url{https://usermanual.wiki/Document/EUROCONTROL20Safetycasedevelopmentmanual.687924513/html}},
volume = {5},
year = {1998}
}

@techreport{AssuranceCaseWorkingGroup,
author = {{OMG}},
title = {{Goal Structuring Notation
Community Standard. Version 3}},
url = {https://scsc.uk/r141C:1?t=1},
year = {2021},
institution = {{OMG}}
}

@misc{standard2011iso,
    author={{ISO}},
    title={{ISO 26262 Road vehicles--Functional Safety, Version 1}},
  year={2011}
}

@book{nipkow2002isabelle,
  title={{Isabelle/HOL: a proof assistant for higher-order logic}},
  author={Nipkow, Tobias and Paulson, Lawrence C and Wenzel, Markus},
  volume={2283},
  year={2002},
  publisher={Springer Science \& Business Media}
}

@inproceedings{ghosh2016arsenal,
  title={ARSENAL: automatic requirements specification extraction from natural language},
  author={Ghosh, Shalini and Elenius, Daniel and Li, Wenchao and Lincoln, Patrick and Shankar, Natarajan and Steiner, Wilfried},
  booktitle={NASA Formal Methods: 8th International Symposium, NFM 2016, Minneapolis, MN, USA, June 7-9, 2016, Proceedings 8},
  pages={41--46},
  year={2016},
  organization={Springer}
}

@inproceedings{gibson2014fdr3,
  title={{FDR3—a modern refinement checker for CSP}},
  author={Gibson-Robinson, Thomas and Armstrong, Philip and Boulgakov, Alexandre and Roscoe, Andrew W},
  booktitle={International conference on tools and algorithms for the construction and analysis of systems},
  pages={187--201},
  year={2014},
  organization={Springer}
}

@article{standard1standard,
  title={Standard 00-55 Part 1 (1997)},
  author={Standard, Defence},
  journal={Requirements for safety related software in defence equipment, Part},
  volume={1},
  year = {1997}
}

@book{Roscoe2010,
author = {Roscoe, Andrew William},
isbn = {1848822588},
publisher = {Springer Science {\&} Business Media},
title = {{Understanding concurrent systems}},
year = {2010}
}

@inproceedings{yan2023automated,
  title={{Automated Compositional Verification for Robotic State Machines using Isabelle/HOL}},
  author={Yan, Fang and Foster, Simon David and Habli, Ibrahim},
  booktitle={27th International Conference on Engineering of Complex Computer Systems},
  year={2023},
  organization={IEEE}
}

@inproceedings{yan2022model,
  title={Model-Based Generation of Hazard-Driven Arguments and Formal Verification Evidence for Assurance Cases},
  author={Yan, Fang and Foster, Simon David and Habli, Ibrahim and Wei, Ran},
  booktitle={10th International Conference on Model-Driven Engineering and Software Development},
  pages={252--263},
  year={2022},
  organization={SciTePress}
}

@article{Miyazawa2019a,
author = {Miyazawa, Alvaro and Ribeiro, Pedro and Li, Wei and Cavalcanti, Ana and Timmis, Jon and Woodcock, Jim},
doi = {10.1007/s10270-018-00710-z},
issn = {16191374},
journal = {Software and Systems Modeling},
number = {5},
pages = {3097--3149},
publisher = {Springer Berlin Heidelberg},
title = {{RoboChart: modelling and verification of the functional behaviour of robotic applications}},
url = {https://doi.org/10.1007/s10270-018-00710-z},
volume = {18},
year = {2019}
}

@inproceedings{jee2010assurance,
  title={Assurance cases in model-driven development of the pacemaker software},
  author={Jee, Eunkyoung and Lee, Insup and Sokolsky, Oleg},
  booktitle={Leveraging Applications of Formal Methods, Verification, and Validation: 4th International Symposium on Leveraging Applications, ISoLA 2010, Heraklion, Crete, Greece, October 18-21, 2010, Proceedings, Part II 4},
  pages={343--356},
  year={2010},
  organization={Springer}
}

@article{Wei2019,
author = {Wei, Ran and Kelly, Tim P. and Dai, Xiaotian and Zhao, Shuai and Hawkins, Richard},
doi = {10.1016/j.jss.2019.05.013},
issn = {01641212},
journal = {Journal of Systems and Software},
month = {Aug},
pages = {211--233},
publisher = {Elsevier Inc.},
title = {{Model based system assurance using the structured assurance case metamodel}},
volume = {154},
year = {2019}
}

@article{Prokhorova2015,
author = {Prokhorova, Yuliya and Laibinis, Linas and Troubitsyna, Elena},
doi = {10.1016/j.infsof.2015.01.001},
issn = {09505849},
journal = {Information and Software Technology},
pages = {51--76},
publisher = {Elsevier B.V.},
title = {{Facilitating construction of safety cases from formal models in Event-B}},
volume = {60},
year = {2015}
}

@inproceedings{Hawkins2015a,
author = {Hawkins, Richard and Habli, Ibrahim and Kolovos, Dimitris and Paige, Richard and Kelly, Tim},
booktitle = {2015 IEEE 16th International Symposium on High Assurance Systems Engineering},
doi = {10.1109/HASE.2015.25},
isbn = {1479981117},
issn = {15302059},
pages = {110--117},
publisher = {IEEE},
title = {{Weaving an Assurance Case from Design: A Model-Based Approach}},
year = {2015}
}

@article{miyazawa2019robochart,
  title={{RoboChart: modelling and verification of the functional behaviour of robotic applications}},
  author={Miyazawa, Alvaro and Ribeiro, Pedro and Li, Wei and Cavalcanti, Ana and Timmis, Jon and Woodcock, Jim},
  journal={Software \& Systems Modeling},
  volume={18},
  number={5},
  pages={3097--3149},
  year={2019},
  publisher={Springer}
}

@inproceedings{kolovos2008epsilon,
  title={{The Epsilon transformation language}},
  author={Kolovos, Dimitrios S and Paige, Richard F and Polack, Fiona AC},
  booktitle={International Conference on Theory and Practice of Model Transformations},
  pages={46--60},
  year={2008},
  organization={Springer}
}

@misc{ObjectManagementGroupOMG,
author = {{OMG}},
number = {Version 2.1 beta},
title = {{Structured Assurance Case Metamodel (SACM), Version 2.1 beta}},
year = {2020}
}

@article{Denney2018,
author = {Denney, Ewen and Pai, Ganesh},
doi = {10.1007/s10515-017-0230-5},
issn = {0928-8910},
journal = {Automated Software Engineering},
number = {3},
pages = {435--499},
publisher = {Springer},
title = {{Tool support for assurance case development}},
volume = {25},
year = {2018}
}

@book{roscoe1998theory,
  title        = {The Theory and Practice of Concurrency},
  author       = {Roscoe, Anthony W.},
  year         = {1998},
  publisher    = {Prentice-Hall},
  series       = {Prentice-Hall Series in Computer Science}
}

@article{foster2021integration,
  title={{Integration of formal proof into unified assurance cases with Isabelle/SACM}},
  author={Foster, Simon and Nemouchi, Yakoub and Gleirscher, Mario and Wei, Ran and Kelly, Tim},
  journal={Formal Aspects of Computing},
  volume={33},
  number={6},
  pages={855--884},
  year={2021},
  publisher={Springer}
}

@article{Nemouchi2019,
author = {Nemouchi, Yakoub and Foster, Simon and Gleirscher, Mario and Kelly, Tim},
doi = {10.1007/978-3-030-34968-4_21},
isbn = {9783030349677},
issn = {16113349},
journal = {Lecture Notes in Computer Science (including subseries Lecture Notes in Artificial Intelligence and Lecture Notes in Bioinformatics)},
pages = {379--398},
title = {{Isabelle/SACM: Computer-Assisted Assurance Cases with Integrated Formal Methods}},
volume = {11918 LNCS},
year = {2019}
}

@article{Calinescu2018b,
author = {Calinescu, Radu and Weyns, Danny and Gerasimou, Simos and Iftikhar, Muhammad Usman and Habli, Ibrahim and Kelly, Tim},
doi = {10.1109/TSE.2017.2738640},
issn = {19393520},
journal = {IEEE Transactions on Software Engineering},
month = {Nov},
number = {11},
pages = {1039--1069},
publisher = {Institute of Electrical and Electronics Engineers Inc.},
title = {{Engineering trustworthy self-adaptive software with dynamic assurance cases}},
volume = {44},
year = {2018}
}

@article{bolton2014automatically,
  title={Automatically generating specification properties from task models for the formal verification of human--automation interaction},
  author={Bolton, Matthew L and Jim{\'e}nez, Noelia and van Paassen, Marinus M and Trujillo, Maite},
  journal={IEEE Transactions on Human-Machine Systems},
  volume={44},
  number={5},
  pages={561--575},
  year={2014},
  publisher={IEEE}
}

@article{akesson2022comprehensive,
  title={A comprehensive survey of industry practice in real-time systems},
  author={Akesson, Benny and Nasri, Mitra and Nelissen, Geoffrey and Altmeyer, Sebastian and Davis, Robert I},
  journal={Real-Time Systems},
  volume={58},
  number={3},
  pages={358--398},
  year={2022},
  publisher={Springer}
}

@techreport{bourbouh2020integration,
  title={Integration and Evaluation of the AdvoCATE, FRET, CoCoSim, and Event-B Tools on the Inspection Rover Case Study},
  author={Bourbouh, Hamza and Farrell, Marie and Mavridou, Anastasia and Sljivo, Irfan},
  year={2020}
}

@inproceedings{Bourbouh2021a,
author = {Bourbouh, Hamza and Farrell, Marie and Mavridou, Anastasia and Sljivo, Irfan and Brat, Guillaume and Dennis, Louise A. and Fisher, Michael},
booktitle = {NASA Formal Methods Symposium},
doi = {10.1007/978-3-030-76384-8_4},
isbn = {9783030763831},
issn = {16113349},
pages = {53--71},
publisher = {Springer},
title = {{Integrating Formal Verification and Assurance: An Inspection Rover Case Study}},
volume = {12673 LNCS},
year = {2021}
}

@inproceedings{giannakopoulou2020formal,
  title={{Formal requirements elicitation with FRET}},
  author={Giannakopoulou, Dimitra and Mavridou, Anastasia and Rhein, Julian and Pressburger, Thomas and Schumann, Johann and Shi, Nija},
  booktitle={International Working Conference on Requirements Engineering: Foundation for Software Quality (REFSQ-2020)},
  isbn={ARC-E-DAA-TN77785},
  year={2020}
}

@article{wei2024access,
  title={ACCESS: Assurance case centric engineering of safety--critical systems},
  author={Wei, Ran and Foster, Simon and Mei, Haitao and Yan, Fang and Yang, Ruizhe and Habli, Ibrahim and O’Halloran, Colin and Tudor, Nick and Kelly, Tim and Nemouchi, Yakoub},
  journal={Journal of Systems and Software},
  volume={213},
  pages={112034},
  year={2024},
  publisher={Elsevier}
}

@article{wei2023automated,
  title={Automated Model-Based Assurance Case Management Using Constrained Natural Language},
  author={Wei, Ran and Jiang, Zhe and Mei, Haitao and Barmpis, Konstantinos and Foster, Simon and Kelly, Tim and Zhuang, Yan},
  journal={IEEE Transactions on Computer-Aided Design of Integrated Circuits and Systems},
  volume={43},
  number={1},
  pages={291--304},
  year={2023},
  publisher={IEEE}
}

@inproceedings{Lindoso2021a,
author = {Lindoso, Waldeck and Nogueira, Sidney C. and Domingues, Renato and Lima, Lucas},
booktitle = {Brazilian Symposium on Formal Methods},
doi = {10.1007/978-3-030-92137-8_3},
isbn = {9783030921378},
pages = {34--52},
publisher = {Springer},
title = {{Visual Specification of Properties for Robotic Designs}},
year = {2021}
}

@phdthesis{kelly1999arguing,
  title={Arguing safety: a systematic approach to managing safety cases},
  author={Kelly, Timothy Patrick},
  year={1999},
  school={University of York York, UK}
}

@inproceedings{sorokin2024towards,
  title={Towards continuous assurance case creation for ads with the evidential tool bus},
  author={Sorokin, Lev and Bouchekir, Radouane and Beyene, Tewodros A and Liao, Brian Hsuan-Cheng and Molin, Adam},
  booktitle={European Dependable Computing Conference},
  pages={49--61},
  year={2024},
  organization={Springer}
}

@inproceedings{xu2024learning,
  title={Learning from failures: Translation of natural language requirements into linear temporal logic with large language models},
  author={Xu, Yilongfei and Feng, Jincao and Miao, Weikai},
  booktitle={2024 IEEE 24th International Conference on Software Quality, Reliability and Security (QRS)},
  pages={204--215},
  year={2024},
  organization={IEEE}
}

@inproceedings{cosler2023nl2spec,
  title={nl2spec: Interactively translating unstructured natural language to temporal logics with large language models},
  author={Cosler, Matthias and Hahn, Christopher and Mendoza, Daniel and Schmitt, Frederik and Trippel, Caroline},
  booktitle={International Conference on Computer Aided Verification},
  pages={383--396},
  year={2023},
  organization={Springer}
}

@article{li2025automatic,
  title={Automatic Generation of Safety-compliant Linear Temporal Logic via Large Language Model: A Self-supervised Framework},
  author={Li, Junle and Tian, Meiqi and Zhong, Bingzhuo},
  journal={arXiv preprint arXiv:2503.15840},
  year={2025}
}

@inproceedings{zhao2024nl2ctl,
  title={Nl2ctl: automatic generation of formal requirements specifications via large language models},
  author={Zhao, Mengyan and Tao, Ran and Huang, Yanhong and Shi, Jianqi and Qin, Shengchao and Yang, Yang},
  booktitle={International Conference on Formal Engineering Methods},
  pages={1--17},
  year={2024},
  organization={Springer}
}

@article{odu2025automatic,
  title={Automatic instantiation of assurance cases from patterns using large language models},
  author={Odu, Oluwafemi and Belle, Alvine B and Wang, Song and Kpodjedo, Segla and Lethbridge, Timothy C and Hemmati, Hadi},
  journal={Journal of Systems and Software},
  volume={222},
  pages={112353},
  year={2025},
  publisher={Elsevier}
}

@inproceedings{meng2021automating,
  title={Automating the assembly of security assurance case fragments},
  author={Meng, Baoluo and Paul, Saswata and Moitra, Abha and Siu, Kit and Durling, Michael},
  booktitle={International Conference on Computer Safety, Reliability, and Security},
  pages={101--114},
  year={2021},
  organization={Springer}
}

@inproceedings{foster2020formal,
  title={Formal model-based assurance cases in {Isabelle/SACM}: An autonomous underwater vehicle case study},
  author={Foster, Simon and Nemouchi, Yakoub and O'Halloran, Colin and Stephenson, Karen and Tudor, Nick},
  booktitle={Proceedings of the 8th International Conference on Formal Methods in Software Engineering},
  pages={11--21},
  year={2020}
}

@misc{pls,
    title = {{PRISM Lab Session, Part B: Mail Delivery Robot}},
        howpublished = {\url{http://www. prismmodelchecker.org/courses/aims1617/deliveryRobot.php}},
             year = {2017},
    note = {online; accessed 5th May, 2022}
}

@article{hilder2012chemical,
  title={Chemical detection using the receptor density algorithm},
  author={Hilder, James A and Owens, Nick DL and Neal, Mark J and Hickey, Peter J and Cairns, Stuart N and Kilgour, David PA and Timmis, Jon and Tyrrell, Andy M},
  journal={IEEE Transactions on Systems, Man, and Cybernetics, Part C (Applications and Reviews)},
  volume={42},
  number={6},
  pages={1730--1741},
  year={2012},
  publisher={IEEE}
}

@inproceedings{bohme2010sledgehammer,
  title={Sledgehammer: judgement day},
  author={B{\"o}hme, Sascha and Nipkow, Tobias},
  booktitle={International Joint Conference on Automated Reasoning},
  pages={107--121},
  year={2010},
  organization={Springer}
}

@article{ye2022probabilistic,
  title={{Probabilistic modelling and verification using RoboChart and PRISM}},
  author={Ye, Kangfeng and Cavalcanti, Ana and Foster, Simon and Miyazawa, Alvaro and Woodcock, Jim},
  journal={Software and Systems Modeling},
  volume={21},
  number={2},
  pages={667--716},
  year={2022},
  publisher={Springer},
doi = {10.1007/s10270-021-00916-8}
}

@inproceedings{kwiatkowska2002prism,
  title={{PRISM: Probabilistic symbolic model checker}},
  author={Kwiatkowska, Marta and Norman, Gethin and Parker, David},
  booktitle={International Conference on Modelling Techniques and Tools for Computer Performance Evaluation},
  pages={200--204},
  year={2002},
  organization={Springer}
}

@Book{Abrial96BBook,
  bibkey	= {abrial:b-book:1996},
  author	= {Jean-Raymond Abrial},
  title		= {The {B-Book}: assigning programs to meanings},
  year		= 1996,
  publisher	= {Cambridge University Press}
}

@article{bouskela2022formal,
  title={Formal requirements modeling for cyber-physical systems engineering: An integrated solution based on FORM-L and Modelica},
  author={Bouskela, Daniel and Falcone, Alberto and Garro, Alfredo and Jardin, Audrey and Otter, Martin and Thuy, Nguyen and Tundis, Andrea},
  journal={Requirements Engineering},
  volume={27},
  number={1},
  pages={1--30},
  year={2022},
  publisher={Springer}
}

@BOOK{Spivey89,
  KEY           = "Spivey89",
  TITLE         = "The Z-Notation - A Reference Manual",
  PUBLISHER     = "Prentice Hall",
  YEAR          = "1989",
  AUTHOR        = "Spivey, M.",
  ADDRESS       = "Englewood Cliffs, N. J.",
  COMMENT       = "NA Referenced in IPTES-PDM-27 PGL"}

@inproceedings{dwyer1999patterns,
  author    = {Dwyer, Matthew B. and Avrunin, George S. and Corbett, James C.},
  title     = {Patterns in Property Specifications for Finite-State Verification},
  booktitle = {Proceedings of the 21st International Conference on Software Engineering},
  pages     = {411--420},
  year      = {1999},
  publisher = {IEEE/ACM}
}

@inproceedings{konrad2005real,
  author    = {Konrad, Sascha and Cheng, Betty H. C.},
  title     = {Real-Time Specification Patterns},
  booktitle = {Proceedings of the 27th International Conference on Software Engineering},
  pages     = {372--381},
  year      = {2005},
  publisher = {ACM}
}

@inproceedings{grunske2008specification,
  author    = {Grunske, Lars},
  title     = {Specification Patterns for Probabilistic Quality Properties},
  booktitle = {Proceedings of the 30th International Conference on Software Engineering},
  pages     = {31--40},
  year      = {2008},
  publisher = {ACM}
}

@article{menghi2019specification,
  author    = {Menghi, Claudio and Tsigkanos, Christos and Pelliccione, Patrizio and Ghezzi, Carlo and Berger, Thorsten},
  title     = {Specification Patterns for Robotic Missions},
  journal   = {IEEE Transactions on Software Engineering},
  volume    = {47},
  number    = {10},
  pages     = {2208--2224},
  year      = {2021},
  publisher = {IEEE}
}

@article{menghi2022mission,
  author    = {Menghi, Claudio and Tsigkanos, Christos and Askarpour, Mehrnoosh and Pelliccione, Patrizio and V{\'a}zquez, Gricel and Calinescu, Radu and Garc{\'i}a, Sergio},
  title     = {Mission Specification Patterns for Mobile Robots: Providing Support for Quantitative Properties},
  journal   = {IEEE Transactions on Software Engineering},
  volume    = {49},
  number    = {4},
  pages     = {2741--2760},
  year      = {2023},
  publisher = {IEEE}
}

@article{autili2015aligning,
  author    = {Autili, Marco and Grunske, Lars and Lumpe, Markus and Pelliccione, Patrizio and Tang, Antony},
  title     = {Aligning Qualitative, Real-Time, and Probabilistic Property Specification Patterns Using a Structured English Grammar},
  journal   = {IEEE Transactions on Software Engineering},
  volume    = {41},
  number    = {7},
  pages     = {620--638},
  year      = {2015},
  publisher = {IEEE}
}

@inproceedings{etumi2025towards,
  author    = {Etumi, Mahdi and Taylor, Hazel M. and Farrell, Marie},
  title     = {Towards A Catalogue of Requirement Patterns for Space Robotic Missions},
  booktitle = {International Workshop on Formal Methods for Autonomous Systems},
  year      = {2025}
}

@inproceedings{viger2020just,
  author    = {Viger, T. and Salay, R. and Selim, G. and Chechik, M.},
  title     = {Just Enough Formality in Assurance Argument Structures},
  booktitle = {International Conference on Computer Safety, Reliability, and Security (SAFECOMP)},
  pages     = {34--49},
  year      = {2020},
  publisher = {Springer}
}

@inproceedings{disandro2020mmint,
  author    = {Di Sandro, A. and Selim, G. and Salay, R. and Viger, T. and Chechik, M. and Kokaly, S.},
  title     = {{MMINT-A} 2.0: Tool Support for the Lifecycle of Model-Based Safety Artifacts},
  booktitle = {Proceedings of the 23rd ACM/IEEE International Conference on Model Driven Engineering Languages and Systems: Companion Proceedings},
  pages     = {1--5},
  year      = {2020},
  publisher = {ACM}
}

@inproceedings{chechik2020uncertainty,
  author    = {Chechik, M. and Kokaly, S. and Rahimi, M. and Salay, R. and Viger, T.},
  title     = {Uncertainty, Modeling and Safety Assurance: Towards a Unified Framework},
  booktitle = {Working Conference on Verified Software: Theories, Tools, and Experiments (VSTTE)},
  pages     = {19--29},
  year      = {2020},
  publisher = {Springer}
}

@inproceedings{carlan2022automating,
  author    = {C{\^a}rlan, C. and Gauerhof, L. and Gallina, B. and Burton, S.},
  title     = {Automating Safety Argument Change Impact Analysis for Machine Learning Components},
  booktitle = {2022 IEEE 27th Pacific Rim International Symposium on Dependable Computing (PRDC)},
  pages     = {43--53},
  year      = {2022},
  publisher = {IEEE}
}

@article{wassyng2015product,
  author    = {Wassyng, A. and Singh, N. K. and Geven, M. and Proscia, N. and Wang, H. and Lawford, M. and Maibaum, T.},
  title     = {Can Product-Specific Assurance Case Templates Be Used as Medical Device Standards?},
  journal   = {IEEE Design \& Test},
  volume    = {32},
  number    = {5},
  pages     = {45--55},
  year      = {2015},
  publisher = {IEEE}
}

@inproceedings{chowdhury2017principles,
  author    = {Chowdhury, T. and Lin, C. W. and Kim, B. and Lawford, M. and Shiraishi, S. and Wassyng, A.},
  title     = {Principles for Systematic Development of an Assurance Case Template from {ISO} 26262},
  booktitle = {2017 IEEE International Symposium on Software Reliability Engineering Workshops (ISSREW)},
  pages     = {69--72},
  year      = {2017},
  publisher = {IEEE}
}

@inproceedings{fung2018mmint,
  author    = {Fung, Nick L. S. and Kokaly, Sahar and Di Sandro, Alessio and Salay, Rick and Chechik, Marsha},
  title     = {{MMINT-A}: A Tool for Automated Change Impact Assessment on Assurance Cases},
  booktitle = {SAFECOMP 2018 Workshops},
  series    = {Lecture Notes in Computer Science},
  volume    = {11094},
  pages     = {60--70},
  publisher = {Springer},
  year      = {2018},
  doi       = {10.1007/978-3-319-99229-7_7}
}
